\documentclass[aps,a4paper,showkeys,nofootinbib,longbibliography,notitlepage]{revtex4-2}
\usepackage[utf8]{inputenc}
\usepackage{filecontents}
\usepackage{natbib}
\usepackage{amsmath,amssymb,bm,amsthm}
\usepackage{subfigure}
\usepackage{graphicx}% Include figure files
\usepackage{dcolumn}% Align table columns on decimal point
\usepackage{bm}% bold math
\usepackage[mathlines]{lineno}% Enable numbering of text and display math
\usepackage{booktabs}
\usepackage{color}
\newcounter{one}
\usepackage{url}
\usepackage{titlesec}
\titleformat{\section}[block]{\normalfont\large\bfseries}{\thesection}{1em}{}
\usepackage{ragged2e}

\usepackage{textcase}

\usepackage{braket}

\usepackage{colortbl}
\usepackage{tabularx}
\usepackage{verbatim}
\usepackage{multirow}

\usepackage[T1]{fontenc} 
\usepackage{lmodern}
\usepackage{bbm}
\usepackage[utf8]{inputenc}
\usepackage{amsfonts}
\usepackage{array}

\usepackage{stmaryrd}

\usepackage{bbm}
\usepackage{enumitem}
\usepackage{umoline}
\usepackage[usenames,svgnames]{xcolor}
\usepackage{natbib}
\usepackage[hyperindex,breaklinks]{hyperref}
\hypersetup{
     colorlinks=true,       		% false: boxed links; true: colored links
     linkcolor=Navy,          	% color of internal links
     citecolor=Navy,            % color of links to bibliography
     filecolor=Navy,      		% color of file links
     urlcolor=Navy,           	% color of external links
    runcolor=cyan,
 }
\usepackage{graphicx}
\usepackage{amsfonts}
\usepackage{amssymb}
\usepackage{amsmath}
\usepackage{bbm}
\usepackage{enumitem}

\newcommand{\tr}[0]{ {\rm tr}}

\newcommand{\half}[1]{{ \rm h}}
\newcommand{\Oorderof}{\mathcal{O}}
\newcommand{\orderof}[1]{\Oorderof(#1)} 

\newcommand{\for}[0]{\quad \textrm{for} \quad}

\newcommand{\dist}{d}

\newcommand{\co}{{\rm c}}

\usepackage{yhmath}

\newcommand{\Cor}{{\rm Cor}}

\newcommand{\ad}{{\rm ad}}

\newcommand{\AND}{\quad {\rm and} \quad}

\def\beq{\begin{equation}}
\def\eeq{\end{equation}}
\def\nbeq{\begin{equation*}}
\def\neeq{\end{equation*}}
\def\<{\langle}
\def\>{\rangle}

\def\tr{{\rm tr}}

\newcommand{\Gc}{\mathcal{G}}

\newcommand{\mD}{d}

\newcommand{\Der}{{\mathcal{D}}}

\newcommand{\ban}{\mathcal{B}}
\newcommand{\mI}{{\mathcal{I}}}

\newcommand{\mQ}{{\mathcal{Q}}}

\newcommand{\mL}{{\mathcal{L}}}
\newcommand{\mA}{{\mathcal{A}}}
\newcommand{\mB}{{\mathcal{B}}}

\newtheorem{theorem}{Theorem}
\newtheorem{subtheorem}{Subtheorem}
\newtheorem{lemma}{Lemma}
\newtheorem{corol}[lemma]{Corollary}
\newtheorem{assump}[lemma]{Assumption} 
\newtheorem{definition}{Definition}  
\newtheorem{prop}[lemma]{Proposition} 
\newtheorem{conj}{Conjecture} 
\newtheorem{conj'}{Conditional version of Conjecture} 

\newcommand{\bal}[2]{#1[#2]}

\newcommand{\br}[1]{\left( #1 \right)}
\newcommand{\brr}[1]{\left[ #1 \right]}
\newcommand{\brrr}[1]{\left\{ #1 \right\}}
 \newcommand{\norm}[1]{\left \|  #1 \right \|}

\newcommand{\abs}[1]{\left| #1 \right|}

\usepackage{mathtools}
\def\multiset#1#2{\ensuremath{\left(\kern-.3em\left(\genfrac{}{}{0pt}{}{#1}{#2}\right)\kern-.3em\right)}}

\definecolor{KKgreen}{RGB}{0,200,0}

\begin{document}

%\twocolumn[ \begin{@twocolumnfalse}

%\title{Approximate quantum Markovianity in steady states of rapidly mixing Liouvillians}
\title{Clustering of Conditional Mutual Information via Quantum \\ Belief-Propagation Channels}

%Polynomial growth of quantum scrambling in long-range...
%Polynomial operator spreading in 
%with locally bounded Hamiltonian

\author{Kohtaro Kato$^1$}
\email{kokato@i.nagoya-u.ac.jp}

\author{Tomotaka Kuwahara$^{2,3,4}$}
\email{tomotaka.kuwahara@riken.jp}
%\altaffiliation{Present address: Mathematical Science Team, RIKEN Center for Advanced Intelligence Project (AIP),1-4-1 Nihonbashi, Chuo-ku, Tokyo 103-0027, Japan}
\affiliation{$^{1}$Department of Mathematical Informatics, Graduate School of Informatics, Nagoya University, Nagoya 464-0814, Japan}

\affiliation{$^{2}$
Analytical Quantum Complexity RIKEN Hakubi Research Team, RIKEN Center for Quantum Computing (RQC), Wako, Saitama 351-0198, Japan
}

\affiliation{$^{3}$
RIKEN Pioneering Research Institute (PRI), Wako, Saitama 351-0198, Japan
}

\affiliation{$^{4}$
PRESTO, Japan Science and Technology (JST), Kawaguchi, Saitama 332-0012, Japan}
%\affiliation{$^{2}$Department of Mathematics, Faculty of Science and Technology, Keio University, 3-14-1 Hiyoshi, Kouhoku-ku, Yokohama 223-8522, Japan}

\begin{abstract}

Conditional mutual information (CMI) has recently attracted significant attention as a key quantity for characterizing quantum correlations in many-body systems. While it is conjectured that CMI decays rapidly in finite-temperature Gibbs states, a complete and general proof remains elusive. 
In this work, we introduce a new formulation of the problem based on the \emph{belief propagation (BP) channel}, namely a completely positive trace-preserving (CPTP) map that realizes local perturbations of the Hamiltonian. 
Within this framework, we prove that establishing the quasi-locality of BP channels implies the decay of CMI, thereby reducing the original conjecture to a more tractable problem. 
We show that such quasi-local BP channels can be constructed under natural physical conditions, such as uniform rapid mixing or uniform clustering. 
Under these assumptions, we obtain conditional proofs of CMI decay valid at all temperatures. Moreover, because these assumptions are automatically satisfied at high temperatures, our results in that regime yield unconditional proofs of CMI decay.
At the same time, in order to better understand the high-temperature behavior of Gibbs states, we revisit the cluster expansion method. 
Contrary to common intuition, we demonstrate that when multipartite correlations such as CMI are considered, the cluster expansion suffers from intrinsic divergence problems rooted in the Baker--Campbell--Hausdorff formula, revealing fundamental limitations of this traditional approach.

%Conditional mutual information (CMI) has recently attracted significant attention as a key quantity for characterizing quantum correlations in many-body systems. While it is conjectured that CMI decays rapidly in finite-temperature Gibbs states, a complete and general proof remains elusive. Previous work addressed this problem in the high-temperature regime using cluster expansion techniques [T. Kuwahara, K. Kato, F.G.S.L. Brand\~ao, Phys. Rev. Lett. 124, 220601 (2020)]; however, flaws in the proof have been pointed out, and the method does not provide a uniformly convergent expansion at arbitrarily high temperatures.
%In this work, we demonstrate that the cluster expansion approach indeed fails to converge absolutely, even at sufficiently high temperatures. To overcome this limitation, we propose a new approach to proving the spatial decay of CMI. Our method leverages the connection between CMI and quantum recovery maps, specifically utilizing the Fawzi-Renner theorem. We show that such recovery maps can be realized through dissipative dynamics, and by analyzing the locality properties of these dynamics, we establish the exponential decay of CMI in high-temperature regimes. As a technical contribution, we also present a new result on the perturbative stability of quasi-local Liouvillian dynamics.
%Our results indicate that, contrary to common intuition, high-temperature Gibbs states can exhibit nontrivial mathematical structure, particularly when multipartite correlations such as CMI are considered.
\end{abstract}

\maketitle

\tableofcontents
%%%%%%%%%%%%%%%%%%%%%%%%%%%%%%%%%%%%%%%%%%%%%%%%%%%%%%%%%%%%%%%%%%%%%%%%%%

\section{Introduction} \label{Sec:intro}

One of the central goals of quantum many-body theory is to uncover universal principles that apply regardless of the microscopic details of individual systems. In recent years, information-theoretic tools have emerged as powerful means to characterize such universal behavior, with quantities like mutual information and quantum entanglement playing critical roles in understanding correlations between subsystems. It is widely recognized that these bipartite information measures exhibit clustering---i.e., exponential decay with spatial separation---away from critical points~\cite{Araki1969,Gross1979,Park1995,ueltschi2004cluster,PhysRevX.4.031019,frohlich2015some,ref:Marco_LD,Netocny2004,Bluhm2022exponentialdecayof,Perez-Garcia2023,Kimura2025}. Although there have been proposals for information-theoretic quantities that exhibit exponential decay even at low temperatures~\cite{PhysRevLett.93.126402,PhysRevLett.117.130401,PhysRevX.12.021022}, clustering is generally expected to break down in this regime due to the emergence of long-range correlations. In contrast, it is widely assumed, although often not explicitly stated, that at sufficiently high temperatures, all physically relevant correlations decay rapidly with distance, reflecting the underlying locality of thermal equilibrium states.
This viewpoint was indeed mathematically formulated in terms of the cluster expansion technique~\cite{Kotecky1986,10.1145/3357713.3384322, KUWAHARA2020168278,PRXQuantum.4.020340,Haah2024,PRXQuantum.5.010305,10756136,tong2025}.

Among the information-theoretic quantities that have attracted increasing attention in recent years, a particularly important one is the \emph{conditional mutual information} (CMI). 
This quantity has emerged as a central tool in quantum information theory~\cite{doi:10.1063/1.1643788,berta2015renyi,sutter2018approximate}, yet many fundamental aspects of its behavior remain poorly understood both in the low-temperature and high-temperature regimes. 
Given a tripartition of a quantum system into regions $A$, $B$, and $C$, and a quantum state $\rho$ defined on the joint Hilbert space of $ABC$, the CMI is defined as
\begin{equation}
    I_\rho(A:C|B): = S_\rho(AB) + S_\rho(BC) - S_\rho(ABC) - S_\rho(B),
\end{equation}
where $S_\rho(X)$ denotes the von Neumann entropy of the reduced state on region $X$.
As a genuinely tripartite quantity, CMI captures correlations beyond pairwise interactions and plays a central role in characterizing many-body correlations such as topological order~\cite{PhysRevLett.96.110404,PhysRevLett.96.110405,PhysRevA.93.022317}. Furthermore, through its connection to \emph{quantum Markovianity}, CMI is deeply linked to the concept of quantum recoverability~\cite{Petz1986,Fawzi2015,PhysRevLett.115.050501,doi101098rspa20150338,Junge2018,Sutter2018}. This connection has led to important applications, including the definition of quantum mixed phases~\cite{Coser2019classificationof,lee2024universalspreadingconditionalmutual,zhang2024nonlocalgrowthquantumconditional,PhysRevX.14.031044,sang2024} and the design of quantum Gibbs sampling algorithms~\cite{PhysRevLett.103.220502,Brandao2019}.

A central open question in this context is the following:
\begin{conj}[CMI decay at arbitrary temperatures] \label{conjecture:1}
For general quantum Gibbs states $\rho_\beta$ at any temperature $\beta$, the conditional mutual information $I_{\rho_\beta}(A:C|B)$ with $A \cup B \cup C = \Lambda$ decays rapidly (e.g., super-polynomially) with the distance between the arbitrary regions $A$ and $C$, where $\Lambda$ denotes the entire system.
\end{conj}
\noindent
This is a quantum analogue of the Hammersley-Clifford theorem in the case of classical or commutative Hamiltonians~\cite{HammersleyClifford1971,brown2012quantum}. 
A conditional version under (uniform) clustering has been discussed in the literature (see, e.g., Brand\~ao–Kastoryano~\cite{Brandao2019}): 
\begin{conj'} \label{conjecture':1}
Under the assumption of the uniform clustering of correlations (or exponential decay of correlations), the quantum Gibbs state exhibits decay of the CMI. 
\end{conj'}
\noindent
We defer the precise definition to Assumption~\ref{def:uniform_clustering}.

At high temperatures, an even stronger form is conjectured:
\begin{conj}[CMI decay at high temperatures] \label{conjecture:2}
For general quantum Gibbs states at sufficiently high temperatures, the conditional mutual information $I_{\rho_\beta}(A:C|B)$ with $A \cup B \cup C \subseteq \Lambda$ decays rapidly with the distance between the regions $A$ and $C$.
\end{conj}
\noindent
These two conjectures capture a hierarchy of decay behavior. When $A \cup B \cup C = \Lambda$, the decay of CMI is expected to be a universal feature, independent of temperature. In contrast, at high temperatures, CMI is expected to exhibit exponential decay even when $A \cup B \cup C$ is strictly contained in $\Lambda$, indicating a stronger form of spatial locality. However, this stronger decay property does not generally hold at low temperatures: explicit counterexamples (e.g., quantum topological order) are known where $I_{\rho_\beta}(A:C|B)$ fails to decay when $A \cup B \cup C \subset \Lambda$ in the low-temperature regime~\cite{PhysRevB.78.155120,PhysRevLett.107.210501}.

The formulation of conjectures concerning the decay of CMI is a relatively recent development. In 2016, a general proof for one-dimensional systems was provided by Kato and Brandão~\cite{kato2016quantum}, marking a significant first step in this direction. For Conjecture~\ref{conjecture:1}, a partial resolution was later achieved in 2024 through the development of a systematic method to construct effective Hamiltonians on subsystems~\cite{kuwahara2024CMI}. This allowed for proofs of the CMI decay at arbitrary temperatures as long as the regions $A$ and $C$ are small. 
Therefore, a key open problem is whether the regions $A$ and $C$ can be taken arbitrarily large, or, equivalently, whether the $|A|, |C|$ dependence of the CMI decay is at most polynomial.

Conjecture~\ref{conjecture:2}, concerning the high-temperature regime, was initially believed to be resolved by the 2020 work of Kuwahara, Kato, and Brandão~\cite{PhysRevLett.124.220601}, who introduced a technique known as the generalized cluster expansion. In this method, physical quantities of interest are expanded perturbatively in terms of Hamiltonian parameters, and the convergence of this expansion is then analyzed. This approach has proven effective in a variety of contexts~\cite{PRXQuantum.4.020340,Haah2024}. However, when applying this technique to CMI, it was later pointed out that the treatment of the logarithm of reduced density matrices involves uncontrolled approximations, which undermines the convergence argument in the original proof. As a result, the applicability of the generalized cluster expansion to establishing CMI decay remains an open question~\cite{PhysRevLett.134.199901}.

In this work, we propose a new approach to establishing the decay of CMI for arbitrarily large subsystems $A$ and $C$, which does not rely on the effective Hamiltonian theory~\cite{kuwahara2024CMI}. Instead, our method is based on the construction of suitable recovery maps for quantum Gibbs states.
To clarify the point, let us consider a tripartite quantum state $\rho_{ABC}$ and examine its marginal $\rho_{AB}$ on the subsystems $A$ and $B$. We then study the possibility of approximately reconstructing $\rho_{ABC}$ from $\rho_{AB}$ via a completely positive trace-preserving (CPTP) map $\tau_{AB \to ABC}$. If such a recovery map can be effectively reduced to a CPTP map $\tau_{B \to BC}$ that acts only on subsystem $B$, then it follows that the conditional mutual information $I_\rho(A:C|B)$ vanishes.
More generally, it is well-known that if $\rho_{ABC}$ can be well-approximated by $\tau_{B \to BC}(\rho_{AB})$, then the CMI $I_\rho(A:C|B)$ must be small~\cite{Fawzi2015}.
In this work, we construct an explicit CPTP map $\tau_{B \to BC}$ that approximately recovers $\rho_{ABC}$ from $\rho_{AB}$.

The central technical component of our approach is the existence of the approximate quasi-local belief-propagation (BP) channel. 
The belief propagation operator transforms the Gibbs state of a full Hamiltonian $H$ into that of a modified Hamiltonian $H + h_i$, where $h_i$ is a local interaction term (e.g., supported near site $i$).
Although the CPTP map in itself does not give the belief propagation operator, we consider a CPTP version of quantum belief propagation, which we call the BP channel.
In detail, we aim to design a local quantum channel that approximately realizes the transformation
\begin{equation}
\frac{e^{\beta H}}{\tr \left( e^{\beta H} \right)} \longrightarrow \frac{e^{\beta (H + h_i)}}{\tr \left( e^{\beta (H + h_i)} \right)},
\end{equation}
and its inverse step can be constructed analogously. 
Remarkably, the sequence of implementations of such a BP channel allows us to derive the CMI decay for arbitrary subsystems $A$ and $C$ (Theorem~\ref{thm:main}).
Therefore, we reduce the challenging CMI decay conjecture to the simpler question of the existence of efficient quantum belief-propagation channels.

The remaining mathematical challenge in our approach lies in constructing the BP channel.
If a quasi-local BP channel exists unconditionally, it leads to the complete resolution of Conjecture~\ref{conjecture:1}, which is still highly challenging. 
Instead, we consider either of the following conditions: i) under the rapid mixing condition (Assumption~\ref{assup:Basic assumptions for quasi-local Liouvillian}), or ii) under the clustering condition (Assumption~\ref{def:uniform_clustering}).
Both conditions can be rigorously verified at high temperatures, while at low temperatures, they are believed to hold only in non-critical regimes.  
Each of the conditions leads to an efficient construction of the BP channel, as shown in Theorems~\ref{thm_high_temp_BP} and \ref{thm_low_temp_BP}. 
Consequently, we resolve the conditional version of Conjecture~\ref{conjecture':1} (under uniform clustering) and Conjecture~\ref{conjecture:2} (at high temperatures) for $A\cup B\cup C=\Lambda$ cases, respectively.

%To be more precise,  
%

Finally, we revisit the cluster expansion technique for the effective Hamiltonian on a subsystem (i.e., $A\cup B\cup C\subset \Lambda$), which plays a critical role in Conjecture~\ref{conjecture:2}. 
Whether the lack of rigorous convergence proof is merely a technical issue or indicates a deeper obstruction has remained a subject of debate. 
When considering reduced density matrices on subsystems, there is in general no guarantee that they can be expressed in the form of Gibbs states. 
As a result, our BP-channel methodology cannot be straightforwardly applied in this setting, and hence, the analyses of the effective Hamiltonian are inevitable. 

In Section~\ref{sec:Divergence of cluster expansion} of this work, we identify that the difficulty of performing a high-temperature expansion of the logarithm of reduced density matrices---which is necessary for computing the CMI---shares essential similarities with the divergence problems encountered in the Baker--Campbell--Hausdorff (BCH) expansion~\cite{BLANES2009151}. This connection suggests that the issue is not simply technical but rather reflects an inherent limitation of the method.
Based on this insight, we are led to the following conjecture:
\begin{conj}[Non-convergence of cluster expansion for CMI] \label{conjecture:3}
The cluster expansion method is not absolutely convergent for the conditional mutual information at any fixed (nonzero) temperature.
\end{conj}
\noindent
A rigorous proof of this conjecture would require a more delicate analysis, potentially along the lines of the techniques developed in Ref.~\cite{10.1063/1.4936209}. 
This observation motivates the development of a completely different approach to proving Conjecture~\ref{conjecture:2}. Since the traditional cluster expansion appears fundamentally limited in its applicability to CMI, a new framework may be necessary to establish its spatial decay in the high-temperature regime.

The rest of this paper is organized as follows:
In Section~\ref{sec:Setup}, we provide a more detailed description of the physical setup and define the class of quantum systems under consideration.
Section~\ref{Sec:Decay of the conditional mutual information} presents an overview of our main results, along with the key ideas behind our approach based on the belief-propagation channel.
In Section~\ref{sec:BP-construction}, we show the existence of approximate quasi-local BP channels under the assumption of the rapid mixing or the clustering of correlations.
Sections~\ref{Pr:thm_high_temp_BP} and \ref{Pr:thm_low_temp_BP} are devoted to the proofs for the quasi-local BP channel in Section~\ref{sec:BP-construction}. 
Section~\ref{sec:Divergence of cluster expansion} discusses the divergence issues that arise when attempting to apply cluster expansion techniques to the logarithm of reduced density matrices.
Finally, in Section~\ref{sec:Conclusion and discussions}, we summarize our results and highlight several open problems and directions for future research.

\section{Setup} \label{sec:Setup}

We study a quantum system located on a graph with $n$ sites, where $\Lambda$ denotes the set of all these sites, thus $|\Lambda| = n$.
We assign a $d$-dimensional Hilbert space $\mathbb{C}^d$ to each of the sites. 
Let $X \subseteq \Lambda$ represent any subset of sites. The number of sites in $X$, called the cardinality, is denoted by $|X|$. The set of sites in $\Lambda$ but not in $X$, called the complementary subset, is represented as $X^\co := \Lambda \setminus X$.
For convenience, the union of two subsets $X$ and $Y$ is often denoted as $XY$ instead of $X \cup Y$.
The distance $\dist_{X,Y}$ between subsets $X$ and $Y$ is defined as the length of the shortest path on the graph that connects a site in $X$ to a site in $Y$. If $X$ and $Y$ intersect, then $\dist_{X,Y} = 0$. For subsets where $X$ contains only one site, say $X = \{i\}$, we simplify $\dist_{\{i\},Y}$ to $\dist_{i,Y}$.

The inner boundary of $X$ is defined as:
\begin{align}
\partial X := \{ i \in X \mid \dist_{i, X^\co} = 1 \}.
\end{align}
We define the extended subset $\bal{X}{r}$ for a subset $X \subseteq \Lambda$ as follows:
\begin{align}
\bal{X}{r} := \{i \in \Lambda \mid \dist_{X, i} \le r\}, \label{def:bal_X_r}
\end{align}
where $\bal{X}{0} = X$, and $r$ is any positive real number ($r \in \mathbb{R}^+$).

We introduce a geometric constant $\gamma$, determined by the lattice structure, such that $\gamma \ge 1$. This constant satisfies:
\begin{align}
\max_{i \in \Lambda} \left|i[r]\right| \le \gamma  r^{D} \label{def:parameter_gamma}
\end{align}
for $r \ge 1$, where $D$ is the spatial dimension of the lattice.

Consider a Hamiltonian $H$ describing short-range interactions on an arbitrary finite-dimensional graph:
\begin{align}
H = \sum_{Z} h_Z, \quad \max_{i \in \Lambda} \sum_{Z : Z \ni i} \|h_Z\| \le g, \label{def:Hamiltonian}
\end{align}
where the decay of interactions is assumed to be finite range $l_H>0$ :
\begin{align}
\sum_{Z : Z \ni \{i,i'\}} \|h_Z\| = 0 \for \dist_{i,i'} >l_H, \label{def_short_range_long_range}
\end{align}
with $\| \cdot \|$ representing the operator norm.

For any operator $O$, the trace norm is $\|O\|_1 := \tr\left(\sqrt{O^\dagger O}\right)$.
The Hamiltonian on a region $L$ and its interaction terms are defined as:
\begin{align}
H_L := \sum_{Z : Z \subset L} h_Z. 
%\quad \widehat{H_L} := \sum_{Z : Z \cup L \neq \emptyset} h_Z = H - H_{L^\co}. 
\label{def:Hamiltonian_subset_L}
\end{align}
The boundary interaction terms on region $L$ are given by:
\begin{align}
\partial h_L \coloneqq H - H_L - H_{L^\co} = \sum_{Z : Z \cap L \neq \emptyset, Z \cap L^\co \neq \emptyset} h_Z. 
%\quad \widehat{H_L} = H_L + \partial h_L. 
\label{def:Ham_surface}
\end{align}

We define the time evolution of any operator $O_1$ under the influence of another Hermitian operator $O_2$ as:
\begin{align}
\label{O_1_O_2,t}
O_1(O_2, t) := e^{iO_2t} O_1 e^{-iO_2t}.
\end{align}
For simplicity, the time evolution of $O_1$ under $H$ is often denoted by $O_1(t)$.

We study the quantum Gibbs state at inverse temperature $\beta$:
\begin{align}
\rho_\beta := \frac{e^{\beta H}}{Z_\beta}, \qquad 
Z_\beta = \tr\!\left(e^{\beta H}\right).
\end{align}
For simplicity, we use $e^{\beta H}$ instead of the standard $e^{-\beta H}$, which does not affect generality.
When we wish to emphasize the underlying Hamiltonian, we will write $\rho_\beta(H)$ explicitly. 
In particular, for a modified Hamiltonian such as $H+h_i$, we denote 
$\rho_\beta(H+h_i) := e^{\beta (H+h_i)}/\tr(e^{\beta (H+h_i)})$.

The reduced density matrix for a region $L$ is defined as:
\begin{align}
\rho_{\beta,L} := \tr_{L^\co}(\rho_\beta)\otimes \hat{1}_{L^\co}, 
\label{def_rho_L_H^ast_L}
\end{align}
where $\tr_{L^\co}(\cdots)$ denotes the partial trace over the complement of $L$.

We introduce the normalized partial trace $\tilde{\tr}_X(O)$ as:
\begin{align}
\tilde{\tr}_X(O) :=  \tr_X(O)\otimes \frac{1}{\tr_X(\hat{1})}\hat{1}_X.
\label{definition_of_tilde_tr_partial}
\end{align}
This operation ensures that $\tilde{\tr}_X(O)$ is supported on $X^\co$ and commutes with any operator supported on $X$, i.e., $[\tilde{\tr}_X(O), O_X] = 0$. Moreover, $\|\tilde{\tr}_X(O)\|$ is always less than or equal to $\|O\|$.

 We define a function $\Theta(x)$ in terms of a variable $x$:
 \begin{align}
 \Theta(x) = \sum_{\sigma= 0, 1} c_{\sigma} x^{\sigma}, 
 \end{align}
 where $0 < c_{\sigma} < \infty$, and these coefficients depend on fundamental parameters listed in Table~\ref{tab:fund_para}.

%The notation $\tO(x)$ is used to indicate an order estimation:
%\begin{align}
%\tO(x) = \mathcal{O}\left(x \cdot \text{polylog}(x)\right).
%\end{align}

\begin{table*}[tt]%The best place to locate the table environment is directly after its first reference in text
 \caption{Fundamental parameters in our statements}
  \label{tab:fund_para} 
\begin{ruledtabular}
\begin{tabular}{lr}
\textrm{\textbf{Definition}}&\textrm{\textbf{Parameters}} 
\\
\colrule
Spatial dimension    
&  $D$ \\
Local Hilbert space dimension 
&  $d$ \\
Constant for spatial structure [see Ineq.~\eqref{def:parameter_gamma}]    
& $\gamma$  \\
One-site energy [see Eq.~\eqref{def:Hamiltonian}]
&  $g$ %\\
\\
Interaction length [see Eq.~\eqref{def_short_range_long_range}]
&  $l_H$ %\\
%Interaction decay [see the definition~\eqref{def_short_range_long_range}] 
%& $\bar{J}(x)$
\end{tabular}
\end{ruledtabular}
\end{table*}

\subsection{Lindblad Liouvillian}  \label{Sec:Lindblad Liouvillian} 

In the subsequent sections, we often consider the dissipative dynamics. 
We provide a brief review of the Lindblad Liouvillian. 

We define the dissipative dynamics governed by the Liouville equation as follows: 
\begin{align}
\frac{d}{dt} \rho(t)= \mL \rho(t),
\end{align}
where $\mL$ and $\rho(t)$ are the Liouvillian, a linear superoperator, and the density matrix at time $t$, respectively. 
We now assume that $\mL$ is also the Lindbladian, which satisfies the following four conditions: i) linear, ii) Markovian, iii) completely positive, and iv) trace-preserving.
Such Lindbladian $\mL$ generally have the following form:
\begin{align}
\label{eq:lindblad_form}
\mL(\rho)=-i [H, \rho]+ \sum_{j} \left( L_{j} \rho   L_{j}^\dagger -\frac{1}{2}\{ L_{j}^\dagger L_{j} ,\rho \}\right),
\end{align}
where $H$ is the Hamiltonian and each of $\{L_j\}_j$ is a jump operator.

For any operator $O$, we denote the Heisenberg picture of the time evolution by $e^{\mathcal{L}^\dagger t}O$, i.e., 
\begin{align}
\tr\brr{O e^{\mL t} \rho} = \tr\brr{\rho e^{\mL^\dagger t} O } 
\end{align}
with 
\begin{align}
\label{eq:lindblad_form_Heisenberg}
\mathcal{L}^\dagger O =i [H, O]+ \sum_{j} \br{ L_{j}^\dagger O L_{j} -\frac{1}{2}\{ L_{j}^\dagger L_{j} ,O\} }.
\end{align}
We note that 
\begin{align}
\mathcal{L}^\dagger O = 0 \for O \quad s.t.\quad  [H,O]=[L_{j},O]=0 .
\end{align}

We consider the $(p\to q)$ norm of the Liouville superoperator which is defined as~\cite{Amasov00,Watrous:2005:NSN:2011608.2011614} 
 \begin{align}
 \label{def:norm_pq}
\|\mL \|_{p\to q} := \sup_{O} \frac{\norm{\mL O}_q}{\norm{O}_p} ,
\end{align}
where $1\leq p,q\leq\infty$ and the supremum is taken for all operators $O$. 
In particular, if we consider $\|\mL \|_{\infty \to \infty}$, we simply denote by
 \begin{align}
 \label{Liouvillian_norm_simplification}
\|\mL\|_{\infty \to \infty} = \|\mL\|
\end{align}
without the index $\infty \to \infty$. 
As a convenient property of the Lindblad operator, we have  
 \begin{align}
 \label{norm_Liouvillian_dybamics}
\|e^{\mL t} \|_{1\to 1} = 1 ,\quad \|e^{\mL^\dagger t} \| \le 1
\end{align}
Note that even though $\mL$ is the Lindbladian, $-\mL$ is generally not, that is,    
\begin{align}
-\mL\rho=i [H, \rho] - \sum_{j} \left( L_{j} \rho   L_{j}^\dagger -\frac{1}{2}\{ L_{j}^\dagger L_{j} ,\rho \}\right) 
\overset{\textrm{not given}}{=} i [H, \rho] + \sum_{j} \left( \bar{L}_{j} \rho  \bar{L}_{j}^\dagger -\frac{1}{2}\{ \bar{L}_{j}^\dagger \bar{L}_{j} ,\rho \}\right) .
\end{align}
by an alternative choice of $\{\bar{L}_{j}\}$. Therefore generally it holds that 
 \begin{align}
\|e^{- \mL t} \|_{1\to 1} > 1. 
\end{align}

\section{Decay of the conditional mutual information} \label{Sec:Decay of the conditional mutual information}

\subsection{Belief propagation (BP) channel}

In this section, we show our main result. 
Instead of relying on the cluster expansion technique, we utilize the Fawzi-Renner theorem~\cite{Fawzi2015} to connect the recovery map and the CMI decay:
\begin{lemma}[Fawzi--Renner inequality~\cite{Fawzi2015}] \label{Fawzi-Renner}
Let $\Lambda = A \cup B \cup C$ be a tripartition of the system, and let $\rho_{ABC}$ be a quantum state with reduced state $\rho_{AB}$. 
Then there exists a completely positive trace-preserving (CPTP) map acting only on subsystem $B$ and producing an output state on $BC$, denoted by $\tau_{B \to BC}$, such that
\begin{align}
\mI_{\rho}(A:C|B) \le 7 \log_2\brr{\min\br{\mathcal{D}_A,\mathcal{D}_C}} \sqrt{\norm{ \tau_{B\to BC} (\rho_{AB}) - \rho }_1 } .
\end{align}
In words, whenever $\rho_{ABC}$ can be approximately recovered from its marginal $\rho_{AB}$ via such a local recovery map on $B$, the conditional mutual information is bounded by the recovery error. 
\end{lemma}

The core idea in our analyses is to utilize the following belief propagation channel. 
It realizes a CPTP map that perturbs the Hamiltonian in the quantum Gibbs states (a quantum analogue of the classical BP). 
Intuitively, one may think of it as a way to ``locally update'' the thermal state when a new interaction term is added to the Hamiltonian, while keeping the rest of the system essentially unchanged.
We define it in the following manner:
\begin{definition}[BP channel and approximate BP channel] \label{def:BP_channel}
Let $H$ be a local Hamiltonian and $h_i$ a local interaction term supported near site $i$.  
We denote by $\rho_\beta(H) := e^{-\beta H}/\tr(e^{-\beta H})$ the Gibbs state at inverse temperature $\beta$.

\begin{itemize}
\item A Belief Propagation (BP) channel is a completely positive trace-preserving (CPTP) map
\begin{equation}
\tau_\beta^{(H \to H+h_i)} : \rho_\beta(H)  \mapsto  \rho_\beta(H+h_i).
\end{equation}

\item For $r > 0$, an approximate BP channel on the ball $i[r]$ is a CPTP map
\begin{equation}
\tilde{\tau}_{\beta,i[r]}^{(H \to H+h_i)} : \rho_\beta(H)  \mapsto  \tilde{\rho}_{\beta,i[r]},
\end{equation}
supported only on $i[r]$, such that
\begin{equation}
\label{approx_BP}
\norm{ \tilde{\tau}_{\beta,i[r]}^{(H \to H+h_i)} \brr{\rho_\beta(H)}- \rho_\beta(H+h_i) }_1 
 \le  \epsilon(\beta,r) ,
\end{equation}
for some error function $\epsilon(\beta,r)$ that typically decays as $r$ increases.
\end{itemize}

In other words, the approximate BP channel realizes the transformation from $\rho_\beta(H)$ to $\rho_\beta(H+h_i)$ up to controllable error, using only operations supported on the finite region $i[r]$.
\end{definition}

{\bf Remark.} 
In the standard formulation of belief propagation~\cite{PhysRevB.76.201102,PhysRevB.86.245116}, one often encounters local positive operators of the form 
$\Phi_i^\dagger e^{-\beta H} \Phi_i$. 
Here, one can prove that the quasi-locality of $\Phi_i$ is ensured by the Lieb--Robinson bound~\cite[Lemma~10 therein]{kuwahara2024CMI}. 
These induce maps of the form 
$\tau(\rho) = A^\dagger \rho A$, 
which are completely positive by construction, since they admit a Kraus representation with a single Kraus operator $A$. 
However, such maps are not necessarily trace preserving unless $A^\dagger A = I$. 
In particular, the conventional belief propagation operator is CP but not TP in general. 
In contrast, in our framework, we explicitly require the construction of a CPTP map that implements the transformation between Gibbs states, and we distinguish it as the BP channel.

\begin{figure}[t]
\centering
\includegraphics[width=0.55\textwidth]{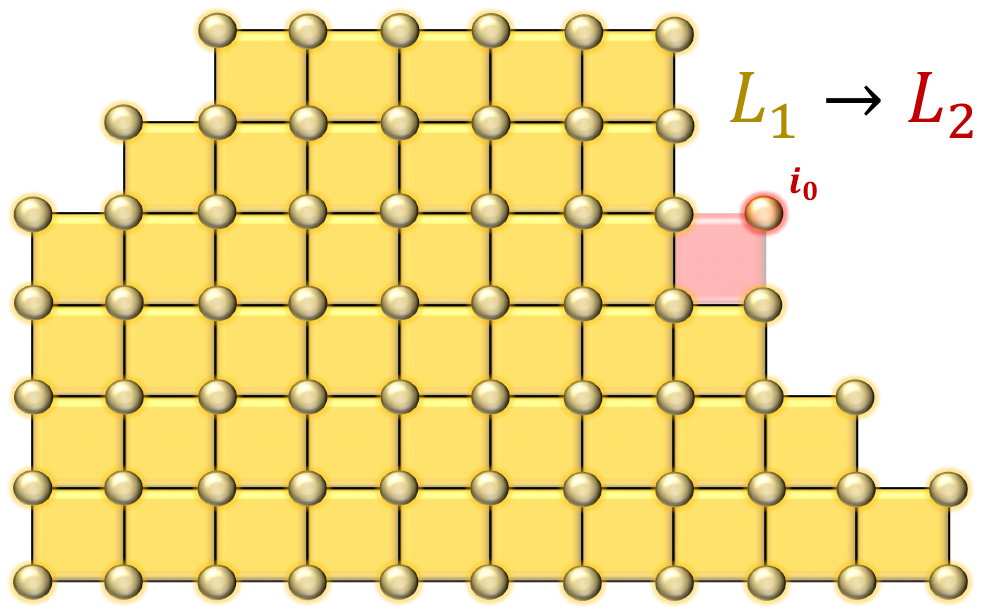}
\caption{
Schematic illustration of a subset Hamiltonian update.
A local region $L_1$ is enlarged by adding a new site $i_0$ (highlighted) 
together with its incident interactions, resulting in an extended region $L_2$.
}
\label{fig:subset_Hamiltonia_update}
\end{figure}

For our purpose, it is not necessary to implement BP channels for all possible local terms in the Hamiltonian.  
Instead, it suffices to consider a restricted class of updates, where a subset Hamiltonian is enlarged by adding exactly one new site and its incident interaction terms.  
We formalize this operation as follows:
\begin{definition}[Subset Hamiltonian update] \label{def:subset-expansion}
Let $H = \sum_Z h_Z$ be a local Hamiltonian and $H_L$ the subset Hamiltonian on $L \subseteq \Lambda$ as in~\eqref{def:Hamiltonian_subset_L}. 
A subset Hamiltonian update refers to the local update
\[
H_{L_1}  \longleftrightarrow  H_{L_2},
\]
where $L_1 \subset L_2$ and $|L_2 \setminus L_1| = 1$. 
That is, the Hamiltonian support is enlarged by one site, together with its incident interaction terms. 
\end{definition}

{\bf Remark.} 
Let $L_2 \setminus L_1 = \{i_0\}$.  
By the finite-range interaction condition~\eqref{def_short_range_long_range},  
the difference between the two subset Hamiltonians, $H_{L_2} - H_{L_1}$ is supported only on the ball $i_0[l_H]$ of radius $l_H$ around $i_0$.
In other words, the additional interaction terms introduced in the expansion are localized near the newly added site $i_0$.

\subsection{Main result}

Using the BP channel formalism, we can prove the main theorem as follows: 
\begin{theorem} \label{thm:main}
Let $A$, $B$, and $C$ constitute a partition of the total system $\Lambda = A \cup B \cup C$.  
Assume that for every subset Hamiltonian update (Definition~\ref{def:subset-expansion}), there exists a BP channel satisfying the approximation property~\eqref{approx_BP}.  
Then there exists a recovery map $\tau_{B \to BC}$ such that
\begin{align}
\label{main_ineq_thm}
\norm{ \tau_{B\to BC} (\rho_{\beta , AB}) - \rho_{\beta, ABC} }_1 
    \le  2|B| \epsilon(\beta,R_0) , \quad R_0 := \frac{R-l_H}{2} ,
\end{align}
where $R = \dist_{A,C}$ denotes the distance between $A$ and $C$, and $\epsilon(\beta,r)$ ($r \in \mathbb{N}$) is the error term associated with the approximate BP channel as in~\eqref{approx_BP}.   

Moreover, by applying the Fawzi--Renner inequality (Lemma~\ref{Fawzi-Renner}), one immediately obtains the following bound on the conditional mutual information:
\begin{align}
\mI_{\rho_\beta}(A:C|B) 
    \le  7 \log_2 \!\Bigl( \min\{\mathcal{D}_A,\mathcal{D}_C\} \Bigr) \sqrt{2|B| \epsilon(\beta,R_0)}.
\end{align}
\end{theorem}

{\bf Remark.} 
An important conceptual contribution of Theorem~\ref{thm:main} is that the proof of Conjecture~1 (stated in the Introduction) can be reduced to the simpler and more tangible problem of proving the existence of quasi-local BP channels.  
In other words, instead of tackling the decay of conditional mutual information directly, it suffices to establish the existence of local CPTP maps implementing the subset Hamiltonian updates.  
This formalism highlights the central role of BP channels and provides a unified framework that, as we discuss later, enables rigorous proofs of CMI decay in both the high-temperature and the low-temperature regimes.

Regarding the $|B|$ dependence, by using a slightly refined analysis in~\eqref{upper_bound_bar_n}, we can replace 
$$|B| \to \gamma^2 l_H^D (R/2)^{D-1}  \min (|\partial A|, |\partial C|)$$ in the inequality~\eqref{main_ineq_thm},  where $\gamma$ has been defined in~\eqref{def:parameter_gamma}. 

\subsection{Proof of Theorem~\ref{thm:main}} \label{Sec:Proof outline}

\begin{figure}[t]
\centering
\includegraphics[width=0.55\textwidth]{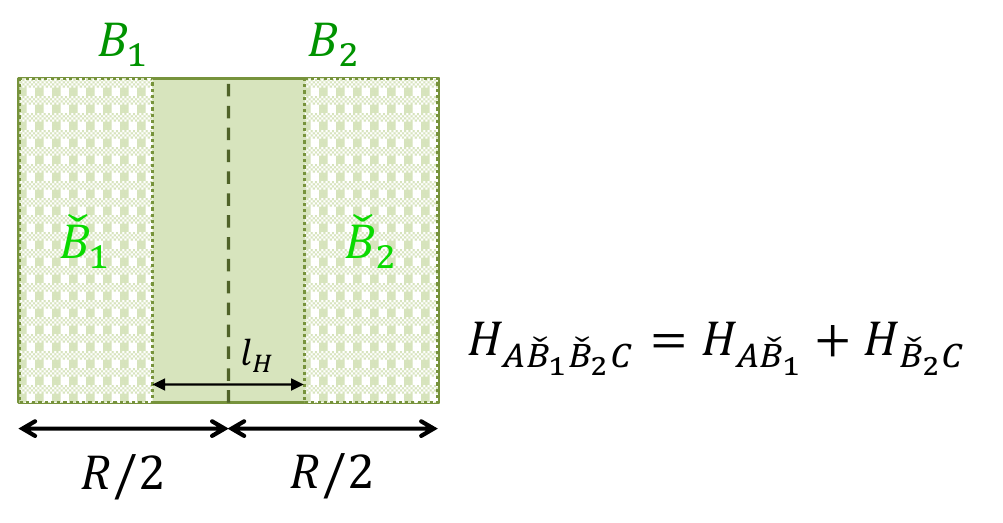}
\caption{
Illustration of the regions $\check{B}_1$ and $\check{B}_2$, obtained from $B_1$ and $B_2$ by removing boundary layers of width $l_H$. 
By construction, there are no interaction terms between $\check{B}_1$ and $\check{B}_2$. 
Consequently, the Hamiltonian factorizes as
$H_{A\check{B}_1\check{B}_2C} = H_{A\check{B}_1} + H_{\check{B}_2C}$,
which will play a key role in the construction of the recovery map.
}
\label{fig:check_B_1B_2}
\end{figure}

We aim to construct a recovery map for an arbitrary decomposition $\Lambda = A \cup B \cup C$ such that
\begin{align}
\tau_{B\to BC}\bigl(\rho_{\beta,AB}\bigr)  \approx  \rho_{\beta,ABC}.
\end{align}
For later use, we define the trimmed regions $\check{B}_1$ and $\check{B}_2$ by removing boundary layers of width $l_H$ from $B_1$ and $B_2$, respectively (see Fig.~\ref{fig:check_B_1B_2}).  
Under the finite-range condition~\eqref{def_short_range_long_range}, no interaction term can connect $\check{B}_1$ and $\check{B}_2$, and hence the subset Hamiltonian on $A \cup \check{B}_1 \cup \check{B}_2 \cup C$ factorizes:
\begin{align}
H_{A\check{B}_1\check{B}_2C}  =  H_{A\check{B}_1} + H_{\check{B}_2C}.
\end{align}
This factorization is the key to the construction below.

\begin{figure}[tt]
\centering
\includegraphics[clip, scale=0.4]{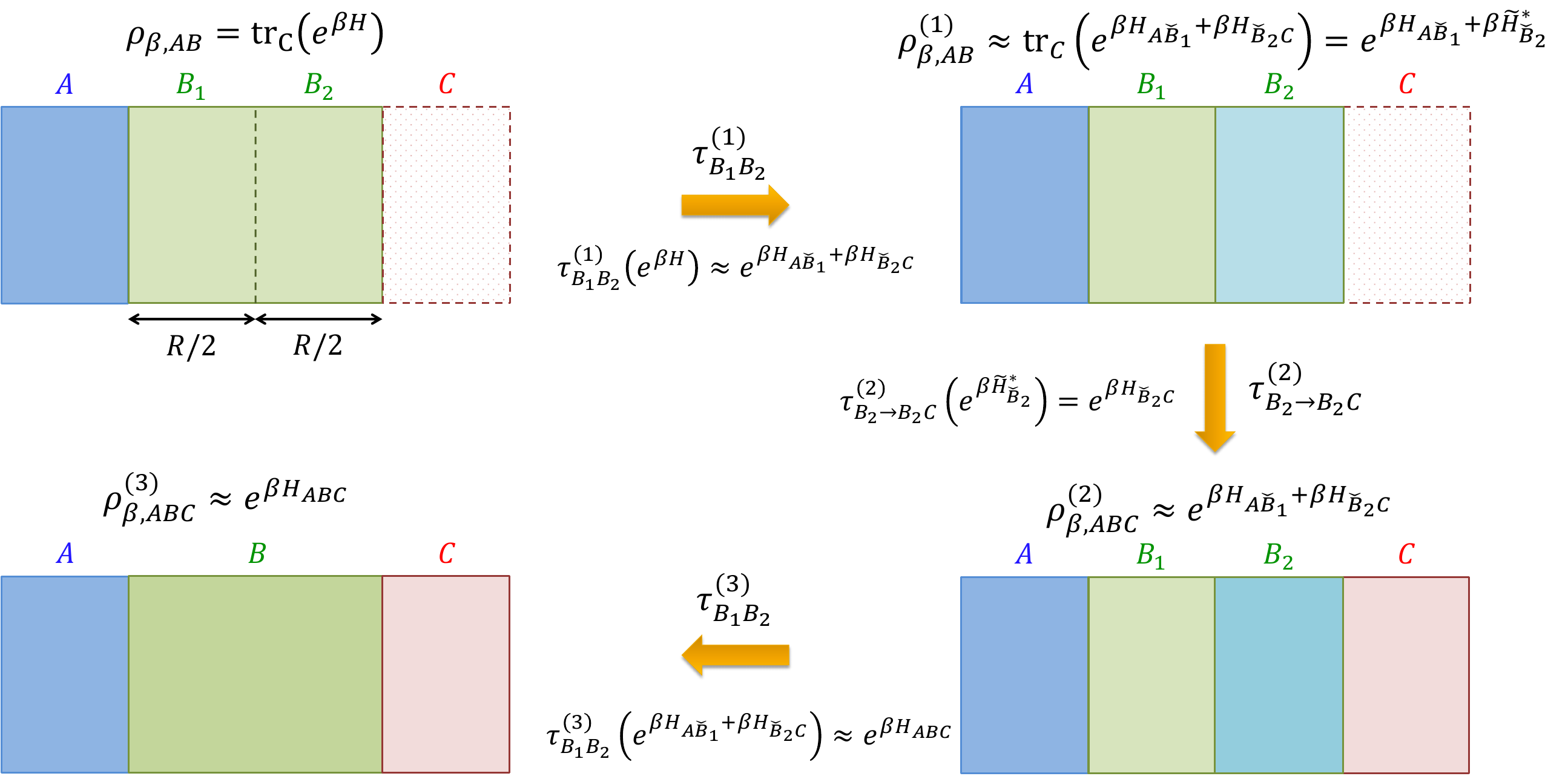}
\caption{Schematic picture of the construction of the recovery map.
}
\label{fig_Recovery_map}
\end{figure}

\medskip
Here we present the construction of the recovery map in three steps (see Fig.~\ref{fig_Recovery_map}). 
For simplicity, we shift Hamiltonians so that each (sub)Gibbs operator used below is normalized to trace one; equivalently, we may write $Z_\beta = 1$ by replacing $H \to H - \beta^{-1}(\log Z_\beta)\hat{1}$.

\begin{enumerate}

\item{} \textit{Decoupling across $B_1|B_2$.}
Decompose $B$ into $B_1$ and $B_2$ with equal width so that $\dist_{A,B_2} = \dist_{A,C}/2$.  
By trimming boundary layers, we obtain $\check{B}_1,\check{B}_2$ and consider a CPTP map $\tau^{(1)}_{B_1B_2}$ that approximately removes the cross interaction across the middle surface and produces the factorized Gibbs operator of
\begin{align}
\label{def:tau_1_B1_B2}
\tau^{(1)}_{B_1B_2} \br{e^{\beta H }} \approx \tilde{\rho}_\beta := e^{\beta H_{A\check{B}_1\check{B}_2C}} = e^{\beta H_{A\check{B}_1}} \otimes e^{\beta H_{\check{B}_2C}}  , 
\end{align}
where we let $\tr\brr{ e^{\beta (H_{AB_1} +H_{B_2C})} }=1$ and the boundary interaction $\partial h_{AB_1}$ is removed from $H$. 
Using the map, we have 
\begin{align}
\label{eq:step1_state}
\rho_{\beta,AB}^{(1)} := \tau^{(1)}_{B_1B_2} \br{\rho_{\beta,AB} }&=  \tr_C \brr{ \tau^{(1)}_{B_1B_2} \br{e^{\beta H }}} \notag\\
&  \approx  
 \tr_C \br{  \tilde{\rho}_\beta }=e^{\beta H_{A\check{B}_1}} \otimes e^{\beta \tilde{H}^\ast_{\check{B}_2}}, 
\end{align}
where we define $e^{\beta \tilde{H}^\ast_{\check{B}_2}}:=  \tr_C \br{e^{\beta H_{\check{B}_2C}}}$.

\item{} \textit{Recovering the Gibbs state on $\check{B}_2C$.} We then consider the state-preparation operation $\tau^{(2)}_{B_2\to B_2C}$ which makes arbitrary input $\sigma_{B_2}$ to 
\begin{align}
\label{Step_2}
\tau^{(2)}_{B_2\to B_2C} \br{\sigma_{B_2}} \propto e^{\beta H_{\check{B}_2C}}.  
\end{align}
Note that it provides 
\begin{align}
\tau^{(2)}_{B_2\to B_2C} \br{ \tr_C \br{  \tilde{\rho}_\beta } }= \tilde{\rho}_\beta, 
\end{align}
where there is no approximation error. 
Applying it to~\eqref{eq:step1_state}, we obtain
 \begin{align}
\label{eq:state_after_step2}
\rho_{\beta,ABC}^{(2)} :=\tau^{(2)}_{B_2\to B_2C}  \tau^{(1)}_{B_1B_2}\br{ \rho_{\beta,AB} } \approx  
\tau^{(2)}_{B_2\to B_2C} \br{ \tr_C \br{  \tilde{\rho}_\beta }}=\tilde{\rho}_\beta .
\end{align}

\item{} \textit{Re-coupling.}
Finally, we apply a local channel $\tau^{(3)}_{B_1B_2}$ that (approximately) reintroduces the removed interaction terms, i.e.,
\begin{align}\label{eq:step3_map}
\tau^{(3)}_{B_1B_2} \br{\tilde{\rho}_\beta}
= \tau^{(3)}_{B_1B_2} \br{e^{\beta H_{A\check{B}_1\check{B}_2C}}}
\approx e^{\beta H}.
\end{align}
By combining the three maps $\tau^{(1)}_{B_1B_2}$, $\tau^{(2)}_{B_2\to B_2C}$ and $\tau^{(3)}_{B_1B_2}$, we obtain the candidate recovery map
\begin{align}
\label{eq:state_after_step3}
\rho_{\beta,ABC}^{(3)}= \tau^{(3)}_{B_1B_2} \tau^{(2)}_{B_2\to B_2C} \tau^{(1)}_{B_1B_2} (\rho_{\beta,AB}) \approx \tau^{(3)}_{B_1B_2} \br{\tilde{\rho}_\beta } \approx \rho_{\beta,ABC} .
\end{align}
\end{enumerate}

The second map $\tau^{(2)}_{B_2\to B_2C}$ is trivially prepared without any error, and hence, the errors of the recovery map stem from $\tau^{(1)}_{B_1B_2}$ and $\tau^{(3)}_{B_1B_2}$. 
To estimate the error, we generally consider  
\begin{align}
\label{recovery_map_arppox_error}
\norm{ \tau^{(3)}_{B_1B_2} \tau^{(2)}_{B_2\to B_2C} \tau^{(1)}_{B_1B_2} (\rho_{\beta,AB}) - \rho_{\beta,ABC}}_1 
&\le \norm{ \tau^{(3)}_{B_1B_2} \tau^{(2)}_{B_2\to B_2C} \tau^{(1)}_{B_1B_2} (\rho_{\beta,AB})- \tau^{(3)}_{B_1B_2}\tilde{\rho}_\beta}_1 + \norm{ \tau^{(3)}_{B_1B_2}  \tilde{\rho}_\beta - e^{\beta H }}_1   \notag \\
&\le \norm{\tau^{(2)}_{B_2\to B_2C} \tau^{(1)}_{B_1B_2} (\rho_{\beta,AB})-\tau^{(2)}_{B_2\to B_2C}  \tr_C \tilde{\rho}_\beta } + \norm{ \tau^{(3)}_{B_1B_2}  \tilde{\rho}_\beta - e^{\beta H }}_1   \notag \\
&\le \norm{ \tr_C \br{ \tau^{(1)}_{B_1B_2} e^{\beta H }- \tilde{\rho}_\beta}}_1 + \norm{ \tau^{(3)}_{B_1B_2}  \tilde{\rho}_\beta - e^{\beta H }}_1   \notag \\
&\le  \norm{\tau^{(1)}_{B_1B_2} e^{\beta H }- \tilde{\rho}_\beta}_1 + \norm{ \tau^{(3)}_{B_1B_2}  \tilde{\rho}_\beta - e^{\beta H }}_1 , 
\end{align}
where we use the fact that the CPTP map does not increase the norm.  
The CPTP map $\tau^{(3)}_{B_1B_2} \tau^{(2)}_{B_2\to B_2C} \tau^{(1)}_{B_1B_2}$ constitutes the desired recovery map $\tau_{B\to BC}$ as in~\eqref{main_ineq_thm}.
Therefore, a sufficient condition for the desired CMI decay is the existence of local CPTP maps that approximate the transformation between $e^{\beta H}$ and $e^{\beta (H_{A\check{B}_1} + H_{\check{B}_2 C})}$.

In the following, we consider the implementation of $\tau^{(3)}_{B_1B_2}$, and the same analyses are applied to $\tau^{(1)}_{B_1B_2}$. 
We aim to construct the subset Hamiltonian update from $H_{A\check{B}_1\check{B}_2 C}$ to $H_{ABC}$

Here, we obtained all the ingredients to prove the main statement. 
We label the sites in $(B_1B_2) \setminus (\check{B}_1\check{B}_2)$ as $\{1,2,3,\ldots, \bar{n}\}$ with $\bar{n}=|(B_1B_2) \setminus (\check{B}_1\check{B}_2)|$. 
We define the subset $B^{(m)}$ as
 \begin{align}
B^{(i)} = (\check{B}_1\check{B}_2) \cup \{1,2,\ldots, i\} ,
\end{align}
and aim to update the Hamiltonian as 
 \begin{align}
H_{A\check{B}_1\check{B}_2 C}= H_{0} \to H_{1} \to H_{2} \to \cdots \to H_{\bar{n}}= H  ,
\end{align}
where we denote the subset Hamiltonians $\{H_{B^{(i)}}\}_{i=1}^{\bar{n}}$ by $\{H_i\}_{i=1}^{\bar{n}}$ for simplicity of notation.

For an arbitrary $i$, we utilize the approximate BP channel~\eqref{approx_BP} to update from $H_{i-1}$ to $H_{i}$, which yields 
 \begin{equation}
\label{approx_BP_application_H_i}
\norm{ \tilde{\tau}_{\beta,i[r]}^{(H_{i-1} \to H_{i})} \brr{\rho_\beta(H_{i-1})}- \rho_\beta(H_{i}) }_1  
 \le  \epsilon(\beta,r) .
\end{equation}
Note that $H_{i}-H_{i-1}$ is supported on the ball region $i[l_H]$. 
To iteratively connect the approximation, we prove the following lemma:
\begin{lemma} \label{iterative_use_of/_corol_lem}
We derive the following error bound:
\begin{align}
\norm{ \prod_{i=1}^{\bar{n}} \tilde{\tau}_{\beta,i[r]}^{(H_{i-1} \to H_{i})} \rho_\beta(H_0) -\rho_\beta(H_{\bar{n}}) }_1\le 
\bar{n} \epsilon(\beta,r)  ,
\label{iterative_use_of/_corol_lem/main}
\end{align}
where the sequence of the CPTP maps $\{ \tilde{\tau}_{\beta,i[r]}^{(H_{i-1} \to H_{i})}\}_{i=1}^{\bar{n}}$ is appropriately ordered. 
\end{lemma}

\textit{Proof of Lemma~\ref{iterative_use_of/_corol_lem}.}
We use the induction method. 
We first prove the case of $\bar{n}=1$, which is immediately obtained by~\eqref{approx_BP_application_H_i} with $i=1$.
Then, for an arbitrary $i_0-1$ $(\le \bar{n})$, we assume the inequality of 
\begin{align}
\norm{  \prod_{i=1}^{i_0-1} \tilde{\tau}_{\beta,i[r]}^{(H_{i-1} \to H_{i})} \rho^{(0)}-\rho^{(i_0-1)}}_1\le 
i_0 \epsilon(\beta,r)    ,
\label{iterative_use_of/_corol_lem/main_induction}
\end{align}
and prove the case of $i_0$. By using the above inequality, we derive
\begin{align}
&\norm{ \prod_{i=1}^{i_0} \tilde{\tau}_{\beta,i[r]}^{(H_i \to H_{i+1})}  \rho^{(0)}-\rho^{(i_0)}}_1 \notag \\
&= \norm{  \tilde{\tau}_{\beta,i_0[r]}^{(H_{i_0-1} \to H_{i_0})}  \prod_{i=1}^{i_0-1} \tilde{\tau}_{\beta,i[r]}^{(H_i \to H_{i+1})}   \rho^{(0)} -\tilde{\tau}_{\beta,i_0[r]}^{(H_{i_0-1} \to H_{i_0})}  \rho^{(i_0-1)} + \tilde{\tau}_{\beta,i_0[r]}^{(H_{i_0-1} \to H_{i_0})}  \rho^{(i_0-1)} - \rho^{(i_0)}}_1 \notag \\
&\le  \norm{\tilde{\tau}_{\beta,i_0[r]}^{(H_{i_0-1} \to H_{i_0})}}_{1\to 1} \norm{  \prod_{i=1}^{i_0-1} \tilde{\tau}_{\beta,i[r]}^{(H_{i-1} \to H_{i})} \rho^{(0)}-\rho^{(i_0-1)}}_1
+   \norm{ \tilde{\tau}_{\beta,i_0[r]}^{(H_{i_0-1} \to H_{i_0})}  \rho^{(i_0-1)} - \rho^{(i_0)}}_1 \notag \\
&\le i_0 \epsilon(\beta,r)   + \epsilon(\beta,r) = (i_0+1)\epsilon(\beta,r)   ,
\end{align}
where, in the second inequality, we use the assumption~\eqref{iterative_use_of/_corol_lem/main_induction} and $\norm{\tilde{\tau}_{\beta,i_0[r]}^{(H_{i_0-1} \to H_{i_0})}}_{1\to 1}=1$.
% and the inequality~\eqref{Liouvillian_le_m+1_approx}. 
This completes the proof. $\square$

{~}

\hrulefill{\bf [ End of Proof of Lemma~\ref{iterative_use_of/_corol_lem}]}

{~}

Now, the constructed recovery map $\prod_{i=1}^{\bar{n}} \tilde{\tau}_{\beta,i[r]}^{(H_{i-1} \to H_{i})}$ is supported within the distance $r$ from the region $(B_1B_2) \setminus (\check{B}_1\check{B}_2)$.
Therefore, as long as $r \le R/2 -l_H/2$, the CPTP map is supported on $B_1 B_2$, and we let 
 \begin{align}
\tau_{B_1B_2}^{(3)} = \prod_{i=1}^{\bar{n}} \tilde{\tau}_{\beta,i[R/2 -l_H/2]}^{(H_{i-1} \to H_{i})} ,
\end{align}
which gives 
 \begin{align}
\norm{ \tau_{B_1B_2}^{(3)} \rho_\beta(H_{A\check{B}_1\check{B}_2 C}) - \rho_\beta(H)} \le \bar{n} \epsilon(\beta,R/2 -l_H/2)  .
\end{align}
By using $\bar{n} \le |B|$, we derive the error by $|B| \epsilon(\beta,R/2 -l_H/2)$, which also upper-bounds the second term of the RHS in~\eqref{recovery_map_arppox_error}.
Applying the same upper bound for $\tau_{B_1B_2}^{(1)}$, we finally reduce the inequality~\eqref{recovery_map_arppox_error} to the desired upper bound~\eqref{main_ineq_thm}. 
This completes the proof. $\square$

{~}

We show a refined estimation of $\bar{n}$, which was upper-bounded by $|B|$ as a trivial bound. 
To improve it, we use the inequality of 
\begin{align}
\bar{n}\le |\partial B_1[l_H]| \le \sum_{i\in \partial B_1} |i[l_H]|  \le  |\partial B_1| \cdot \gamma l_H^D ,
\end{align}
where we use the inequality~\eqref{def:parameter_gamma}. 
On the size  $|\partial B_1|$, it is smaller than $\min (|\partial A[R/2]|, |\partial C[R/2]|)$, i.e., 
\begin{align}
|\partial B_1|\le \min (|\partial A[R/2]|, |\partial C[R/2]|) \le  \gamma (R/2)^{D-1} \min (|\partial A|, |\partial C|) .
\end{align}
By combining the above two inequalities, we have 
\begin{align}
\label{upper_bound_bar_n}
\bar{n}\le \gamma^2 l_H^D (R/2)^{D-1} \min (|\partial A|, |\partial C|)  .
\end{align}

\section{Implementation of the approximate BP channel} 
\label{sec:BP-construction}

As already stated in Theorem~\ref{thm:main}, the essential step in understanding the structure of conditional mutual information (CMI) lies in whether one can efficiently implement an approximate belief-propagation (BP) channel. In order to completely resolve the most important conjecture on CMI decay (Conjecture~1, presented in the Introduction), it is necessary to carry out this implementation unconditionally. In this work, we demonstrate that the existence of the approximate quasi-local BP channel can be rigorously established under either of the following assumptions:  
i) uniform rapid mixing, or ii) uniform clustering.  
Here, the terminology ``uniform'' is adopted from Ref.~\cite{Brandao2019}, and it refers not only to the Gibbs state of the full system, $e^{\beta H}$, but also to the Gibbs states of subsystems, $e^{\beta H_L}$ for $L \subseteq \Lambda$.

\subsection{Approximate BP channel under uniform rapid mixing}

\subsubsection{Assumptions for the Liouvillian}
In order to discuss the rapid mixing condition, we first show the assumption on the Liouvillian form in Eq.~\eqref{eq:lindblad_form}. 
\begin{assump}[Basic assumptions for quasi-local Liouvillian] \label{assup:Basic assumptions for quasi-local Liouvillian}
Let $\mL^{(H_L)}$ be a Lindblad Liouvillian with $\rho_\beta(H_L)$ its steady state for $\forall X\subseteq \Lambda$. 
We then assume the following properties for $\mL^{(H_L)}$ for $\forall L \subseteq \Lambda$:
\begin{enumerate}
\item{} (Frustration-free Lindbladian) The $\mL^{(H_L)}$ is decomposed as 
\begin{align}
\label{FF_Lindbladian_deef}
\mL^{(H_L)} = \sum_{i \in X} \mathfrak{L}^{(H_L)}_{i},  \quad  \mathfrak{L}^{(H_L)}_{i}\rho_\beta(H_L) = 0,\quad \norm{\mathfrak{L}^{(H_L)}_{i}}_{1\to 1} \le \mathfrak{g} , 
\end{align}
where each of $\{\mathfrak{L}^{(H_L)}_{i}\}_{i\in \Lambda}$ is Lindbladian, and $\mathfrak{g}$ is an $\orderof{1}$ constant.
%In the CKG Liouvillian, the norm of the local Liouvillian was upper-bounded by $3$ as in~\eqref{Liouvillian_norm_local}. 

\item{} (Quasi-locality) There is a decomposition of $\mL^{(H_L)}$ into sum of strictly local terms
\begin{align}
\label{Quasi_loacality_decomp}
\mL^{(H_L)} = \sum_{i \in L} \sum_{\ell=0}^\infty \delta \mathfrak{L}^{(H_L)}_{i[\ell]}, 
\end{align}
such that 
 \begin{align}
 \label{quasi/local_function/mathcal_J_0}
\sum_{\ell > r_1}  \norm{\delta \mathfrak{L}^{(H_L)}_{i[\ell]} }_{1\to 1} \le \mathcal{J}_0(r_1) \for \forall i\in \Lambda ,
\end{align}
where $ \mathcal{J}_0(r_1)$ is a monotonically decaying function. Note that the decomposed Liouvillian $\delta \mathfrak{L}_{i[\ell]}$ is not assumed to be given by the Lindblad form.
\item{} (Subset Liouvillian is Lindbladian) 
For any given subsets $X$ and $X'$ such that $X \subseteq X'$. The Liouvillian 
 \begin{align}
 \label{subset_Lindbladian/prop}
 \sum_{i \in X} \sum_{\ell:i[\ell]\subset X'} \delta \mathfrak{L}^{(H_L)}_{i[\ell]} 
\end{align}
is given by the Lindblad form. In particular, for $X'=X$, we denote the above one by $\mL^{(H_L)}_{X}$.

\item{} (Quasi-local stability of the Liouvillian) 
Let us define $L'=L \oplus \{i_0\}$ with $i_0 \in \Lambda \setminus L$.
Then, the difference between the Liouvillians $\mL^{(H_L)}$ and $\mL^{(H_{L'})}$ is quasi-local in the sense that 
\begin{align}
\label{error_between_L_i_L'_i_2}
\norm{ \mathfrak{L}^{(H_L)}_{i}- \mathfrak{L}^{(H_{L'})}_{i}}_{1\to 1}  \le  c_0\mathcal{J}_0(\dist_{i,i_0}) ,  
\end{align}
where $\mathfrak{L}^{(H_L)}_{i}$ and $\mathfrak{L}^{(H_{L'})}_{i}$ are decomposed terms in $\mL^{(H_L)}$ and $\mL^{(H_{L'})}$, respectively [see Eq.~\eqref{Quasi_loacality_decomp}]. 
It means that $\mathfrak{L}^{(H_L)}_{i}$ and $\mathfrak{L}^{(H_{L'})}_{i}$ are almost equal to each other as the distance $\dist_{i,i_0}$ increases:

\item{} (Uniform rapid mixing) 
For any quantum state $\sigma$, the Liouvillian $\mL^{(H_L)}$ satisfies the rapid mixing condition in the sense that 
\begin{align}
\label{assup:mixing_ineq}
\norm{e^{\mL^{(H_L)} t} \sigma - \rho_\beta(H_L)}_1 \le  C_0 |L|^\nu e^{-t \Delta}  ,
\end{align}
where $C_0$, $\nu$ and $\Delta$ are $\orderof{1}$ constant. 
\end{enumerate}

\end{assump}
{\bf Remark.} As shown in Appendix~\ref{Review:CKG}, the CKG Liouvillian~\eqref{Dissipative Dynamics_Liuou_def} satisfies the properties (1)-(4) in Assumption~\ref{assup:Basic assumptions for quasi-local Liouvillian}. The first property has been ensured as in Ref.~\cite{chen2023efficient}. 
The second-to-fourth properties will been given in Lemmas~\ref{lem:quasi-locality_mL},~\ref{lem:subset_Liouvillian}, and \ref{lem:Perturbed Liouvillian_locality}, respectively. 
%The third property has been proved in Lemma~\ref{lem:subset_Liouvillian}. The fourth one will be proven in Lemma~\ref{lem:Perturbed Liouvillian_locality}
From the inequality~\eqref{mL_decay/func_sum}, it is sufficient to consider the form of 
\begin{align}
\mathcal{J}_0(r) =
\Theta(1) e^{-\mu r} .
\label{choice_of_mathcal/J_0_r}
\end{align}
Note that we can let $\mu=1$ in \eqref{mL_decay/func_sum}. 

On the last property of the rapid mixing condition, to be more precise, it is enough to consider $\sigma=e^{\beta H_{L'}}$ with $|L'\setminus L| = 1$ for the subset Hamiltonian update (Def.~\ref{def:subset-expansion}).
The rapid mixing itself is not straightforward to verify in general; so far, only a specific cases can be proven. 
At high temperatures, the condition universally holds as shown in Corollary~\ref{corol:chi_2_divergenece_bound}. 
Another interesting case is weakly interacting fermions at arbitrary temperatures, which has been recently shown in Ref.~\cite{h1dx-ps5p,smid2025}. 

As a relevant remark, the adiabatic preparation for the purified quantum Gibbs state is often used to prepare a quantum Gibbs state on a quantum computer.
However, as shown in Appendix~\ref{Sec:Why Lindblad dynamics is required}, we have to treat the dissipative dynamics without relying on the purification.

Under the above assumptions, we prove the existence of the approximate BP channel.  
We prove the following theorem:
\begin{theorem} \label{thm_high_temp_BP}
Let us consider two subsets $H_{L}$ and $H_{L'}$ with their quantum Gibbs states $\rho_{\beta}(H_L)$ and  $\rho_{\beta}(H_{L'})$, respectively. 
Then, under the properties in Assumption~\ref{assup:Basic assumptions for quasi-local Liouvillian} with $\mathcal{J}_0(r)$ in Eq.~\eqref{choice_of_mathcal/J_0_r}, there exists an approximate BP channel $ \tilde{\tau}_{\beta,i_0[r]}^{(H_L \to H_{L'})}$ satisfying 
\begin{align}
\norm{ \tilde{\tau}_{\beta,i_0[r]}^{(H_L \to H_{L'})}\brr{ \rho_\beta(H_L)} - \rho_\beta(H_{L'})}_1\le  \epsilon(\beta,r),
\end{align}
with
\begin{align}
\label{thm_high_temp_BP_main}
\epsilon(\beta,r) \le \frac{n^\nu \Theta\br{r^{2D+1}}}{\Delta} e^{-\Theta(1) (r\Delta)^{1/(D+3)}},
\end{align}
where explicit $\beta$ dependence is absorbed to $\Delta$, the rate of the rapid mixing~\eqref{assup:mixing_ineq}. 
\end{theorem}

\subsection{Approximate BP channel under uniform clustering} \label{sec:BP channel_low-temperature}

We then consider the existence of the quasi-local BP channel under uniform clustering conditions as follows:
\begin{assump}[Uniform Clustering Property]\label{def:uniform_clustering}
Let $H_L$ be an arbitrary subset Hamiltonian defined in Eq.~\eqref{def:Hamiltonian_subset_L}.
Then, for $\forall L\subseteq \Lambda$, the quantum Gibbs state $\rho_{\beta}(H_L):= e^{-\beta H_L}/\tr(e^{-\beta H_L})$ satisfies the clustering condition as follows:
\begin{align}
\label{def:uniform_clustering_inq}
&\abs{\Cor_{\rho_{\beta}(H_L)}(O_X,O_Y) } \le C_1 \min( |X|, |Y| ) e^{- \dist_{X,Y}/\xi } ,
\end{align}
with 
\begin{align}
& \Cor_{\rho_{\beta}(H_L)}(O_X,O_Y) := \tr \brr{ \rho_{\beta}(H_L) O_X O_Y } - \tr \brr{\rho_{\beta}(H_L) O_X}  \tr \brr{\rho_{\beta}(H_L) O_Y} 
\end{align}
for $X,Y \subseteq L$,
where we set $\norm{O_X} = \norm{O_Y} =1$. 
\end{assump}

Under the uniform clustering~\ref{def:uniform_clustering}, one can prove the following theorem:
\begin{theorem} \label{thm_low_temp_BP}
Let $\rho_{\beta}(H_L)$ and  $\rho_{\beta}(H_{L'})$ be the quantum Gibbs states with $|L' \setminus L|=1$. 
Then, under Assumption~\ref{def:uniform_clustering}, there exists an approximate BP channel $ \tilde{\tau}_{\beta,i_0[r]}^{(H_L \to H_{L'})}$ satisfying 
\begin{align}
\norm{ \tilde{\tau}_{\beta,i_0[r]}^{(H_L \to H_{L'})}\brr{ \rho_\beta(H_L)} - \rho_\beta(H_{L'})}_1\le  \epsilon(\beta,r),
\end{align}
with
\begin{align}
\label{thm_low_temp_BP_main}
\epsilon(\beta,r) \le e^{\Theta(\beta) - \Theta(1) \kappa_\beta (r/\xi_\beta)^{1/D}}  + \Theta(n) e^{-\Theta(r)/\tilde{\xi}_\beta},
\end{align}
where $\kappa_\beta=\min(1/\beta,1/\xi)$, and $\xi_\beta$ is a constant which depends on $\beta$.
\end{theorem}

\section{Proof of Theorem~\ref{thm_high_temp_BP}}\label{Pr:thm_high_temp_BP}

% based on Corollaries~\ref{corol:chi_2_divergenece_bound} and \ref{corol:approx_local_Liou}
\subsection{Proof strategy} \label{sec:Proof of Subtheorem_subthm:local_perturbation}

In the proof, for simplicity of notations, we denote 
\begin{align}
\label{special_notations}
&H_L \to H, \quad H_{L'} \to H' , \quad \rho_{\beta}(H_L) \to \rho_0, \quad \rho_{\beta}(H_{L'}) \to \rho',  \notag \\
&\mL^{(H_L)} \to \mL,\quad  \mL^{(H_{L'})} \to \mL',    
\end{align}
and 
\begin{align}
\label{unperturbed/Liou}
\mL = \sum_{i \in \Lambda} \mathfrak{L}_{i} , \quad  \mL'=\sum_{i \in \Lambda} \mathfrak{L}'_{i} . 
\end{align}
Here, for $L' \setminus L=\{i_0\}$, Assumption~\ref{assup:Basic assumptions for quasi-local Liouvillian} implies that the difference between $\mathfrak{L}_{i}$ and $\mathfrak{L}'_{i}$ becomes smaller as the distance $\dist_{i,i_0}$ increases:
\begin{align}
\label{error_between_L_i_L'_i_2}
\norm{ \mathfrak{L}_{i}- \mathfrak{L}'_{i}}_{1\to 1}  \le  c_0\mathcal{J}_0(\dist_{i,i_0}) . 
\end{align}

For the proof of Theorem~\ref{thm_high_temp_BP}, we consider the convergence of the quantum state $\rho$ to $\rho'$ by the dissipative dynamics $e^{\mL' t}$.
Using the inequality~\eqref{assup:mixing_ineq}, we have 
\begin{align}
\label{rapid_mixing_apply}
\norm{ e^{\mL' t} \rho_0 - \rho'}_1  \le C n^\nu e^{-t \Delta} ,
\end{align}
where we use $|L|, |L'| \le |\Lambda|=n$. 
Therefore, by choosing $t$ smaller than $\log(n)/\Delta$, one can prove that the quantum states $e^{\mL' t} \rho_0$ and $\rho'$ are sufficiently close to each other. 

Then, the primary challenge here is the local reduction of the Liouville dynamics $e^{\mL' t}$. 
By proving that $e^{\mL' t}$ is approximated by a local CPTP map $\tau_{i_0[r]} (t)$ supported on a subset $i_0[r]$, we are able to prove the main theorem.

\subsection{Dynamics by the perturbed Liouvillian} \label{sec:Dynamics by the perturbed Liouvillian}

To achieve this, we make use of dissipative dynamics. Specifically, we show that if a suitably defined Liouvillian satisfies the rapid-mixing condition, then one can construct a local dissipative evolution that connects the thermal states of $H$ and $H + h_i$.
Crucially, the high-temperature assumption in our setting plays an essential role in ensuring that the Liouvillian indeed exhibits rapid mixing. This property underpins the locality and convergence behavior of the recovery maps we construct.

A key mathematical challenge in our approach lies in approximating short-time Liouville dynamics by a local CPTP map. More precisely, suppose we are given a Liouvillian $\mathcal{L}$ and its steady state $\rho_0$:
\begin{align}
e^{\mathcal{L} t} \rho_0 = \rho_0 .
\end{align}
We then consider a quasi-local perturbation $\delta \mathcal{L}_{i_0}$ supported near site $i_0$, and study the perturbed generator $\mathcal{L} + \delta \mathcal{L}_{i_0}$:
\begin{align}
\label{perturbed_dybamics_eq}
e^{\mathcal{L}' t} \rho_0=e^{(\mathcal{L}+\delta \mathcal{L}_{i_0}) t} \rho_0  
\end{align}
with $\mathcal{L}' =\mathcal{L}+\delta \mathcal{L}_{i_0}$. 

It is expected that $\rho_0$ remains unchanged in regions far away from the perturbation. 
This leads us to the following fundamental question:
\begin{center}
\textbf{Question.} 
\RaggedRight
Can we approximate the dynamics by using a local Liouvillian $\mathcal{L}'_{i_0[r]}$ around the site $i_0$, where $\mathcal{L}_{i_0[r]}$ is the local approximation onto the ball region $i_0[r]$ with radius $r$ centered at the site $i_0$.
That is, our problem is to answer 
\begin{align}
\label{perturbed_dybamics_eq_question}
e^{\mathcal{L}' t} \rho_0
\overset{?}{\approx}e^{\mathcal{L}'_{i_0[r]} t} \rho_0  .
\end{align}
\end{center}

In the case where the Liouvillian is exactly local and frustration-free, i.e., 
\begin{align}
\mathcal{L}=\sum_{Z:|Z|\le k} \mathfrak{L}_Z ,\quad \mathcal{L}'=\mathcal{L}+ \sum_{Z:Z\ni i_0} \mathfrak{L}'_Z  
\end{align}
with $\mathfrak{L}_Z\rho_0=0$, we can easily prove the relation~\eqref{perturbed_dybamics_eq_question} using similar analysis to the Liouvillain Lieb--Robinson bound~\cite{PhysRevLett.104.190401,PhysRevLett.108.230504} (see also Ref.~\cite[Lemmas~12 and 13]{10.1063/1.4932612}).
However, when the Liouvillian becomes quasi-local, the analyses turned out to be highly challenging. 
One of the technical contributions of this work is to provide a general and rigorous answer to this question.
We establish a universal approximation result for quasi-local Liouvillian perturbations (see Subtheorem~\ref{thm:approx_local_Liou} in Section~\ref{sec:Dynamics by the perturbed Liouvillian}).
%n the proof of the clustering theorem for the CMI, the optimal estimation of the approximation error for~\eqref{perturbed_dybamics_eq_question} plays a crucial role.

\subsubsection{Critical difference from the unitary dynamics }
 In what follows, we first discuss the challenge of the local approximation compared to unitary dynamics and then prove the equation~\eqref{perturbed_dybamics_eq_question}. 
 
In the unitary dynamics, we can write
\begin{align}
e^{(\mathcal{L}+\delta \mathcal{L}_{i_0}) t} \rho_0 = e^{-i(H+v_{i_0})t} \rho_0e^{i(H+v_{i_0})t} ,
\end{align}
where $v_{i_0}$ is a quasi-local operator around the site $i_0$.
By decomposing the unitary operator as 
\begin{align}
e^{i(H+v_{i_0})t} = e^{iHt} \mathcal{T} e^{i \int_0^t v_{i_0}(H,-x)dx}  ,
\end{align}
we have
\begin{align}
e^{-i(H+v_{i_0})t} \rho_0e^{i(H+v_{i_0})t} = \br{\mathcal{T} e^{i \int_0^t v_{i_0}(H,-x)dx}  }^\dagger  \rho_0 \mathcal{T} e^{i \int_0^t v_{i_0}(H,-x)dx}  ,
\end{align}
where we use $e^{-iHt}\rho_0e^{iHt}=\rho_0$.
Then, the Lieb--Robinson bound immediately yields the local approximation of the dynamics by $v_{i_0}(H_0,-x)= e^{-iH_0x} v_{i_0} e^{iH_0x}$.
On the other hand, we have 
\begin{align}
e^{\mathcal{L}' t} =  \mathcal{T} e^{ \int_0^t e^{\mathcal{L}x}\delta \mathcal{L}_{i_0} e^{-\mathcal{L}x}dx}  e^{\mathcal{L}t} ,
\end{align}
but the quasi-locality of $e^{\mathcal{L}x}\delta \mathcal{L}_{i_0}  e^{-\mathcal{L}x}$ cannot be treated by the standard Lieb--Robinson bound.
We need to rely on the standard expansion 
\begin{align}
 e^{\mathcal{L}x}\delta \mathcal{L}_{i_0} e^{-\mathcal{L}x} = \sum_{m=0}^\infty \frac{x^m}{m!} \ad_{\mathcal{L}}^m (\delta \mathcal{L}_{i_0}) ,
\end{align}
which, similar to the imaginary time evolution, diverges beyond a threshold of $x>0$. 
Even though the above expansion converges, we have another problem: the Liouvillian $ e^{\mathcal{L}x}\delta \mathcal{L}_{i_0} e^{-\mathcal{L}x} $ is no longer given by the Lindbladian.

\subsection{Local reduction of perturbed dynamics: main technical theorem}

In this section, we generally prove that the dynamics~\eqref{perturbed_dybamics_eq} can be approximated by local Lindblad dynamics (see Section~\ref{proof_Thm_approx_local_Liou} below for the proof):
\begin{subtheorem} \label{thm:approx_local_Liou} 
% and we assume that $ \mathcal{J}_0(r)$ decays faster than $r^{-3D}$.  
Let us assume that the Liouvillian satisfies the properties (1)-(4) in Assumption~\ref{assup:Basic assumptions for quasi-local Liouvillian}. 
Under the notations of~\eqref{special_notations}, \eqref{unperturbed/Liou} and \eqref{error_between_L_i_L'_i_2}, 
one can construct a local CPTP map $\tau_{i_0[r]} (t)$ on $i_0[r]$ that approximates the dynamics $e^{t\mL'}\rho_0$ up to an error of 
\begin{align}
\label{thm:approx_local_Liou_main} 
\norm{\brr{ e^{\mathcal{L}'t} - \tau_{i_0[r]} (t)}  \rho_0} _1\le t \Theta\br{r^{2D}/\ell_0^{D}} e^{-\mu \ell_0}, 
\end{align}
where the length $\ell_0$ is chosen as follows:
\begin{align}
\label{Eq:choice_of_ell_0}
\ell_0 = \Theta(1) \br{\frac{r}{\mu t}}^{1/(D+2)}.
\end{align}
Note that $\mu$ has been defined in Eq.~\eqref{choice_of_mathcal/J_0_r}.
\end{subtheorem}

From the subtheorem, one can prove the local approximation of $e^{\mathcal{L}'t}$ onto a local region $i_0[r]$ with a sub-exponentially decaying error.

\subsection{Completing the proof of Theorem~\ref{thm_high_temp_BP}}

We now have all the ingredients for the proof. 
We begin with the triangle inequality of 
\begin{align}
\norm{ \tau_{i_0[r]} (t) \rho_0 - \rho'}_1 
&=\norm{ \tau_{i_0[r]} (t) \rho_0 - e^{\mathcal{L}'t} \rho_0 + e^{\mathcal{L}'t} \rho_0 - \rho'}_1 \notag \\
&\le\norm{  \brr{e^{\mathcal{L}'t}  - \tau_{i_0[r]} (t)} \rho_0} + \norm{e^{\mathcal{L}^{(H')}t}  \rho_0- \rho_\beta(H')}_1 .
\end{align}
Then, by combining the inequality~\eqref{rapid_mixing_apply} and Subtheorem~\ref{thm:approx_local_Liou}, we reduce the above inequality to 
\begin{align}
\label{Final_error_t_not_chosen}
\norm{ \tau_{i_0[r]} (t) \rho_0 - \rho'}_1 
\le  t \Theta\br{r^{2D}} e^{-\Theta(1) (r/t)^{1/(D+2)}}+ C n^\nu e^{-t \Delta} . 
\end{align}

Finally, choosing $t=t_0$ such that 
 \begin{align}
\exp\brr{- \Theta(1) \br{\frac{r}{t_0}}^{1/(D+2)}} = e^{-t_0 \Delta}
\longrightarrow   t_0 = \Theta(1) r^{1/(D+3)} \Delta^{-(D+2)/(D+3)}  ,
\end{align}
we obtain 
 \begin{align}
\textrm{RHS of~\eqref{Final_error_t_not_chosen}} \le \frac{n^\nu \Theta\br{r^{2D+1}}}{\Delta} e^{-\Theta(1) (r\Delta)^{1/(D+3)}}.
\end{align}
Therefore, by choosing $\tau_{i_0[r]} (t_0) $ as $\tilde{\tau}_{\beta,i_0[r]}^{(H\to H')}$, we prove the main inequality~\eqref{thm_high_temp_BP_main}
This completes the proof. $\square$

\subsection{Proof of Subtheorem~\ref{thm:approx_local_Liou}} \label{proof_Thm_approx_local_Liou}

 \begin{figure}[tt]
\centering
\includegraphics[clip, scale=0.35]{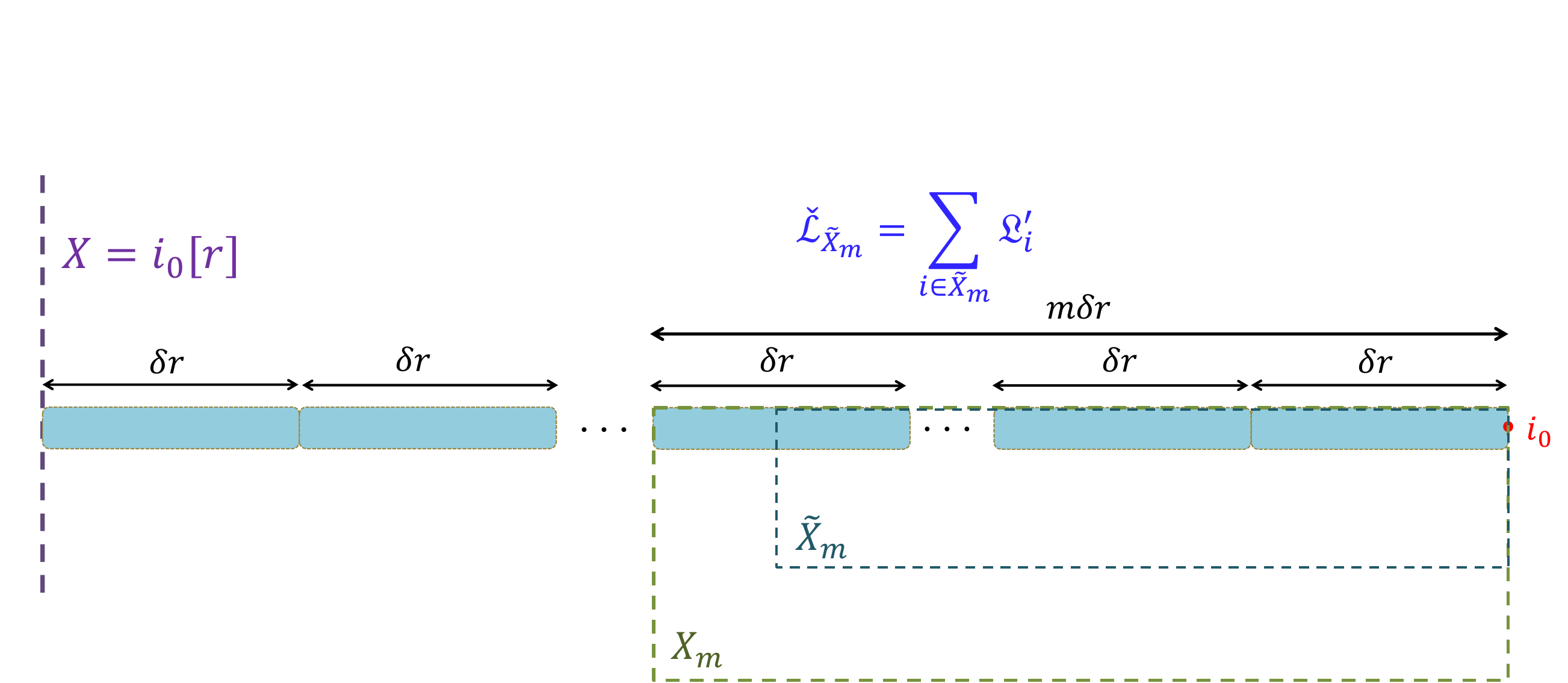}
\caption{Schematic picture of the decomposition of the system (1D case).
The target region is now given by $X=i_0[r]$, and we define $X_m$ as $i_0[r_m]$ with $r_m=m\delta r$
 and consider the step-by-step approximations. In the time evolution of the first piece $\delta t$, we approximate the dynamics in the region $X_1$ [see~\eqref{error_estimation_starting}]. In the second piece of the time evolution, we approximate it in the region $X_2$ [see~\eqref{second_step_approx}].
 By slightly extending the dynamics, we generally approximate the dynamics in the region $X_M=i_0[r]$ up to an approximation error~\eqref{approx_M_mL_delta t}. 
The approximation error in each of the time step is evaluated in Proposition~\ref{prop:main_ineq_short_time}. 
}
\label{fig_Stability}
\end{figure}

A primary challenge for the proof originates from the fact that the frustration-free Liouvillian does not necessarily satisfy the exact locality.
Under the assumption of strict $k$-locality and frustration freeness, one can derive the statement~\eqref{thm:approx_local_Liou_main} by following the same analyses as in Ref.~\cite[Lemmas~12 and 13]{10.1063/1.4932612}.
However, in extending the quasi-local cases, we suffered from the errors originating from
\begin{align}
\norm{\br{\mathfrak{L}_i -  \tilde{\mathfrak{L}}_{i[\ell]}} \rho_0 }_1 \for \forall \ell <\infty  .
\end{align}
Unlike the cases of the Lieb--Robinson bound, this error norm makes the analyses of the approximation~\eqref{thm:approx_local_Liou_main} significantly more complicated even for small $t=\orderof{1}$\footnote{A simple application of the techniques in Ref.~\cite{10.1063/1.4932612} leads to the upper bound as $\mathcal{J}_0(r)e^{\Omega(r^{D})}$, which is meaningless for all $r$ except in 1D case.}. 

To treat the approximation error~\eqref{thm:approx_local_Liou_main}, we adopt the decomposition technique which has been employed in Refs.~\cite{Kuwahara_2016_njp,PhysRevX.10.031010,kuwahara2020absence,PhysRevLett.127.070403,kuwahara2022optimal}.
For the purpose, we decompose the total time $t$ and length $r$ into $M$ pieces, 
and letting
  \begin{align}
  \label{delta_t_delta_r_def}
\delta t = \frac{t}{M},\quad  \delta r = \frac{r}{M+1} ,
\end{align}
where the number $M$ is determined afterward\footnote{We chose $\delta r\equiv r/(M+1)$ instead of $r/M$ so that  $\dist_{X_M,X^\co} \ge \delta r$.
This condition will be used in the inequality~\eqref{upper_bund_summation_J_0}.}.
Moreover, we define the subsets $\{X_m\}$ as follows (see Fig.~\ref{fig_Stability}):
 \begin{align}
\label{subset/definitions/X2}
&X_m= i_0[ r_m] , \quad r_m=m \delta r ,\notag \\
&\tilde{X}_m= i_0[\tilde{r}_m] ,  \quad \tilde{r}_m= \br{m-\frac{1}{2}} \delta r .
\end{align}
In particular, we denote $X$ by
 \begin{align}
\label{subset/definitions/X_M}
X= X_{M+1} = i_0[r].
\end{align}

Then, we start from the approximation of $e^{\mL' t}\rho_0$. 
For this purpose, we consider the triangle inequality as follows:
\begin{align}
&\norm{\br{ e^{\mathcal{L}'\delta t} - e^{\bar{\mL}_{X_1} \delta t}  }  \rho_0} _1
\le \norm{\br{ e^{ \check{\mL}_{\tilde{X}_1}\delta t }  - e^{\bar{\mL}_{X_1} \delta t} } \rho_0} _1 +\norm{\br{ e^{\mathcal{L}'\delta t} - e^{ \check{\mL}_{\tilde{X}_1} \delta t}}  \rho_0} _1 ,
\label{error_estimation_starting}
\end{align}
where $ \check{\mL}_{\tilde{X}_1}$ and $\bar{\mL}_{X_1}$ are defined as 
\begin{align}
\label{definition_check_mL_tildeX_1}
\check{\mL}_{\tilde{X}_1}:=   \sum_{i \in \tilde{X}_1} \mathfrak{L}'_{i}= \sum_{i \in \tilde{X}_1}  \sum_{\ell=0}^\infty \delta \mathfrak{L}'_{i[\ell]} , 
\end{align}
and 
\begin{align}
\label{definition_bar_mL_X_1}
\bar{\mL}_{X_1}:= \sum_{i \in \tilde{X}_1} \sum_{\ell: i[\ell] \subseteq X_1} \delta \mathfrak{L}'_{i[\ell]}  .
\end{align}
We adopt the same definitions for $\check{\mL}_{\tilde{X}_m}$ and $\bar{\mL}_{X_m}$ for $m\in [1,M]$. 
We defer the estimation of the norms in the RHS of~\eqref{error_estimation_starting}. 

In the next step, we consider the approximation of 
\begin{align}
\label{second_step_approx}
\norm{\br{ e^{2\mathcal{L}'\delta t} - e^{\bar{\mL}_{X_2} \delta t} e^{\bar{\mL}_{X_1} \delta t}  }  \rho_0} _1 
&=\norm{ e^{\mathcal{L}' \delta t} \br{e^{\mathcal{L}'\delta t} -e^{\bar{\mL}_{X_1} \delta t}  }  \rho_0 
+ \br{e^{\mathcal{L}'\delta t} -e^{\bar{\mL}_{X_2} \delta t}} e^{\bar{\mL}_{X_1} \delta t}   \rho_0 
} _1 \notag \\
&\le 
\norm{\br{e^{\mathcal{L}'\delta t} -e^{\bar{\mL}_{X_1} \delta t}  }  \rho_0 }_1
+ \norm{\br{e^{\mathcal{L}'\delta t} -e^{\bar{\mL}_{X_2} \delta t}} \rho_{X_1}  
} _1 ,
\end{align}
where $\rho_{X_1}:= e^{\bar{\mL}_{X_1} \delta t}   \rho_0 $, and we use  $\norm{e^{\mathcal{L}'\delta t}}_{1\to 1}\le 1$. 
By repeating the same processes, we get 
\begin{align}
\label{approx_M_mL_delta t}
\norm{\br{ e^{M\mathcal{L}'\delta t} - e^{\bar{\mL}_{X_M} \delta t} e^{\bar{\mL}_{X_{M-1}} \delta t} \cdots e^{\bar{\mL}_{X_1} \delta t}  }  \rho_0} _1 
&\le \sum_{m=1}^{M}  
\norm{\br{e^{\mathcal{L}'\delta t} -e^{\bar{\mL}_{X_m} \delta t}  }  \rho_{X_{m-1}} } _1 ,
\end{align}
with 
\begin{align}
\label{Eq_rho_X_m-1}
\rho_{X_{m-1}}= e^{\bar{\mL}_{X_{m-1}} \delta t} e^{\bar{\mL}_{X_{m-2}} \delta t} \cdots e^{\bar{\mL}_{X_1} \delta t} \rho_0 .
\end{align}

The primary technical ingredient is the following statement:
\begin{prop} \label{prop:main_ineq_short_time}
Let $\ell_0$ be an arbitrary positive integer such that $\ell_0 \le \delta r/2$.
We then choose the integer $M$ so that $\delta t$ may satisfy 
\begin{align}
\delta t = \frac{t}{M}\le  \frac{1}{e \zeta_{\ell_0}}=\frac{1}{2e \gamma (2\ell_0)^D \mathfrak{g}} , 
\end{align}
we get the upper bound of 
\begin{align}
\label{prop:main_ineq_short_time/main;ineq}
\norm{\br{e^{\mathcal{L}'\delta t} -e^{\bar{\mL}_{X_m} \delta t}  }  \rho_{X_{m-1}} } _1 
\le \Theta(r^D \delta t )  \brr{r^D \delta t  \mathcal{J}_0(\ell_0)  +e^{-\delta r/ (4\ell_0)} } ,
\end{align}
where we adopt the notation of Eq.~\eqref{definition_bar_mL_X_1} for the approximate Liouvillian $\bar{\mL}_{X_m} $.
\end{prop}

By applying Proposition~\ref{prop:main_ineq_short_time} to the inequality~\eqref{approx_M_mL_delta t} with $M=t/\delta t$, we prove the main inequality~\eqref{thm:approx_local_Liou_main} as
\begin{align}
\label{approx_M_mL_delta t_2}
\norm{\br{ e^{M\mathcal{L}'\delta t} - e^{\bar{\mL}_{X_M} \delta t} e^{\bar{\mL}_{X_{M-1}} \delta t} \cdots e^{\bar{\mL}_{X_1} \delta t}  }  \rho_0} _1 
&\le \frac{t}{\delta t} \Theta(r^D \delta t )  \brr{r^D \delta t  \mathcal{J}_0(\ell_0)  +e^{-\delta r/ (4\ell_0)} } \notag \\
&\le t \Theta\br{r^{2D}/\ell_0^{D}} \mathcal{J}_0(\ell_0)  ,
\end{align}
where we set $\ell_0$ as in Eq.~\eqref{Eq:choice_of_ell_0} to make $e^{-\delta r/ (4\ell_0)}\le \mathcal{J}_0(\ell_0)$. 
Because of 
\begin{align}
\label{choice_M_delta_r}
M\propto t\ell_0^D \AND \delta r \propto r/(t\ell_0^D) ,
\end{align}
this condition for $l_0$ is derived by inserting $\mathcal{J}_0(\ell_0) =\Theta(1) e^{-\mu \ell_0}$ and then taking the logarithm
 \begin{align}
 \frac{r}{t\ell_0^{D+1}} \ge \Theta(1) \mu \ell_0 ,
\end{align}
which gives
\begin{align}
\ell_0 = \Theta(1) \br{\frac{r}{\mu t}}^{1/(D+2)}.
\end{align}
This completes the proof. $\square$

\subsubsection{Proof of Proposition~\ref{prop:main_ineq_short_time}}

We adopt the two-step approximation as in the inequality~\eqref{error_estimation_starting}:
\begin{align}
&\norm{\br{ e^{\mathcal{L}'\delta t} - e^{\bar{\mL}_{X_m} \delta t}  }  \rho_{X_{m-1}}} _1
\le \norm{\br{ e^{ \check{\mL}_{\tilde{X}_m}\delta t }  - e^{\bar{\mL}_{X_m} \delta t} } \rho_{X_{m-1}}} _1 
+\norm{\br{ e^{\mathcal{L}'\delta t} - e^{ \check{\mL}_{\tilde{X}_m} \delta t}}  \rho_{X_{m-1}}} _1 ,
\label{error_estimation_starting_m}
\end{align}
We start from the estimation of the first term in the RHS of \eqref{error_estimation_starting_m}. 
By using $\norm{e^{\bar{\mL}_{X_m} t}}_{1\to 1}\le 1$ and $\norm{e^{\check{\mL}_{\tilde{X}_m}t}}_{1\to 1}\le 1$ with the decomposition of 
\begin{align}
\label{decomp_dynamical_Liouvillian}
e^{ \check{\mL}_{\tilde{X}_m}\delta t} 
&=  e^{\bar{\mL}_{X_m} \delta t} + \int_0^{\delta t}   \frac{d}{d t_1} e^{\bar{\mL}_{X_m} (\delta t-t_1)} e^{ \check{\mL}_{\tilde{X}_m}t_1} dt_1\notag \\
&=  e^{\bar{\mL}_{X_m} \delta t} + \int_0^{\delta t}  e^{\bar{\mL}_{X_m} (\delta t-t_1)}  (\check{\mL}_{\tilde{X}_m}-\bar{\mL}_{X_m}) e^{ \check{\mL}_{\tilde{X}_m}t_1} dt_1 ,
\end{align}
we can derive 
\begin{align}
\norm{e^{ \check{\mL}_{\tilde{X}_m}\delta t} -e^{\bar{\mL}_{X_m}\delta t} }_{1\to 1} 
&\le  \delta t \norm{ \check{\mL}_{\tilde{X}_m}-\bar{\mL}_{X_m}}_{1\to 1}  \notag \\
&\le \delta t \sum_{i \in \tilde{X}_m}  \sum_{\ell: \ell > \delta r/2}\norm{\delta \mathfrak{L}'_{i[\ell]} }_{1\to 1} 
\le  \delta t |i_0[r]| \mathcal{J}_0(\delta r/2)  \le \delta t \Theta(r^D) \mathcal{J}_0(\delta r/2) , 
\label{error_estimation_upper_1}
\end{align}
where we use the condition~\eqref{quasi/local_function/mathcal_J_0}, $|\tilde{X}_m| \le |X_M|\le|i_0[r]|$, and the fact that 
$i[\ell] \cap X_m^\co=i[\ell] \cap i_0[r_m]^\co \neq \emptyset$ is satisfied for $\ell >\delta r/2$ as long as $i \in \tilde{X}_m=i_0[r_m-\delta r/2]$ (see also Fig.~\ref{fig_Stability}).  

We next consider the second term in the RHS of \eqref{error_estimation_starting_m}. 
we use the same decomposition as Eq.~\eqref{decomp_dynamical_Liouvillian} to obtain 
\begin{align}
e^{\mL' \delta t}
&=e^{(\check{\mL}_{\tilde{X}_m} +\check{\mL}_{X^\co}) \delta t}+  \int_0^{\delta t} e^{ \mL' (\delta t-t_1)}  (\mL' -\check{\mL}_{\tilde{X}_m} -\check{\mL}_{X^\co}) e^{(\check{\mL}_{\tilde{X}_m} +\check{\mL}_{X^\co}) t_1}  dt_1 \notag \\
&= e^{(\check{\mL}_{\tilde{X}_m} +\check{\mL}_{X^\co}) \delta t}+  \sum_{i_1\in X\setminus \tilde{X}_m }  \int_0^{\delta t} e^{ \mL' (\delta t-t_1)}  \mathfrak{L}'_{i_1} e^{(\check{\mL}_{\tilde{X}_m} +\check{\mL}_{X^\co})  t_1}  dt_1 ,
\label{exp_mL_t_decomposition}
\end{align}
where we defined $X:= i_0[r]=X_M$ as in Eq.~\eqref{subset/definitions/X_M} and $\check{\mL}_{X^c}:=   \sum_{i \in X^c} \mathfrak{L}'_{i}$. 
Using the above decomposition and the triangle inequality, we obtain 
\begin{align}
\label{exp_mL_t_decomposition_approx}
&\norm{\br{ e^{\mathcal{L}'\delta t} - e^{ \check{\mL}_{\tilde{X}_m} \delta t}}  \rho_{X_{m-1}}} _1 \notag \\
&\le  
\norm{\br{e^{ \check{\mL}_{\tilde{X}_m} \delta t} - e^{(\check{\mL}_{\tilde{X}_m} +\check{\mL}_{X^\co}) \delta t}}  \rho_{X_{m-1}}} _1
+ 
 \norm{ \left( \sum_{i_1\in X\setminus \tilde{X}_m } \int_0^{\delta t} e^{\mL'(\delta t-t_1)}\mathfrak{L}'_{i_1}e^{(\check{\mL}_{\tilde{X}_m} +\check{\mL}_{X^\co})t_1}dt_1 \right)\rho_{X_{m-1}} }_{1}
 \notag \\
&\le 
\norm{\br{e^{ \check{\mL}_{\tilde{X}_m} \delta t} - e^{(\check{\mL}_{\tilde{X}_m} +\check{\mL}_{X^\co}) \delta t}}  \rho_{X_{m-1}}} _1
+ 
\sum_{i_1\in X\setminus \tilde{X}_m }  \int_0^{\delta t}  \norm{\mathfrak{L}'_{i_1}e^{(\check{\mL}_{\tilde{X}_m} +\check{\mL}_{X^\co}) t_1} \rho_{X_{m-1}}}_1dt_1 \notag \\
&\le \norm{\br{e^{ \check{\mL}_{\tilde{X}_m} \delta t} - e^{(\check{\mL}_{\tilde{X}_m} +\check{\mL}_{X^\co}) \delta t}}  \rho_{X_{m-1}}} _1 \notag \\
&+ \sum_{i_1\in X\setminus \tilde{X}_m }  \int_0^{\delta t}  
\br{ \norm{  \mathfrak{L}'_{i_1}}_{1\to1} \cdot\norm{\br{e^{ \check{\mL}_{\tilde{X}_m}t_1} - e^{(\check{\mL}_{\tilde{X}_m} +\check{\mL}_{X^\co}) t_1}}  \rho_{X_{m-1}}} _1
+\norm{ \mathfrak{L}'_{i_1} e^{\check{\mL}_{\tilde{X}_m} t_1}  \rho_{X_{m-1}}}_1 }dt_1. 
\end{align}

To reduce the above upper bound, we need to prove 
\begin{align}
\label{exp_mL_t_decomposition_approx_term1}
\norm{\br{e^{ \check{\mL}_{\tilde{X}_m} \delta t} - e^{(\check{\mL}_{\tilde{X}_m} +\check{\mL}_{X^\co}) \delta t}}  \rho_{X_{m-1}}} _1
\le \delta t \Theta(r^D) \mathcal{J}_0(\delta r/2)  
\end{align}
and 
\begin{align}
\label{exp_mL_t_decomposition_approx_term2}
\sum_{i_1\in X\setminus \tilde{X}_m }  \int_0^{\delta t} \norm{ \mathfrak{L}'_{i_1} e^{\check{\mL}_{\tilde{X}_m} t_1}  \rho_{X_{m-1}}}_1 dt_1
\le \Theta(r^D \delta t )  \brr{r^D \delta t  \mathcal{J}_0(\ell_0)  +e^{-\delta r/ (4\ell_0)} } ,
\end{align}
separately. 
By applying the inequalities~\eqref{exp_mL_t_decomposition_approx_term1} and \eqref{exp_mL_t_decomposition_approx_term2} to~\eqref{exp_mL_t_decomposition_approx}, we prove 
\begin{align}
\label{exp_mL_t_decomposition_approx_fin}
&\norm{\br{ e^{\mathcal{L}'\delta t} - e^{ \check{\mL}_{\tilde{X}_m} \delta t}}  \rho_{X_{m-1}}} _1\le  
\Theta(r^D \delta t )  \brr{r^D \delta t  \mathcal{J}_0(\ell_0)  +e^{-\delta r/ (4\ell_0)} } ,
\end{align}
where we have chosen $\ell_0$ so that $\ell_0 \le \delta r/2$. This choice is indeed satisfied in~\eqref{choice_M_delta_r}.
By combining the upper bounds~\eqref{error_estimation_upper_1} and~\eqref{exp_mL_t_decomposition_approx_fin}, 
we prove the main inequality~\eqref{prop:main_ineq_short_time/main;ineq}.
This completes the proof of the proposition~\ref{prop:main_ineq_short_time}.
$\square$

{~}\\

\noindent
{\bf [Proof of the inequality~\eqref{exp_mL_t_decomposition_approx_term1}]}

For this purpose, we first utilize the approximation 
\begin{align}
\check{\mL}_{\tilde{X}_m} +\check{\mL}_{X^\co} \approx \bar{\mL}_{X_m} + \bar{\mL}_{X_m^\co},
\end{align}
where $\bar{\mL}_{X_m}$ was defined by Eq.~\eqref{definition_bar_mL_X_1} and we define $\bar{\mL}_{X_m^\co}$ for $m\in [1,M]$ as 
\begin{align}
\label{definition_bar_mL_X_m^co}
\bar{\mL}_{X_m^\co}:= \sum_{i \in X^\co} \sum_{\ell: i[\ell] \subseteq X_m^\co} \delta \mathfrak{L}'_{i[\ell]}  .
\end{align}
Using the same inequality as~\eqref{error_estimation_upper_1}
\begin{align}
\norm{e^{(\check{\mL}_{\tilde{X}_m} +\check{\mL}_{X^\co}) \delta t} -
e^{(\bar{\mL}_{X_m} + \bar{\mL}_{X_m^\co}) \delta t} }_{1\to 1} 
&\le  \delta t \norm{ \check{\mL}_{\tilde{X}_m}-\bar{\mL}_{X_m}}_{1\to 1} 
+ \delta t \norm{\check{\mL}_{X^\co} - \bar{\mL}_{X_m^\co} }_{1\to 1} 
 \notag \\
&\le \delta t \sum_{i \in \tilde{X}_m}  \sum_{\ell: \ell > \delta r/2}\norm{\delta \mathfrak{L}'_{i[\ell]} }_{1\to 1} 
+\delta t \sum_{i \in X^\co}  \sum_{\ell: \ell > \dist_{i,X_m}}\norm{\delta \mathfrak{L}'_{i[\ell]} }_{1\to 1}  \notag \\
&\le  \delta t |i_0[r]| \mathcal{J}_0(\delta r/2) + 
 \delta t \sum_{i \in X^\co} \mathcal{J}_0(\dist_{i,X_m})  
\le \delta t \Theta(r^D) \mathcal{J}_0(\delta r/2) , 
\label{error_estimation_upper_1_rre}
\end{align}
where, in the last inequality, we use the definitions of $X_m=i_0[r_m]$ and $X=i_0[r]$ to obtain 
\begin{align}
\sum_{i \in X^\co} \mathcal{J}_0(\dist_{i,X_m}) 
&\le 
\sum_{s=1}^\infty \sum_{i \in \partial i_0[r+s]} \mathcal{J}_0(\delta r + s) \notag \\
&\le 
\sum_{s=1}^\infty (r+s)^{D-1} \mathcal{J}_0(\delta r + s) 
\le \Theta(r^D) \mathcal{J}_0(\delta r) .
\label{upper_bund_summation_J_0}
\end{align}
Note that as long as $m\le M$, we have $r-r_m \ge \delta r$ from the definitions~\eqref{delta_t_delta_r_def} and \eqref{subset/definitions/X2}. 

By applying the inequality~\eqref{error_estimation_upper_1_rre} to the LHS of~\eqref{exp_mL_t_decomposition_approx_term1}, we obtain 
\begin{align}
\label{error_estimation_upper_1_rre_00}
\norm{\br{e^{ \check{\mL}_{\tilde{X}_m} \delta t} - e^{(\check{\mL}_{\tilde{X}_m} +\check{\mL}_{X^\co}) \delta t}}  \rho_{X_{m-1}}} _1
\le \delta t\Theta(r^D) \mathcal{J}_0(\delta r/2) + \norm{ \br{e^{ \check{\mL}_{\tilde{X}_m} \delta t} - e^{\bar{\mL}_{X_m} \delta t} e^{\bar{\mL}_{X_m^\co} \delta t}  }\rho_{X_{m-1}}} _1 ,
\end{align}
where we use $[\bar{\mL}_{X_m},\bar{\mL}_{X_m^\co}]=0$ to get $e^{(\bar{\mL}_{X_m} + \bar{\mL}_{X_m^\co}) \delta t}=e^{\bar{\mL}_{X_m} \delta t} e^{\bar{\mL}_{X_m^\co} \delta t}$. 
For the second term in the RHS of the above inequality, we consider 
\begin{align}
\label{error_estimation_upper_1_rre_1}
 \norm{ \br{e^{ \check{\mL}_{\tilde{X}_m} \delta t} - e^{\bar{\mL}_{X_m} \delta t} e^{\bar{\mL}_{X_m^\co} \delta t}  }\rho_{X_{m-1}}} _1 
 &=\norm{ e^{ \bar{\mL}_{X_m} \delta t}\br{
 1-e^{\bar{\mL}_{X_m^\co} \delta t}  }\rho_{X_{m-1}}
 +
 \br{e^{ \check{\mL}_{\tilde{X}_m} \delta t} - e^{\bar{\mL}_{X_m} \delta t}  }\rho_{X_{m-1}}
 } _1  \notag \\
 &\le \norm{\br{ e^{\bar{\mL}_{X_m^\co} \delta t} -1 }\rho_{X_{m-1}}} _1 +  \delta t \Theta(r^D) \mathcal{J}_0(\delta r/2)  ,
\end{align}
where in the last inequality, we use the upper bound~\eqref{error_estimation_upper_1}. 
For the first term, using the form of Eq.~\eqref{Eq_rho_X_m-1}, we obtain 
\begin{align}
\label{error_estimation_upper_1_rre_2}
\norm{\br{ e^{\bar{\mL}_{X_m^\co} \delta t} -1 }\rho_{X_{m-1}}} _1  
&= \norm{e^{\bar{\mL}_{X_{m-1}} \delta t} e^{\bar{\mL}_{X_{m-2}} \delta t} \cdots e^{\bar{\mL}_{X_1} \delta t} \br{ e^{\bar{\mL}_{X_m^\co} \delta t} -1 } \rho_0} _1  \notag \\
&\le  \norm{ \br{ e^{\bar{\mL}_{X_m^\co} \delta t} -1 } \rho_0} _1   
=  \norm{ \br{ e^{\check{\mL}_{X^\co} \delta t} - e^{\delta t \sum_{i\in X^\co}\mathfrak{L}_{i}}} \rho_0} _1   ,
 \end{align}
 where in the last equation, we use $\mathfrak{L}_{i}\rho_0=0$ for $\forall i\in \Lambda$. 
Finally, from the first inequality in~\eqref{error_estimation_upper_1_rre}, we can derive 
\begin{align}
\label{error_estimation_upper_1_rre_3}
\norm{ \br{ e^{\check{\mL}_{X^\co} \delta t} - e^{\delta t \sum_{i\in X^\co}\mathfrak{L}_{i}}} \rho_0} _1  
&\le \delta t  \sum_{i\in X^\co} \norm{\mathfrak{L}'_i - \mathfrak{L}_{i}} _{1\to1}  \notag  \\
&\le c_0 \delta t  \sum_{i\in X^\co} \mathcal{J}_0(\dist_{i,i_0})
\le  c_0 \delta t \Theta(r^D) \mathcal{J}_0(r),
 \end{align}
 where we use the condition~\eqref{error_between_L_i_L'_i_2} and the inequality~\eqref{upper_bund_summation_J_0}. 
By combining the inequalities~\eqref{error_estimation_upper_1_rre_00} and \eqref{error_estimation_upper_1_rre_3}, we prove the inequality~\eqref{exp_mL_t_decomposition_approx_term1} as follows:
\begin{align}
\label{error_estimation_upper_1_final}
 \norm{ \br{e^{ \check{\mL}_{\tilde{X}_m} \delta t} - e^{(\check{\mL}_{\tilde{X}_m} +\check{\mL}_{X^\co}) \delta t}  }\rho_{X_{m-1}}} _1 
 &\le \delta t \Theta(r^D) \mathcal{J}_0(\delta r/2)  ,
\end{align}
where we use $r\ge \delta r$, which gives $ \mathcal{J}_0(r) \le  \mathcal{J}_0(\delta r)$.

{~}\\

\noindent
{\bf [Proof of the inequality~\eqref{exp_mL_t_decomposition_approx_term2}]}

In the following, we define 
\begin{align}
\mathfrak{L}'_{i_1,\ell_0} := \sum_{\ell=0}^{\ell_0} \delta \mathfrak{L}'_{i_1[\ell]}  , \quad 
\check{\mL}_{\tilde{X}_m,\ell_0}:=   \sum_{i \in \tilde{X}_m} \sum_{\ell=0}^{\ell_0} \delta \mathfrak{L}'_{i[\ell]} , \quad 
\label{mathcal_L_L_X_1_error}
\end{align}
We recall that $i_1\in X\setminus \tilde{X}_m $. 
By using $\norm{e^{ \mL' (t-t_1)}}_{1\to1},\norm{e^{\check{\mL}_{\tilde{X}_m,\ell_0} t_1}  }_{1\to1}\le 1$ and $\norm{e^{\check{\mL}_{\tilde{X}_m} t_1}  }_{1\to1}\le 1$, we upper-bound the LHS of \eqref{exp_mL_t_decomposition_approx_term2} by
\begin{align}
&\norm{  \mathfrak{L}'_{i_1} e^{\check{\mL}_{\tilde{X}_m} t_1} \rho_{X_{m-1}}}_1  
\le \norm{  \mathfrak{L}'_{i_1,\ell_0} e^{\check{\mL}_{\tilde{X}_m} t_1} \rho_{X_{m-1}}}_1 
+ \norm{  \br{ \mathfrak{L}'_{i_1}  -\mathfrak{L}'_{i_1,\ell_0} }e^{\check{\mL}_{\tilde{X}_m} t_1} \rho_{X_{m-1}}}_1 
 \notag \\
&\le \norm{ 
 \mathfrak{L}'_{i_1,\ell_0} e^{\check{\mL}_{\tilde{X}_m,\ell_0} t_1} \rho_{X_{m-1}}}_1  
 + \norm{ \mathfrak{L}'_{i_1,\ell_0}   \br{ e^{\check{\mL}_{\tilde{X}_m,\ell_0} t_1} - e^{\check{\mL}_{\tilde{X}_m} t_1} } \rho_{X_{m-1}}}_1 
 + \norm{ \mathfrak{L}'_{i_1}  -\mathfrak{L}'_{i_1,\ell_0} }_{1\to 1} \notag \\
&=
 \norm{  e^{\check{\mL}_{\tilde{X}_m,\ell_0} t_1}
e^{-\check{\mL}_{\tilde{X}_m,\ell_0} t_1}  \mathfrak{L}'_{i_1,\ell_0}  e^{\check{\mL}_{\tilde{X}_m,\ell_0} t_1} \rho_{X_{m-1}}}_1   \notag \\
&\quad  + \norm{ \mathfrak{L}'_{i_1,\ell_0}  \br{ e^{\check{\mL}_{\tilde{X}_m,\ell_0} t_1} - e^{\check{\mL}_{\tilde{X}_m} t_1} } \rho_{X_{m-1}}}_1 
  + \norm{ \mathfrak{L}'_{i_1} -\mathfrak{L}'_{i_1,\ell_0} }_{1\to 1}  \notag \\
 &\le 
 \norm{e^{-\check{\mL}_{\tilde{X}_m,\ell_0} t_1} \mathfrak{L}'_{i_1,\ell_0} e^{\check{\mL}_{\tilde{X}_m,\ell_0} t_1} \rho_{X_{m-1}}}_1
 +  \delta t\norm{\mathfrak{L}'_{i_1,\ell_0} }_{1\to 1} \cdot \norm{ \check{\mL}_{\tilde{X}_m}-\check{\mL}_{\tilde{X}_m,\ell_0} }_{1\to 1}+ \norm{ \mathfrak{L}'_{i_1} -\mathfrak{L}'_{i_1,\ell_0} }_{1\to 1} , 
 \label{main_ineq_for_small_time_approx}
\end{align}
where, in the last inequality, we use the first inequality in~\eqref{error_estimation_upper_1} and $t_1\le \delta t$ from Eq.~\eqref{exp_mL_t_decomposition}. 
Using similar inequalities to \eqref{error_estimation_upper_1} with the condition~\eqref{quasi/local_function/mathcal_J_0}, we obtain 
 \begin{align}
  \label{third_term_upper_boud}
\norm{ \mathfrak{L}'_{i_1} -\mathfrak{L}'_{i_1,\ell_0} }_{1\to 1}
 \le \mathcal{J}_0(\ell_0) , 
\end{align}
and
 \begin{align}
 \delta t\norm{\mathfrak{L}'_{i_1,\ell_0} }_{1\to 1} \cdot \norm{ \check{\mL}_{\tilde{X}_m}-\check{\mL}_{\tilde{X}_m,\ell_0} }_{1\to 1}
 \le \delta t \Theta(r^D) \mathcal{J}_0(\ell_0) , 
 \label{second_term_upper_boud}
\end{align}
which reduce the inequality~\eqref{main_ineq_for_small_time_approx} to 
\begin{align}
&\norm{   \mathfrak{L}'_{i_1} e^{\check{\mL}_{\tilde{X}_m} t_1} \rho_{X_{m-1}}}_1  
\le  \norm{e^{-\check{\mL}_{\tilde{X}_m,\ell_0} t_1} \mathfrak{L}'_{i_1,\ell_0} e^{\check{\mL}_{\tilde{X}_m,\ell_0} t_1} \rho_{X_{m-1}}}_1
 + \brr{1+ \delta t \Theta(r^D) }\mathcal{J}_0(\ell_0)  . 
 \label{main_ineq_for_small_time_approx__2}
\end{align}

%To estimate the first term in the RHS of~\eqref{main_ineq_for_small_time_approx}, we define 
%\begin{align}
%\Delta X_m := \tilde{X}_m \setminus X_{m-1}, 
%\end{align}
%and upper-bound 
%\begin{align}
%& \norm{e^{-\check{\mL}_{\tilde{X}_m,\ell_0} t_1} \mathfrak{L}'_{i_1,\ell_0} e^{\check{\mL}_{\tilde{X}_m,\ell_0} t_1} \rho_{X_{m-1}}}_1 \notag \\
%& \le  
% \norm{e^{-\check{\mL}_{\Delta X_m,\ell_0} t_1} \mathfrak{L}'_{i_1,\ell_0} e^{\check{\mL}_{\Delta X_m,\ell_0} t_1} \rho_{X_{m-1}}}_1 
% +   \norm{e^{-\check{\mL}_{\tilde{X}_m,\ell_0} t_1} \mathfrak{L}'_{i_1,\ell_0} e^{\check{\mL}_{\tilde{X}_m,\ell_0} t_1}- e^{-\check{\mL}_{\Delta X_m,\ell_0} t_1} \mathfrak{L}'_{i_1,\ell_0} e^{\check{\mL}_{\Delta X_m,\ell_0} t_1} }_{1\to 1}  .
%\end{align}
To estimate the first term in~\eqref{main_ineq_for_small_time_approx__2}, we utilize $i_1[\ell_0] \cap X_{m-1}=\emptyset $, which is derived from $i_1\in X\setminus \tilde{X}_m $ and  $\ell_0 \le \delta r/2$. 
We then first evaluate 
\begin{align}
\norm{\mathfrak{L}'_{i_1,\ell_0}  \rho_{X_{m-1}}}_1 
&= \norm{e^{\bar{\mL}_{X_{m-1}} \delta t} e^{\bar{\mL}_{X_{m-2}} \delta t} \cdots e^{\bar{\mL}_{X_1} \delta t} \mathfrak{L}'_{i_1,\ell_0}  \rho_0 }_1 \notag \\
&\le \norm{\mathfrak{L}'_{i_1,\ell_0}  \rho_0 }_1 \le \norm{\mathfrak{L}'_{i_1}-\mathfrak{L}'_{i_1,\ell_0}}_{1\to 1} + 
\norm{\mathfrak{L}'_{i_1}  \rho_0 }_1 \notag \\
&\le  \norm{\mathfrak{L}'_{i_1}-\mathfrak{L}'_{i_1,\ell_0}}_{1\to 1} + 
\norm{\mathfrak{L}_{i_1} - \mathfrak{L}'_{i_1}}_{1\to1}+\norm{\mathfrak{L}_{i_1}  \rho_0 }_1 
 \notag \\
 &\le  \mathcal{J}_0(\ell_0) +c_0\mathcal{J}_0(\dist_{i_1,i_0}) ,
\end{align}
where we use the inequality~\eqref{third_term_upper_boud} and the condition~\eqref{error_between_L_i_L'_i_2} in the last inequality.
The condition of $i_1[\ell_0] \cap X_{m-1}=\emptyset $ implies $ \ell_0\le \dist_{i_1,X_{m-1}}\le \dist_{i_1,i_0}$, and hence $\mathcal{J}_0(\dist_{i_1,i_0})\le 
\mathcal{J}_0(\ell_0)$, which yields 
\begin{align}
\norm{\mathfrak{L}'_{i_1,\ell_0}  \rho_{X_{m-1}}}_1 
\le(1+c_0)  \mathcal{J}_0(\ell_0).
\end{align}
Using the above property, we utilize the following lemma (see Section~\ref{Proof of Lemma_lem:approx_FF_img} for the proof):
\begin{lemma} \label{lem:approx_FF_img}
Let $H$ be an arbitrary Hamiltonian in the form of 
\begin{align}
H=\sum_{i\in \Lambda} h_{i[\ell]} ,\quad \norm{h_{i[\ell]}} \le \mathfrak{g},  
\end{align}
where $h_{i[\ell]}$ acts on the subset $i[\ell]$. 
For a quantum state $\ket{\psi_X}$, we also assume that each of the interaction terms $\{h_{i,\ell}\}_{i \in \Lambda}$ satisfy
\begin{align}
\norm{h_{i[\ell]} \ket{\psi_X}} \le \epsilon_\ell \for i[\ell] \subset X^\co .
\end{align}
We obtain 
\begin{align}
 \label{lem:approx_FF_img_main_ineq}
\norm{e^{-\tau H} h_{i_0[\ell]} e^{\tau H} \ket{\psi_X}} \le 
\frac{1}{1-\zeta_\ell  |\tau|} \brr{ \epsilon_\ell +\br{\zeta_\ell |\tau|}^{\dist_{i_0,X}/(2\ell)} }
\end{align}
with
\begin{align}
 \label{lem:approx_FF_img_main_zeta_ell}
\zeta_\ell := 2 \gamma (2\ell)^D \mathfrak{g}.
\end{align}
\end{lemma}

We here use Lemma~\ref{lem:approx_FF_img} with the choices of 
\begin{align}
\ell\to \ell_0, 
\quad h_{i[\ell]} \to \mathfrak{L}'_{i,\ell_0} , 
\quad H\to  \check{\mL}_{\tilde{X}_m,\ell_0} ,  
\quad \epsilon_\ell \to (1+c_0)  \mathcal{J}_0(\ell_0)
\end{align}
Note that the norm of the local Liouvillian has been upper-bounded by $\mathfrak{g}$ as in~\eqref{FF_Lindbladian_deef}. 
Then, we obtain 
\begin{align}
\norm{e^{-\check{\mL}_{\tilde{X}_m,\ell_0} t_1} \mathfrak{L}'_{i_1,\ell_0} e^{\check{\mL}_{\tilde{X}_m,\ell_0} t_1} \rho_{X_{m-1}}}_1
\le \frac{1}{1-\zeta_{\ell_0}  |\delta t|} \brr{ \epsilon_{\ell_0} +\br{\zeta_{\ell_0}  |\delta t|}^{\dist_{i_1,X_{m-1}}/(2\ell_0)} }
\end{align}
with
\begin{align}
\zeta_{\ell_0} := 2 \gamma (2\ell_0)^D \mathfrak{g} ,\quad \epsilon_{\ell_0}=(1+c_0)  \mathcal{J}_0(\ell_0)
\end{align}
Using the condition for $\delta t$ of
\begin{align}
\delta t =\frac{1}{e \zeta_{\ell_0}}=\frac{1}{2e \gamma (2\ell_0)^D \mathfrak{g}} , 
\end{align}
we obtain 
\begin{align}
\norm{e^{-\check{\mL}_{\tilde{X}_m,\ell_0} t_1} \mathfrak{L}'_{i_1,\ell_0} e^{\check{\mL}_{\tilde{X}_m,\ell_0} t_1} \rho_{X_{m-1}}}_1
\le 2 \brr{ (1+c_0)  \mathcal{J}_0(\ell_0) +e^{-\delta r/ (4\ell_0)} } ,
\end{align}
where we use $\dist_{i_1,X_{m-1}}\ge \delta r/2$ for $i_1 \in X\setminus \tilde{X}_m$. 
We thus reduce the inequality~\eqref{main_ineq_for_small_time_approx__2} to 
\begin{align}
\norm{ e^{ \mL' (t-t_1)}  \mathfrak{L}'_{i_1} e^{\check{\mL}_{\tilde{X}_m} t_1} \rho_{X_{m-1}}}_1  
\le  \brr{\Theta(1)+ \delta t \Theta(r^D) }\mathcal{J}_0(\ell_0)  +2e^{-\delta r/ (4\ell_0)} 
\end{align}
By using the above inequality, we can obtain the upper bound as follows:
\begin{align}
\sum_{i_1\in X\setminus \tilde{X}_m }  \int_0^{\delta t} \norm{ \mathfrak{L}'_{i_1} e^{\check{\mL}_{\tilde{X}_m} t_1}  \rho_{X_{m-1}}}_1 dt_1
\le \gamma \delta t r^D \brrr{\brr{\Theta(1)+ \delta t \Theta(r^D) }\mathcal{J}_0(\ell_0)  +2e^{-\delta r/ (4\ell_0)} } ,
\end{align}
where we use $|X|=|i_0[r]|\le \gamma r^D$. 
We thus prove the inequality~\eqref{exp_mL_t_decomposition_approx_term2}.

\subsubsection{Proof of Lemma~\ref{lem:approx_FF_img}} \label{Proof of Lemma_lem:approx_FF_img}
In order to analyze the multi-commutator, we define the subset $\Lambda_{i_0,i_1,i_2,\ldots , i_m}$ as 
\begin{align}
\Lambda_{i_0,i_1,i_2,\ldots , i_m}=\bigcup_{s=0}^m i_s[2\ell]  ,
\end{align}
and obtain 
\begin{align}
&\ad_{H} \br{h_{i_0[\ell]}}  = \sum_{i_1\in \Lambda_{i_0}} \ad_{h_{i_1[\ell]}} \br{h_{i_0[\ell]}} ,\quad 
\ad_{H}^2 \br{h_{i_0[\ell]}}  = \sum_{i_2\in \Lambda_{i_0,i_1}}  \sum_{i_1\in \Lambda_{i_0}} \ad_{h_{i_2[\ell]}}  \ad_{h_{i_1[\ell]}} \br{h_{i_0[\ell]}}, \notag\\
&\ad_{H}^s \br{h_{i_0[\ell]}}  = \sum_{i_s\in \Lambda_{i_0,i_1,i_2,\ldots , i_{s-1}}} \cdots  \sum_{i_2\in \Lambda_{i_0,i_1}}  \sum_{i_1\in \Lambda_{i_0}}
 \ad_{h_{i_s[\ell]}} \cdots  \ad_{h_{i_2[\ell]}} \ad_{h_{i_1[\ell]}} \br{h_{i_0[\ell]}}.
\end{align}
From the above equation, by using 
\begin{align}
|\Lambda_{i_0,i_1,i_2,\ldots , i_m}| \le m \gamma (2\ell)^D  , \quad 
\norm{ \ad_{h_{i_m[\ell]}} \cdots \ad_{h_{i_2[\ell]}}  \ad_{h_{i_1[\ell]}} (h_{i_0[\ell]})} \le 2^m \mathfrak{g}^{m+1}  ,
\end{align}
we can derive the upper bound of 
\begin{align}
\label{upper_bound_multi_com_psi_out_X}
\norm{\ad_{H}^m \br{h_{i_0[\ell]}}}\le  \mathfrak{g} m!  \brr{2 \gamma (2\ell)^D \mathfrak{g}}^m .
\end{align}

On the other hand, in the case where $i_s[\ell]\subset X^\co $ for $\forall s\in[0, m]$, we have 
\begin{align}
\norm{h_{i_0[\ell]} h_{i_1[\ell]} \cdots h_{i_m[\ell]} \ket{\psi_X} }\le
\mathfrak{g}^{m-1} \epsilon_\ell  ,
\end{align}
Therefore, under the condition of 
\begin{align}
\Lambda_{i_0,i_1,i_2,\ldots , i_m} \cap X =\emptyset \longrightarrow 2\ell m < \dist_{i_0,X},
\end{align}
we have an upper bound:
\begin{align}
\label{upper_bound_multi_com_psi_X}
\norm{\ad_{H}^m \br{h_{i_0[\ell]}}\ket{\psi_X} }\le  m!  \brr{2 \gamma (2\ell)^D \mathfrak{g}}^m \epsilon_\ell .
\end{align}
By combining the inequalities~\eqref{upper_bound_multi_com_psi_out_X} and \eqref{upper_bound_multi_com_psi_X}, we obtain 
\begin{align}
\norm{e^{-\tau H} h_{i_0[\ell]} e^{\tau H} \ket{\psi_X}} 
&\le  \sum_{m=0}^\infty \frac{|\tau|^m}{m!}
\norm{\ad_{H}^m \br{h_{i_0[\ell]}}  \ket{\psi_X}}\notag \\
&\le  \sum_{m<\dist_{i_0,X}/(2\ell)} \frac{|\tau|^m}{m!} m!  \brr{2 \gamma (2\ell)^D \mathfrak{g}}^m \epsilon_\ell 
+ \sum_{m\ge \dist_{i_0,X}/(2\ell)} \frac{|\tau|^m}{m!}  \mathfrak{g} m!  \brr{2 \gamma (2\ell)^D \mathfrak{g}}^m  \notag \\
&\le \frac{1}{1-2 \gamma (2\ell)^D \mathfrak{g}|\tau|} \brr{ \epsilon_\ell +\br{2 \gamma (2\ell)^D \mathfrak{g}|\tau|}^{\dist_{i_0,X}/(2\ell)} }.
\end{align}
This gives the main inequality~\eqref{lem:approx_FF_img_main_ineq} under the definition of Eq.~\eqref{lem:approx_FF_img_main_zeta_ell}.
This completes the proof. $\square$

\section{Proof of Theorem~\ref{thm_low_temp_BP}} \label{Pr:thm_low_temp_BP}

The key ingredient is the local indistinguishability of the quantum Gibbs state, which has been defined in Ref.~\cite{Brandao2019}.
Let $i_0$ be $L' \setminus L$. 
Then, one can prove the following lemma:
\begin{prop} [Local indistinguishability] \label{lem:local_indistinguish}
Let us consider the reduced density matrices on $i_0[\ell]^\co$:  
\begin{align}
 \rho_{\beta,i_0[\ell]^\co}(H):= \tr_{i_0[\ell]} \brr{ \rho_{\beta}(H) } ,\quad 
  \rho_{\beta,i_0[\ell]^\co}(H'):= \tr_{i_0[\ell]} \brr{ \rho_{\beta}(H') } .
\end{align}
Under the uniform clustering condition, the two reduced states $ \rho_{\beta,i_0[\ell]^\co}(H)$ and $ \rho_{\beta,i_0[\ell]^\co}(H')$ are close to each other in the sense that 
\begin{align}
\label{lem:local_indistinguish_main}
\norm{ \rho_{\beta,i_0[\ell]^\co}(H)-  \rho_{\beta,i_0[\ell]^\co}(H') }_1  \le  e^{\Theta(\beta) - \Theta(1) \kappa_\beta \ell}  ,
\end{align}
where $\kappa_\beta= \min(1/\xi, 1/\beta)$. 
\end{prop}

\textit{Proof of Proposition~\ref{lem:local_indistinguish}.}
Let us denote $H'$ by $H'=H+v_{i_0}$ with 
\begin{align}
\label{eq_belief_propagation_operator}
v_{i_0} = \sum_{Z: Z\ni i_0} h_Z.
\end{align}
We here introduce the quantum belief propagation~\cite[Lemma~8 therein]{kuwahara2024CMI}]: 
\begin{align}
e^{\beta H'} = \Phi_{i_0}^\dagger  e^{\beta H}   \Phi_{i_0},
\end{align}
with
\begin{align}
\label{sup_Def:Phi_0_phi}
&\Phi_{i_0}:= \mathcal{T} e^{\int_0^{1} \phi_{\mB,x} dx} , \notag \\
&\phi_{i_0,x}:= \frac{\beta}{2}  \int_{-\infty}^\infty f_\beta(t)v_{i_0}(H+x v_{i_0},t) dt, 
\end{align}
where we use the notation~\eqref{O_1_O_2,t}. 
Then, using the Lieb--Robinson bound, one can prove from Ref.~\cite[Corollary~11 therein]{kuwahara2024CMI}:
\begin{align}
\label{BP_approx}
\norm{ e^{\beta H'} - \tilde{\Phi}_{i_0[\ell_1]}^\dagger  e^{\beta H} \tilde{\Phi}_{i_0[\ell_1]}}_1 \le \tr\br{e^{\beta H'}}  e^{c_0 \beta g - c_1 \kappa_\beta \ell_1} , \quad \norm{\tilde{\Phi}_{i_0[\ell_1]}}\le e^{\beta g/2},
\end{align}
where $\kappa_\beta=\min(1/\beta, 1/\xi)$, $c_0$ and $c_1$ ($\ge 1$) are $\orderof{1}$ constants, and $\tilde{\Phi}_{i_0[\ell_1]}$ is an appropriate local approximation for $\Phi_{i_0}$ that is supported on $i_0[\ell_1]$.

Next, the definition of the trace norm gives
\begin{align}
\label{target_quantity}
\norm{ \rho_{\beta,i_0[\ell]^\co}(H)-  \rho_{\beta,i_0[\ell]^\co}(H') }_1  
&= \sup_{O_{i_0[\ell]^\co} : \norm{O_{i_0[\ell]^\co}}=1} \tr_{i_0[\ell]^\co}  \brrr{ O_{i_0[\ell]^\co} \brr{\rho_{\beta,i_0[\ell]^\co}(H)-  \rho_{\beta,i_0[\ell]^\co}(H') } } \notag \\
&=  \sup_{O_{i_0[\ell]^\co} : \norm{O_{i_0[\ell]^\co}}=1} \tr \brrr{ O_{i_0[\ell]^\co} \brr{\rho_{\beta}(H)-  \rho_{\beta}(H') } } .
\end{align}
By using the approximate belief propagation operator $\tilde{\Phi}_{i_0[\ell_1]}$, we have 
\begin{align}
 \tr \brr{ O_{i_0[\ell]^\co} \rho_{\beta}(H')  } 
 &= \frac{1}{\tr\br{e^{\beta H'}}}  \tr \brr{ O_{i_0[\ell]^\co}  \tilde{\Phi}_{i_0[\ell_1]}^\dagger  e^{\beta H} \tilde{\Phi}_{i_0[\ell_1]} } 
 + \frac{1}{\tr\br{e^{\beta H'}}}  \tr \brr{ O_{i_0[\ell]^\co} \br{ e^{\beta H'} - \tilde{\Phi}_{i_0[\ell_1]}^\dagger  e^{\beta H} \tilde{\Phi}_{i_0[\ell_1]} } } \notag \\
 &= \frac{\tr\br{e^{\beta H}}}{\tr\br{e^{\beta H'}}} \brrr{ \Cor_{\rho_\beta(H)}  (O_{i_0[\ell]^\co}, \tilde{\Phi}_{i_0[\ell_1]} \tilde{\Phi}_{i_0[\ell_1]}^\dagger  ) + 
   \tr \brr{ O_{i_0[\ell]^\co} \rho_\beta(H) }    \tr \brr{  \tilde{\Phi}_{i_0[\ell_1]}^\dagger   \rho_\beta(H) \tilde{\Phi}_{i_0[\ell_1]} }    } \notag \\
   &+ \frac{1}{\tr\br{e^{\beta H'}}}  \tr \brr{ O_{i_0[\ell]^\co} \br{ e^{\beta H'} - \tilde{\Phi}_{i_0[\ell_1]}^\dagger  e^{\beta H} \tilde{\Phi}_{i_0[\ell_1]} } }  .
\end{align}
By applying the clustering condition~\eqref{def:uniform_clustering_inq} and the inequality~\eqref{BP_approx}, we have 
\begin{align}
\label{target_quantity_2}
&\abs{ \tr \brrr{ O_{i_0[\ell]^\co} \brr{\rho_{\beta}(H)-  \rho_{\beta}(H') } } } \notag \\
\le&  \frac{\tr\br{e^{\beta H}}}{\tr\br{e^{\beta H'}}} C_1 \norm{\tilde{\Phi}_{i_0[\ell_1]} \tilde{\Phi}_{i_0[\ell_1]}^\dagger}  |i_0[\ell_1]| e^{-(\ell-\ell_1)/\xi} + 
\abs{1-  \frac{\tr\br{e^{\beta H}}}{\tr\br{e^{\beta H'}}}  \tr \brr{  \tilde{\Phi}_{i_0[\ell_1]}^\dagger   \rho_\beta(H) \tilde{\Phi}_{i_0[\ell_1]} } }+e^{c_0 \beta g - c_1 \kappa_\beta \ell_1} \notag \\
\le & C_1 \gamma \ell_1^D e^{2\beta g -(\ell-\ell_1)/\xi} + 2e^{c_0 \beta g - c_1 \kappa_\beta \ell_1} ,
\end{align}
where we use $|i_0[\ell_1]|\le \gamma \ell_1^D$, 
\begin{align}
\abs{1-  \frac{\tr\br{e^{\beta H}}}{\tr\br{e^{\beta H'}}}  \tr \brr{  \tilde{\Phi}_{i_0[\ell_1]}^\dagger   \rho_\beta(H) \tilde{\Phi}_{i_0[\ell_1]} } }
&= \frac{1}{\tr\br{e^{\beta H'}}}\abs{\tr\br{e^{\beta H'}}-    \tr \brr{  \tilde{\Phi}_{i_0[\ell_1]}^\dagger  e^{\beta H} \tilde{\Phi}_{i_0[\ell_1]} } } \le e^{c_0 \beta g - c_1 \kappa_\beta \ell_1} ,
\end{align}
and 
\begin{align}
\tr\br{e^{\beta H}} = \tr\br{e^{\beta (H' - v_{i_0}}} \le   \tr\br{e^{\beta H'} e^{-\beta v_{i_0}}} \le e^{\beta \norm{v_{i_0}}}  \tr\br{e^{\beta H'} } \le e^{\beta g} \tr\br{e^{\beta H'} }.
\end{align}
By choosing $\ell_1=\ell/2$ and apply~\eqref{target_quantity_2} to \eqref{target_quantity}, we prove the main inequality~\eqref{lem:local_indistinguish_main}.

{~}

\hrulefill{\bf [ End of Proof of Proposition~\ref{lem:local_indistinguish}]}

{~}

We then consider a recovery map $\tau_{i_0[\ell]^\co \to \Lambda}$ from $\rho_{\beta,i_0[\ell]^\co}(H')$ to $\rho_\beta(H')$.
By letting $A=i_0[\ell]$, $B=i_0[r]\setminus i_0[\ell]$ and $C= i_0[r]^\co$, we can write 
\begin{align}
\rho_{\beta,i_0[\ell]^\co}(H')= \rho_{\beta,BC}(H')= \tr_{A} \br{e^{\beta H'}} . 
\end{align}
We then consider a local recovery map that achieves 
\begin{align}
\tau_{B \to AB} \brr{ \rho_{\beta,BC}(H') }  \approx \rho_{\beta}(H') .
\end{align}
Once we can find it, we utilize it to convert 
\begin{align}
\rho_{\beta}(H) \to \rho_{\beta}(H')  ,
\end{align}
because of 
\begin{align}
\label{Ineq_1_Global_Markov}
\norm{ \tau_{B \to AB} \tr_A \brr{ \rho_{\beta}(H)} - \rho_{\beta}(H')  }_1 
\le &\norm{ \tau_{B \to AB} \tr_A \brr{ \rho_{\beta}(H) -  \rho_{\beta}(H')}  } + \norm{\tau_{B \to AB} \tr_A \brr{  \rho_{\beta}(H')}  - \rho_{\beta}(H')  }_1 \notag \\
\le &\norm{\tr_A \brr{ \rho_{\beta}(H) -  \rho_{\beta}(H')}  } +  \norm{\tau_{B \to AB} \tr_A \brr{  \rho_{\beta}(H')}  - \rho_{\beta}(H')  }_1 \notag \\
\le &e^{\Theta(\beta) - \Theta(1) \kappa_\beta \ell}  +  \norm{\tau_{B \to AB} \tr_A \brr{  \rho_{\beta}(H')}  - \rho_{\beta}(H')  }_1,
\end{align}
where we use Proposition~\ref{lem:local_indistinguish} in the last inequality. 

Finally, we estimate the recovery map for $H'$. Here, the point is that the region $A$ is small in the sense that $|A| = i_0[\ell] \propto \ell^D$. 
Hence, one can utilize the CMI decay for the small region (or the local Markov property):
\begin{lemma}[Corollary III.2 in Ref.~\cite{chen-rouze}]
\label{lem:local_markov}
At any temperature, the quantum Gibbs state $\rho_\beta(H')$ show a CMI decay as 
\begin{align}
\mI_{\rho_\beta(H')}(A:C|B) \le \Theta(1) |A|\cdot |C|e^{\Theta(1) \min(|A|,|C|) - \dist_{A,C} /\tilde{\xi}_\beta} , 
\end{align}
where $\tilde{\xi}_\beta$ is a constant which depends on $\beta$ non-trivially. 
\end{lemma}
By combining Lemma~\ref{lem:local_markov} with Fawzi--Renner inequality~\cite{Fawzi2015}, we ensure that there exists a CPTP map $\tau_{B \to AB}$ such that 
\begin{align}
\label{Ineq_2_Global_Markov}
 \norm{\tau_{B \to AB} \tr_A \brr{  \rho_{\beta}(H')}  - \rho_{\beta}(H')  }_1 \le \Theta(1) |A|\cdot |C|e^{\Theta(1) \min(|A|,|C|) - \dist_{A,C} /\tilde{\xi}_\beta},
\end{align}
which reduces the inequality~\eqref{Ineq_1_Global_Markov} to 
\begin{align}
\label{Ineq_3_Global_Markov}
\norm{ \tau_{B \to AB} \tr_A \brr{ \rho_{\beta}(H)} - \rho_{\beta}(H')  }_1 
\le  &e^{\Theta(\beta) - \Theta(1) \kappa_\beta \ell}  + \Theta(n) e^{\Theta(1) \ell^D - (r-\ell) /\tilde{\xi}_\beta} ,
\end{align}
where we use the definitions of $A,B,C$ above, which gives $ \dist_{A,C} =r-\ell$. 
Note that the CPTP map $\tau_{B \to AB} $ is now supported on $i_0[r]$.

Finally, by choosing $\ell$ such that $\Theta(1) \ell^D = (r-\ell) /\tilde{\xi}_\beta/2$ or $\ell\propto (r/\xi_\beta)^{1/D}$, we reduce the inequality~\eqref{Ineq_3_Global_Markov} to the main inequality~\eqref{thm_low_temp_BP_main}. 
This completes the proof. $\square$

\section{Divergence of cluster expansion} \label{sec:Divergence of cluster expansion}

\subsection{Overview}

We demonstrate that the cluster expansion technique encounters fundamental difficulties when applied directly to the analysis of CMI decay.
To see this point, we here consider a simple 1D Hamiltonian with nearest-neighbor interactions: 
\begin{align}
H= \sum_{i=1}^{n-1} h_{i,i+1} + \sum_{i=1}^n h_i  .
\end{align}
Our goal is to compute the effective Hamiltonian corresponding to the reduced density matrix on the subsystem ${1,2,\ldots,n-1}$, given by:
\begin{align}
\log\brr{ \tr_n \br{e^{\beta H} } },
\end{align}
where $\tr_n\br{\cdots}$ is the partial trace with respect to the right-end site $n$. 

In the generalized cluster expansion technique~\cite{PhysRevLett.124.220601,PRXQuantum.4.020340}, we parameterize the Hamiltonian as 
\begin{align}
\label{parameterized_Hamiltonian}
H_{\vec{a}}= \sum_{i=1}^{n-1} a_{i,i+1} h_{i,i+1}+ \sum_{i=1}^n a_i h_i  .
\end{align}
and consider the expansion of 
\begin{align}
\label{parameterized_Hamiltonian_log}
\tilde{H}_{\vec{a}} := \log\brr{ \tr_n \br{e^{\beta H_{\vec{a}}} } } = \sum_{m=0}^\infty \frac{\beta^m}{m!} \frac{d^m}{d\vec{a}^m} \log\brr{ \tr_n \br{e^{\beta H_{\vec{a}}} } } .
\end{align}
In general, the multi-derivative of the operator logarithm has a complicated form, as shown in Appendix~\ref{app:Multi-derivative of the operator logarithm}.
We here denote $w$ by a choice of $\vec{a}$, e.g., $w=\{a_{1,2},a_3, a_{4},a_{5,6}\}$. 
We define the set $\Gc_{n,w}$ to be the collection of $w$ such that all the indices in $\vec{a}$ are connected with each other and at least one index includes $\{n-1,n\}$; for example, $w=\{a_{n},a_{n-1}, a_{n-1,n}, a_{n-2,n-1}\}$ is included in $\Gc_{n,w}$, while $w'=\{a_n, a_{n-1}, a_{n-1,n-2}\}$ ($\{n-1,n\}$ is not included) or $w'=\{a_n, a_{n-1}, a_{n,n-1}, a_{n-2,n-3}\}$ ($a_{n-2,n-3}$ is isolated from the others) are not included in $\Gc_{n,w}$.

As has been proven in Ref.~\cite[Propositon~3 therein]{PhysRevLett.124.220601}, we obtain 
\begin{align}
\label{tilde_H_a_expansion}
\tilde{H}_{\vec{a}} = \sum_{m=0}^\infty \frac{\beta^m}{m!} \sum_{w: w\in \Gc_{n,w}, |w|=m }\frac{d^m}{d\vec{a}^m} \log\brr{ \tr_n \br{e^{\beta H_{\vec{a}}} } } \biggl |_{\vec{a}=\vec{0}},
\end{align}
where $|w|$ means the number of elements in $w$. 
The cluster expansion method aims to prove the convergence of 
 \begin{align}
 \label{Cluster_absolute}
\sum_{m>\bar{m}}^\infty \frac{\beta^m}{m!} \sum_{w: w\in \Gc_{n,w}, |w|=m } \norm{ \frac{d^m}{d\vec{a}^m} \log\brr{ \tr_n \br{e^{\beta H_{\vec{a}}} } } \biggl |_{\vec{a}=\vec{0}}}  
\end{align}
at sufficiently high temperatures.
Note that as long as we take the terms of $m\le m_0$ in the expansion~\eqref{tilde_H_a_expansion}, the approximated effective Hamiltonian has an interaction length at most $m_0$ from the site $n$.

\subsection{Divergence problem}  \label{Seec:Divergence problem}

To simplify the analysis, we consider a lower bound of
 \begin{align}
 &\label{Cluster_absolute_2}
\sum_{m>\bar{m}}^\infty \frac{\beta^m}{m!} \sum_{w: w\in \Gc_{n,w}, |w|=m } \norm{ \frac{d^m}{d\vec{a}^m} \log\brr{ \tr_n \br{e^{\beta H_{\vec{a}}} } } \biggl |_{\vec{a}=\vec{0}}}   \notag \\
&\ge \bar{C}_{\bar{m}} +  \sum_{m=0}^\infty \frac{\beta^m}{m!} \sum_{w: w\in \Gc_{n,w}, |w|=m } \norm{ \frac{d^m}{d\vec{a}^m} \log\brr{ \tr_n \br{e^{\beta H_{\vec{a}}} } } \biggl |_{\vec{a}=\vec{0}}} \notag \\
&\ge \bar{C}_{\bar{m}} +  \sum_{m=0}^\infty \frac{\beta^m}{m!}  \norm{   \frac{d^m}{da^m}\log\brr{ \tr_n \br{e^{\beta H_{a}}} } \biggl|_{a=0 } } 
\end{align}
with 
\begin{align}
\label{Ham_decompo}
H_{a}=a h_{n-1,n} + \sum_{i=1}^{n-2} h_{i,i+1} + \sum_{i=1}^n h_i =a h_{n-1,n}+H_0 ,
\end{align}
where we use the fact that the operator norm satisfies subadditivity, i.e., $\norm{O_1} + \norm{O_2} \ge \norm{O_1+O_2}$. 
In the following, we are going to demonstrate that the first order in $a$ leads to the divergence in the thermodynamic limit $(n\to \infty)$. 

For this purpose, using the notation of Eq.~\eqref{Ham_decompo}, we obtain 
\begin{align}
e^{\beta H_a} = e^{\beta \br{H_0 +a  h_{n-1,n} }} 
&=  e^{\beta H_0} + e^{\beta H_0} \cdot  a \beta \int_0^1  e^{-x H_0}  h_{n-1,n} e^{x H_0} dx+ \Omega(a^2)  \notag \\
&= e^{\beta H_0} + a \beta \int_0^1 e^{x H_0}  h_{n-1,n} e^{-x H_0} dx \cdot e^{\beta H_0} + \Omega(a^2)  .
\label{e^beta_H_decomp}
\end{align}
Because of 
\begin{align}
e^{\beta H_0} =  e^{\beta H_{\le n-1}} \otimes e^{\beta h_n} ,
\end{align}
we have the partial trace of Eq.~\eqref{e^beta_H_decomp} as follows:
\begin{align}
\tr_n\br{e^{\beta H}}  
&= G e^{\beta H_{\le n-1}} \brr{ 1 + \frac{a \beta}{G} \int_0^1 \tr_n \br{ e^{\beta h_n} e^{-x H_0} h_{n-1,n} e^{x H_0}} dx} \notag \\
&= G \brr{1 +  \frac{a \beta}{G}  \int_0^1  \tr_n \br{e^{\beta h_n}   e^{x H_0} h_{n-1,n} e^{-x H_0} } dx } e^{\beta H_{\le n-1}}
\label{e^beta_H_decomp2},
\end{align}
where we define $G:= \tr_n\br{e^{\beta h_n}}$. 
Therefore, by defining 
\begin{align}
\partial \tilde{h}_n := \frac{1}{2G} \int_0^1 \tr_n \br{ e^{\beta h_n} e^{-x H_0} h_{n-1,n} e^{x H_0}} dx ,
\label{e^beta_H_decomp_def_tilde_h}
\end{align}
we obtain 
\begin{align}
\label{reduced_density_matrix_site_n}
\tr_n\br{e^{\beta H}}  
&= G e^{\beta a \br{\partial \tilde{h}_n}^\dagger}e^{\beta H_{\le n-1}} e^{\beta a \partial \tilde{h}_n} + \Omega(a^2) .
\end{align}

We then utilize the following general decomposition~\cite[Eq. (2.7) therein]{Scharf_1988}: 
\begin{align}
\label{BCH_exp_sp}
\log \br{e^{\beta a \mB^\dagger } e^{\beta \mA }  e^{\beta a \mB}} 
=  \beta \mA + \beta a \sum_{m=1}^\infty \frac{\beta^m B_m}{m!} \brr{ \ad_{\mA}^m(\mB)+{ \rm h.c.} } + \Omega(a^2)  , 
\end{align}
where $B_m$ is the Bernoulli number, which increases as $B_{2j}\approx (-1)^{j+1} 4\sqrt{\pi j} \br{\frac{j}{\pi e}}^{2j}$. 
By applying the above decomposition to Eq.~\eqref{reduced_density_matrix_site_n}, we derive
\begin{align}
\log\brr{\tr_n\br{e^{\beta H}} }  - \log(G) 
&= \beta H_{\le n-1} + \beta a\sum_{m=1}^\infty \frac{\beta^m B_m}{m!}  \brr{ \ad_{H_{\le n-1}}^m\br{\partial \tilde{h}_n} + {\rm h.c.} } + \Omega(a^2) .
\end{align}

The above expression~\eqref{BCH_exp_sp} shows that the cluster expansion method for the effective subsystem Hamiltonian is closely related to the Baker-Campbell-Hausdorff formula. It is well-known that this expansion is not absolutely convergent unless $\norm{\beta H}$ is below a certain threshold~\cite{SBlanes_1998,Moan2008,BLANES2009151}. 
Indeed, the norm $\ad_{H_{\le n-1}}^m\br{\partial \tilde{h}_n}$ is estimated to scale as $(Cm)^m$ with $C=\orderof{1}$, and hence the rough estimation gives 
\begin{align}
\norm{ \frac{\beta^m B_m}{m!}  \brr{ \ad_{H_{\le n-1}}^m\br{\partial \tilde{h}_n} + {\rm h.c.} } } \propto (C'm \beta)^m , 
\end{align}
which leads to divergence for $m\to \infty$ in the thermodynamic limit\footnote{More precisely, for a finite system, we have $\norm{\ad_{H_{\le n-1}}^m\br{\partial \tilde{h}_n}} \le \min\brr{(Cm)^m, (Cn)^m}$, so convergence occurs only if $\beta \lesssim 1/n$.}.

Nevertheless, we can prove its conditional convergence using the method in Ref.~\cite[Lemma~18 therein]{kuwahara2024CMI}, where the partial trace $\tr_n\br{\cdots}$ is shown to yield a quasi-local effective interaction centered around site $n$. When the traced-out region becomes large, however, the degree of quasi-locality depends on the region size~\cite[Theorem~2 therein]{kuwahara2024CMI}. 

A natural direction for future work is to refine the method under high-temperature conditions to elucidate the fine structures of the effective subsystem Hamiltonians.

%
%The above expression~\eqref{BCH_exp_sp} shows that the cluster expansion method for the effective subsystem Hamiltonian is closely related to the Baker-Campbell-Hausdorff formula. It is well-known that the expansion is not absolutely convergent as long as $\norm{\beta H}$ is smaller than a threshold~\cite{SBlanes_1998,Moan2008,BLANES2009151}. 
%
%Still, we can still prove the conditional convergence using the method in Ref.~\cite[Lemma~18 therein]{kuwahara2024CMI}, and the partial trace $\tr_n\br{\cdots}$ is proven to yield only a quasi-local effective interaction around the site $n$. 
%However, when the region of the partial trace becomes large, the quasi-locality depends on the region size~\cite[Theorem~2 therein]{kuwahara2024CMI}. 
%An intriguing open problem is to refine the method with the high-temperature condition to elucidate the fine structures of the subsystem's effective Hamiltonians.   

\subsection{Numerical calculations}
\begin{figure}[tt]
\centering
\includegraphics[clip, scale=0.8]{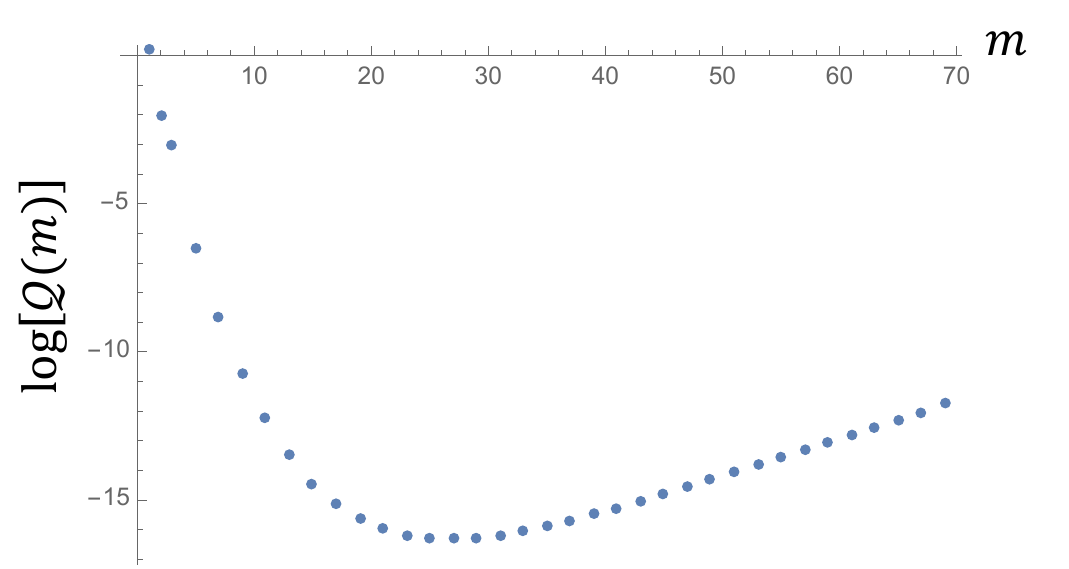}
\caption{Numerical calculations for the norm~\eqref{norm_cluster_exapnasion_num}. From the plots, the divergence starts around $m\approx 25$. 
}
\label{fig_BCH_plot}
\end{figure}

We here consider the XYZ Heisenberg model as 
\begin{align}
h_{i,i+1} = \frac{1}{6} \br{ 3 \sigma_i^x \otimes  \sigma_{i+1}^x +2 \sigma_i^y \otimes  \sigma_{i+1}^y+ \sigma_i^z \otimes  \sigma_{i+1}^z}
,\quad h_i= \frac{1}{3}\br{ \sigma_i^x+ \sigma_i^y+ \sigma_i^z}. 
\end{align}
We also set $\beta=1/2$ and $n=10$.
Then, we calculate the function $\mQ(m)$ as 
\begin{align}
\label{norm_cluster_exapnasion_num}
\mQ(m) := \norm{ \beta  \frac{\beta^m B_m}{m!}  \brr{ \ad_{H_{\le n-1}}^m\br{\partial \tilde{h}_n} + {\rm h.c.} }}_F , 
\end{align}
up to $m=69$, where $\norm{\cdots}$ means the Frobenius norm. 
The numerical plots for the logarithm of the above quantity and the order degree $m$ are given in Fig.~\ref{fig_BCH_plot}:
In the simulation, to avoid error accumulation, we calculate with a precision of 500 digits.

\section{Conclusion and discussions} \label{sec:Conclusion and discussions}

In this work, we developed a new framework for constructing recovery maps based on the belief-propagation-channel formalism (Definition~\ref{def:BP_channel}).
This allows us to prove the spatial decay of conditional mutual information (CMI) with polynomial dependence of the subsystem sizes (Theorem~\ref{thm:main}). 
We believe this approach holds potential for reaching the complete resolution of the CMI decay conjecture (Conjecture~\ref{conjecture:1}).
The technically involved part of our analysis lies in proving the quasi-locality of the BP channels (Theorems~\ref{thm_high_temp_BP} and \ref{thm_low_temp_BP}).

%The most technically involved part of our analysis lies in Subtheorem~\ref{thm:approx_local_Liou}, where we approximate the dynamics of a perturbed Liouvillian by a local CPTP map. This approximation plays a key role in our proof, and it would be an interesting challenge to simplify or optimize this result in future work.

Despite the success of our method, many open problems remain.  
Unconditional proof for the quasi-locality of the BP channels is one of the most important open problems since it is a sufficient condition for proving Conjecture~\ref{conjecture:1}.  
From the physical perspective, the existence itself of BP channels is fundamentally intriguing, as it is closely related to the question of whether local quantum circuits can simulate perturbations of Hamiltonians.  
Exploring such alternative applications of BP channels, therefore, constitutes an interesting future research direction beyond the specific problem of CMI decay.

Even within the high-temperature regime, when considering CMI decay on subsystems---i.e., for $A \cup B \cup C \subset \Lambda$---a major obstacle arises: the reduced state $\rho_{\beta,ABC}$ is no longer guaranteed to be the Gibbs state of a quasi-local Hamiltonian. Consequently, the recovery map construction illustrated in Fig.~\ref{fig_Recovery_map} is no longer directly applicable.
Resolving this issue would require establishing that the reduced density matrix on a subsystem can still be approximated by a Gibbs state of some quasi-local effective Hamiltonian. However, due to the non-convergence of cluster expansions in this setting, the problem remains particularly challenging. While only limited progress has been made so far, it may be expected that the methodology developed in Ref.~\cite{kuwahara2024CMI}, when combined with high-temperature conditions, may offer a promising direction. Importantly, this problem is deeply connected to the question of how well open quantum systems preserve Markovianity, and thus represents a fundamental challenge in understanding the stability of quantum mixed phases~\cite{sang2024}.

High-temperature quantum systems are often regarded as intuitively simple and mathematically tractable. Indeed, high-temperature Gibbs states can be generated efficiently by low-depth quantum circuits~\cite{rouze2024efficient,10756136}, reinforcing the view that correlations and complexity are minimal in this regime. However, our results reveal that this simplicity can be deceptive: the analysis of effective Hamiltonians on subsystems uncovers richer and more intricate structures than previously anticipated.
This work not only sheds light on such hidden structures that lie beyond the reach of traditional high-temperature expansions but also lays the groundwork for new methodologies and future developments in quantum many-body physics.

\section{Acknowledgment}

T. K. acknowledges the Hakubi projects of RIKEN.
T. K. was supported by JST PRESTO (Grant No. JPMJPR2116), ERATO (Grant No. JPMJER2302), and JSPS Grants-in-Aid for Scientific Research (No. JP23H01099, JP24H00071), Japan. 
K. K. acknowledges support from JSPS Grant-in-Aid for Early-Career Scientists, No. 22K13972; from MEXT-JSPS Grant-in-Aid for Transformative Research Areas (B), No. 24H00829.
We are grateful to Samuel Scalet, Dominic Wild, and \'Alvaro Alhambra for bringing to our attention a flaw in the proof of Ref.~\cite{PhysRevLett.124.220601} in 2022, which led to renewed interest in understanding the behavior of CMI at high temperatures.

{~} \\

\textit{Note Added.}
When this manuscript was nearing completion, we became aware of a concurrent and independent result by Chi-Fang Chen and Cambyse Rouz\'e, which proves the decay of CMI in quantum Gibbs states using similar methods based on dissipative dynamics~\cite{chen-rouze}, whose technique has been utilized for proving Theorem~\ref{thm_low_temp_BP} in updating the manuscript. 
We thank them for letting us know about their concurrent and independent work.

\section{Declarations}
\subsection*{Conflict of interest}
The authors declare that there is no conflict of interest.

\subsection*{Data availability}
The numerical data generated in this study are available from the authors on request.

\appendix

\section{Review of the CKG Liouvillian} \label{Review:CKG}
To construct the CPTP map from $e^{\beta H_0}$ to $e^{\beta (H_0+h_{i_0})}$, we adopt the dissipative dynamics that are governed by the Liouvillian introduced by Chen, Kastoryano, and Gily\'en (CKG)~\cite{chen2023efficient}\footnote{As another candidate, we can use the heat-bath generator~\cite{Kastoryano2016}. At high temperatures, we can prove the quasi-locality, but the CKG Liouvillian is more appropriate in treating general interaction forms (e.g., power-law decaying interactions).}, which satisfies convenient properties addressed in Assumption~\ref{assup:Basic assumptions for quasi-local Liouvillian}.  
Let $\{A_{i,a}\}_{a}$ be the Hermitian operator bases on the site $i$ (e.g., the Pauli matrices).  
Then, the CKG Liouvillian $\mL_{\beta H}$ for the quantum Gibbs state $e^{\beta H}$ is defined by 
\begin{align}
\mL^{(H)}  =  \sum_{i\in \Lambda} \mathfrak{L}_i=  \sum_{i\in \Lambda} \sum_{a=1}^{d^2-1}  \mathfrak{L}_{i,a}, 
\end{align}
and 
\begin{align}
\label{Dissipative Dynamics_Liuou_def}
\mathfrak{L}_{i,a}\rho :=- i[\mB_{i,a}, \rho] +  \int_{-\infty}^\infty \gamma(\omega) 
\brr{\mA_{i,a}(\omega) \rho \mA_{i,a}(\omega)^\dagger 
- \frac{1}{2} \brrr{\mA_{i,a}(\omega)^\dagger \mA_{i,a}(\omega), \rho }  }d\omega ,
\end{align}
where $\mA_{i,a}(\omega)$ is defined by 
\begin{align}
&\mA_{i,a}(\omega) :=\frac{1}{\sqrt{2\pi}} \int_{-\infty}^\infty \mA_{i,a}(H,t) e^{-i\omega t} \frac{e^{-t^2/\beta^2}}{\sqrt{\beta\sqrt{\pi/2}}} dt,
 \notag \\
 &\gamma(\omega) := \exp\brr{-\frac{(\beta \omega +1)^2}{2}}  , 
\end{align}
and $\mB_{i,a} $ is defined by
\begin{align}
\label{mB_i_a_Def}
\mB_{i,a} =\int_{-\infty}^\infty b_1(t) e^{-i\beta Ht } \br{\int_{-\infty}^\infty b_2(t') 
 \mA_{i,a}(H,\beta t')  \mA_{i,a}(H, - \beta t')dt' } e^{i\beta Ht } dt
\end{align}
with
\begin{align}
\label{def_b_1_b_2_t}
&b_1(t) = 2\sqrt{\pi} e^{1/8} \int_{-\infty}^\infty \frac{\sin(-t+s)e^{-2(t-s)^2}}{\cosh(2\pi s)} ds,  \notag \\
&b_2(t)=\frac{1}{2\pi^{3/2}} e^{-4t^2 -2it} .
\end{align}

The norm of the coherence term $\mB_{i,a}$ satisfies 
\begin{align}
\norm{\mB_{i,a}} \le \int_{-\infty}^\infty |b_1(t)|  \int_{-\infty}^\infty |b_2(t')| dt' dt \le \frac{e^{1/8}}{4\sqrt{2}} .
\end{align}

The jump operators $\{ \mA_{i,a}(\omega)\}$ satisfy
\begin{align}
\int_{-\infty}^\infty \gamma(\omega) \norm{\mA_{i,a}(\omega)} \cdot \norm{\mA_{i,a}(\omega)} d\omega 
\le \frac{\beta}{(2\pi)^{1/2}}  \int_{-\infty}^\infty e^{-(\beta \omega +1)^2/2}  d\omega =1 ,
\end{align}
where we use 
\begin{align}
\norm{\mA_{i,a}(\omega)} \le \frac{1}{\sqrt{2\pi}} \int_{-\infty}^\infty  \frac{e^{-t^2/\beta^2}}{\sqrt{\beta\sqrt{\pi/2}}} = 
\frac{\beta^{1/2}}{(2\pi)^{1/4}}.
\end{align}
We thus obtain the norm of the Liouvillian as follows:
\begin{align}
\label{Liouvillian_norm_local}
\norm{\mathfrak{L}_{i,a}}_{1\to 1}
& \le 2 \norm{\mB_{i,a}}  + 2\int_{-\infty}^\infty \gamma(\omega) \norm{\mA_{i,a}(\omega)} \cdot \norm{\mA_{i,a}(\omega)} d\omega   \notag \\
&\le \frac{e^{1/8}}{2\sqrt{2}} + 2 \le 3.
\end{align}

\subsection{Quasi-locality of the CKG Liouvillian}

\subsubsection{Lieb--Robinson bound}

\begin{lemma}[Lieb--Robinson bound by local unitary dynamics] \label{lem:quasi-locality_Lieb_Robinson}
Let $O_i$ be an arbitrary local operator on the site $i$ and $O_i(t)$ be locally approximated onto the ball region $i[r]$ by $O^{(t)}_{i[r]} := \tilde{\tr}_{i[r]^\co} \brr{O_i(t) }$. Then it holds that
\begin{align}
\label{lem:quasi-locality_mL/main}
\norm{ O^{(t)}_{i[r]}  - O^{(t)}_{i[r-1]} } \le \min\brr{ 2, \mathcal{F}(r,t) } ,
\end{align}
where $\mathcal{F}(r,t)$ is given by 
\begin{align}
\label{Lieb_Robinson/func}
\mathcal{F}(r,t) = C \br{ \frac{vt}{r/l_H}}^r 
\end{align}
where we have assumed $H$ has the finite-range interactions as in~\eqref{def_short_range_long_range}. 
\end{lemma}

\textit{Proof of Lemma~\ref{lem:quasi-locality_Lieb_Robinson}.}
We start from the standard Lieb--Robinson bound~\cite{ref:Hastings2006-ExpDec,ref:Nachtergaele2006-LR,PhysRevA.80.052104} as 
\begin{align}
\label{LR_ineq_communtator}
\norm{ \brr{O_i(H,t), u_{X}}} \le 
C \br{ \frac{vt}{\dist_{i,X}/l_H}}^{\dist_{i,X}}. 
\end{align}
By using the unitary expression of the normalized partial trace as 
\begin{align}
\label{lunitary_exp}
\tilde{\tr}_{X} \br{O} = \int \mu(u_X) u_X^\dagger  O u_X  ,
\end{align}
where $u_X:=\bigotimes_{i\in X}u_i$ and $\mu(u_X):=\prod\mu(u_i)$ is the Haar measure, we obtain 
\begin{align}
\norm{ O^{(t)}_{i[r]} - O^{(t)}_{i[r-1]} }%&=\norm{ \int \mu(u_{(i[r-1])^\co})\left( u_{(i[r])^\co}O_i(t)u_{(i[r])^\co}^\dagger-u_{(i[r-1])^\co}O_i(t)u_{(i[r-1])^\co}^\dagger\right)  }\\
&=\norm{ \int \mu(u_{\partial(i[r])})\left( O^{(t)}_{i[r]}-u_{\partial(i[r])}O^{(t)}_{i[r]}u_{\partial(i[r])}^\dagger\right)  } 
\le\norm{ \int \mu(u_{\partial (i[r])})\left( O_i(t)-u_{i[r-1]}O_i(t)u_{i[r-1]}^\dagger\right)  } \notag\\
&\le\norm{ \int \mu(u_{\partial (i[r])})\left( O_i(t)-u_{i'}O_i(t)u_{i'}^\dagger\right)  }+\norm{ \int \mu(u_{\partial (i[r])})\left( u_{i'}O_i(t)u_{i'}^\dagger-u_{j'}u_{i'}O_i(t)u_{i'}^\dagger u_{j'}^\dagger\right)  }+\cdots\notag\\
%&\le \norm{ \int \mu(u_{\partial (i[r])})\left[O_i(t),u_{i[r-1]}\right]u_{i[r-1]}^\dagger  }\le \norm{ \int \mu(u_{\partial (i[r])})\left[O_i(t),u_{i[r-1]}\right] }\\
&\le \sum_{i'\in  \partial (i[r])}\norm{ \int \mu(u_{\partial (i[r])})\left( O_i(t)-u_{i'}O_i(t)u_{i'}^\dagger\right)  }\le \sum_{i'\in  \partial (i[r])}\max_{u_{i'}} \norm{ \brr{O_i(t), u_{i'}}} ,\label{lunitary_exp__3}
\end{align}
where $i',j'$ in the second line are sites in $\partial(i[r])$. 
By combining the above expressions with the standard Lieb--Robinson bound~\eqref{LR_ineq_communtator} for the commutators, we prove the main inequality~\eqref{lem:quasi-locality_mL/main}. 
This completes the proof. $\square$

\subsubsection{Quasi-locality lemma}
 
Using the Lieb--Robinson bound in Lemma~\ref{lem:quasi-locality_Lieb_Robinson}, we can prove the quasi-locality of the CKG Liouvillian by following the analyses in Ref.~\cite{rouze2024efficient}.
We consider the decomposition of 
\begin{align}
\mL^{(H)} = \sum_{i\in \Lambda} \sum_{\ell=0}^\infty \delta \mathfrak{L}_{i[\ell]}  ,
\end{align}
with
\begin{align}
\delta \mathfrak{L}_{i[0]}:=\tilde{\mathfrak{L}}_{i[0]} , \quad \delta \mathfrak{L}_{i[\ell]}:=\tilde{\mathfrak{L}}_{i[\ell]}- \tilde{\mathfrak{L}}_{i[\ell-1]}  
\end{align}
for $\ell \ge 1$.
Here, the Liouvillian $\tilde{\mathfrak{L}}_{i[\ell]}$ is constructed by replacing $\mA_{i,a}(H,t)$ in the definition of CKG Liouvillian by local operators  $\tilde{\tr}_{i[\ell]^\co}\brr{\mA_{i,a}(H,t)}$ as in~\eqref{lem:quasi-locality_mL/main}. 

As an important notice, each of the decomposed Liouvillians $\{\delta \mathfrak{L}_{i[\ell]}\}_{\ell}$ is NOT given by the Lindblad form. 
Instead, we can only ensure that the following summation, 
\begin{align}
\label{Lind_blad_sum_r_Liou_0}
\sum_{\ell\le \ell_0}\delta \mathfrak{L}_{i[\ell]}  = \tilde{\mathfrak{L}}_{i[\ell_0]}, 
\end{align}
is Lindbladian for $\forall i\in \Lambda$ and $\ell_0 \in \mathbb{N}$ because $\tilde{\mathfrak{L}}_{i[\ell]}$ is Lindbladian. 
Using the above fact, we can prove the following lemma to ensure that the subset Liouvillian is given by the Lindblad form.
%This lemma plays an important role in proving the Lieb--Robinson bound below (see Lemma~\ref{lem:Liouvillian Lieb--Robinson bound}).
\begin{lemma} \label{lem:subset_Liouvillian}
Let us define the subset Liouvillian $\mL^{(H)}_{X_0}$ as 
\begin{align}
\mL^{(H)}_{X_0}= \sum_{i\in X_0} \sum_{\ell: i[\ell] \subseteq X_0} \delta \mathfrak{L}_{i[\ell]}  .
\end{align}
Then, the subset Liouvillian is still Lindbladian, and hence 
\begin{align}
\label{norm_decreasing_partial/Liouvillian/lem}
\norm{ e^{t \mL^{(H)}_{X_0}} }_{1\to1} \le 1   ,\quad \forall t\geq0.
\end{align}
\end{lemma}

\textit{Proof of Lemma~\ref{lem:subset_Liouvillian}.}
For the proof, we use that the condition $i[\ell] \subseteq X_0$ is satisfied for $i\in X_0$ and $\ell \le \dist_{i,X^c_0}-1$, and hence 
\begin{align}
\mL^{(H)}_{X_0}=\sum_{i\in X_0} \sum_{\ell \le \dist_{i,X_0^\co}-1}\delta \mathfrak{L}_{i[\ell]}  =\sum_{i\in X_0}\tilde{\mathfrak{L}}_{i[\dist_{i,X_0^\co}-1]} .
\end{align}
Because the Liouvillian in the form of Eq.~\eqref{Lind_blad_sum_r_Liou_0} is Lindbladian, we can ensure that 
$\mL^{(H)}_{X_0}$ is also Lindbladian. 
We thus prove the inequality~\eqref{norm_decreasing_partial/Liouvillian/lem}. 
This completes the proof. $\square$

{~}

\hrulefill{\bf [ End of Proof of Lemma~\ref{lem:subset_Liouvillian}]}

{~}

Second, we prove the quasi-locality of each of the local Liouvillian $\mathfrak{L}_i$: 
\begin{lemma} \label{lem:quasi-locality_mL}
Under the Lieb--Robinson bound in Lemma~\ref{lem:quasi-locality_Lieb_Robinson}, the CKG Liouvillian $\mathfrak{L}_{i,a}$ is approximated onto the region $i[\ell]$ with an error of 
\begin{align}
\norm{  \delta \mathfrak{L}_{i[\ell]}  }_{1\to1} \le \Theta(1) 
\br{\frac{\Theta(\beta l_H)}{\sqrt{\ell}}}^{\ell} \le \Theta(1)  e^{-\ell} , 
\label{mL_decay/func}
\end{align}
where we use the notation of Eq.~\eqref{Liouvillian_norm_simplification} for the Liouvillian's norm. 
Note that $e^{{\rm const.} (\beta l_H)^2}=\Theta(1)$ under the assumption $\beta \le \beta_c=1/(4gk)$ (see the statement of Theorem~\ref{thm_high_temp_BP}).
After a simple calculation, we also obtain 
\begin{align}
\sum_{\ell>r} \norm{  \delta \mathfrak{L}_{i[\ell]}  }_{1\to1} \le
 \Theta(1)  e^{-\ell}  .
\label{mL_decay/func_sum}
\end{align}
\end{lemma}

{~} \\

\textit{Proof of Lemma~\ref{lem:quasi-locality_mL}.}
The proof is the same as in Ref.~\cite{rouze2024efficient}, and hence, we only show the essence.
For simplicity, we estimate the quasi-locality of $\mB_{i,a}$ in Eq.~\eqref{mB_i_a_Def}, and the other terms in Eq.~\eqref{Dissipative Dynamics_Liuou_def} can be treated in the same way. 
Because of $\norm{A_{i,a}}=1$, the quasi-locality of $\mB_{i,a}$ is characterized by
\begin{align}
&\int_{-\infty}^\infty |b_1(t)| dt \int_{-\infty}^\infty |b_2(t')| \cdot
 \norm{ \mA_{i[r],a}^{(-\beta t \pm \beta t')} - \mA_{i[r-1],a}^{(-\beta t \pm \beta t')} } dt' \notag \\
&\le \int_{-\infty}^\infty |b_1(t)| dt \int_{-\infty}^\infty |b_2(t')| 
\min\brr{ 2, \mathcal{F}(r, |\beta t| + |\beta t'|) }  dt' ,
\end{align}
where $\mA_{i[r],a}^{(-\beta t \pm \beta t')}$ is the approximation of $\mA_{i,a}(H,\beta t + \beta t')$ onto the ball region $i[r]$. 

Now, the time-dependence of $\mathcal{F}(r, t)$ is given by $C (vl_H t/r)^r$ from Eq.~\eqref{Lieb_Robinson/func}.
We then need to estimate 
\begin{align}
&\int_{-\infty}^\infty |b_1(t)| dt \int_{-\infty}^\infty |b_2(t')| \min\brr{ 2, \mathcal{F}(r, |\beta t| + |\beta t'|) } dt' \notag \\
&\le 
C\br{\frac{\beta vl_H}{r}} ^r  \int_{-\infty}^\infty |b_1(t)|dt \int_{-\infty}^\infty |b_2(t')| \cdot |t+t'|^{r}dt' \notag \\
&\le 
C\br{\frac{\beta vl_H}{r}} ^r  \int_{-\infty}^\infty |b_1(t)|dt \int_{-\infty}^\infty |b_2(t')| \cdot 2^r \br{t^r+t'^r}dt' .
\end{align}
Finally, because $b_1(t)$ and $b_2(t')$ decays as $e^{-\Omega(t^2)}$ from the definition~\eqref{def_b_1_b_2_t}, we obtain 
\begin{align}
 \int_{-\infty}^\infty |b_1(t)|dt \int_{-\infty}^\infty |b_2(t')| \cdot 2^r \br{t^r+t'^r}dt' \le (\tilde{C}_b r)^{r/2} ,
\end{align}
where $\tilde{C}_b$ is a constant which does not depend on the length $r$ and the Hamiltonian parameters. 
By combining the above estimations, we derive the main inequality~\eqref{mL_decay/func}.  
$\square$

{~}

\hrulefill{\bf [ End of Proof of Lemma~\ref{lem:quasi-locality_mL}]}

{~}

%
%\begin{equation}
%\label{H'_definition}
%H' = H + v_{i_0}, 
%\end{equation}
%such that
%\begin{align}
%\label{v_i_0_definition}
%v_{i_0}= \sum_{Z:Z\ni i_0} v_Z , \quad  \norm{v_{i_0}} \le \sum_{Z:Z\ni i_0} \norm{v_Z} \le g_0 . 
%\end{align}

%\begin{equation}
%\label{H'_definition}
%H' = H + v_{i_0}, 
%\end{equation}
%where $v_{i_0}$ is a local interaction supported on $i_0[l_H]$ and given by
%\begin{align}
%\label{v_i_0_definition}
%v_{i_0}= \sum_{Z:Z\ni i_0} v_Z , \quad  \norm{v_{i_0}} \le \sum_{Z:Z\ni i_0} \norm{v_Z} \le g_0 . 
%\end{align}

\begin{lemma}[Perturbed Liouvillian]
\label{lem:Perturbed Liouvillian_locality}
Let us consider two Liouvillians $\mL^{(H)}$ and $\mL^{(H')}$ such that $H'= H + v_{i_0}$,
where $v_{i_0}$ is a local interaction supported on $i_0[l_H]$ and given by
\begin{align}
\label{v_i_0_definition}
v_{i_0}= \sum_{Z:Z\ni i_0} v_Z , \quad  \norm{v_{i_0}} \le \sum_{Z:Z\ni i_0} \norm{v_Z} \le g_0 . 
\end{align}
Then, by denoting each of the Liouvillians $\mL^{(H)}$ and $\mL^{(H')}$ as 
\begin{align}
\label{definition/_delta_mL_i_0}
\mL^{(H)} &= \sum_{i\in \Lambda}\mathfrak{L}_i ,\quad 
\mL^{(H')} = \sum_{i\in \Lambda}\mathfrak{L}'_i  ,
\end{align}
we have 
\begin{align}
\norm{\mathfrak{L}_i- \mathfrak{L}'_i}_{1\to 1} \le g_0 \Theta(1) e^{-r}  .
\label{lem:Perturbed Liouvillian_locality/main}
\end{align}
\end{lemma}

{~} \\

\textit{Proof of Lemma~\ref{lem:Perturbed Liouvillian_locality}.}
%We second estimate the second term in the RHS of \eqref{decomposition_delta_mL_i}. 
In order to estimate the closeness between $\mathfrak{L}_i$ and $\mathfrak{L}'_i$ in Eq.~\eqref{definition/_delta_mL_i_0}, 
we analyze the Liouvillian~\eqref{Dissipative Dynamics_Liuou_def}.
For this purpose, we generally consider 
\begin{align}
 \mA_{i,a}(H,t_1) O \mA_{i,a}(H, t_2) -  \mA_{i,a}(H',t_1) O  \mA_{i,a}(H', t_2) , 
\end{align}
where $O$ is chosen as $\rho$ or $\hat{1}$.
By defining 
\begin{align}
\label{def_tilde_u_i_0_decop}
e^{iH't} =e^{i(H+v_{i_0})t} =\mathcal{T} e^{i\int_0^t v_{i_0}(H,x)dx} e^{iHt}  =: \tilde{u}_{i_0}^{(t)} e^{iHt},
\end{align}
we have 
\begin{align}
\mA_{i,a}(H',t_1) O  \mA_{i,a}(H', t_2) 
&=  \tilde{u}_{i_0}^{(t_1)}  \mA_{i,a}(H,t_1 ) \tilde{u}_{i_0}^{(t_1)\dagger}  
O  \tilde{u}_{i_0}^{(t_2)}  \mA_{i,a}(H, t_2)   \tilde{u}_{i_0}^{(t_2)\dagger}  \notag \\
&= \brr{ \tilde{u}_{i_0}^{(t_1)} , \mA_{i,a}(H,t_1 )}\tilde{u}_{i_0}^{(t_1)\dagger} O  \tilde{u}_{i_0}^{(t_2)} 
\mA_{i,a}(H, t_2)   \tilde{u}_{i_0}^{(t_2)\dagger}    +  \mA_{i,a}(H,t_1 ) O \tilde{u}_{i_0}^{(t_2)} 
\mA_{i,a}(H,  t_2)   \tilde{u}_{i_0}^{(t_2)\dagger}  \notag \\
&=\brr{ \tilde{u}_{i_0}^{(t_1)} , \mA_{i,a}(H,t_1 )}\tilde{u}_{i_0}^{(t_1)\dagger} O  \tilde{u}_{i_0}^{(t_2)} 
\mA_{i,a}(H, t_2)   \tilde{u}_{i_0}^{(t_2)\dagger}    +  \mA_{i,a}(H,t_1 ) O\brr{ \tilde{u}_{i_0}^{(t_2)} ,
\mA_{i,a}(H,  t_2) }  \tilde{u}_{i_0}^{(t_2)\dagger}  \notag \\
&\quad + \mA_{i,a}(H,t_1 )O\mA_{i,a}(H,  t_2) .
\end{align}
We therefore derive 
\begin{align}
&\norm{\mA_{i,a}(H',t_1) O  \mA_{i,a}(H', t_2)- \mA_{i,a}(H,t_1 )O\mA_{i,a}(H,  t_2) } \notag\\ 
&\le \brrr{\norm{ \brr{ \tilde{u}_{i_0}^{(t_1)} , \mA_{i,a}(H,t_1 )} } + \norm{\brr{\mA_{i,a}(H,  t_2),\tilde{u}_{i_0}^{(t_2)} } }} \norm{\mA_{i,a} } \notag \\
&\le \int_0^{|t_1|} \norm{ \brr{ v_{i_0} (H,x) , \mA_{i,a}}} dx +  \int_0^{|t_2|} \norm{ \brr{ v_{i_0} (H,x) , \mA_{i,a}}} dx ,
\end{align}
where we use from Eq.~\eqref{def_tilde_u_i_0_decop} 
\begin{align}
\norm{ \brr{ \tilde{u}_{i_0}^{(t_0)} , \mA_{i,a}(H,t_0)} } 
&= \norm{ \brr{\mathcal{T} e^{i\int_0^{t_0} v_{i_0}(H,x)dx}, \mA_{i,a}(H,t_0)} }  \notag \\
&\le \int_0^{t_0} \norm{ \brr{ v_{i_0}(H,x), \mA_{i,a}(H,t_0)} }dx \notag \\
&= \int_0^{t_0} \norm{ \left(\brr{ v_{i_0}(H,x-t_0), \mA_{i,a}}\right)(H,t_0) }dx =
\int_0^{t_0} \norm{ \brr{ v_{i_0}(H,x), \mA_{i,a}} }dx  ,
\end{align}
which holds for an arbitrary $t_0$.  Note that $\norm{\brr{\mathcal{T} e^{i\int_0^{t_0} A_x dx}, B}} \le \int_0^{t_0} \norm{ \brr{A_x,B}} dx$ for arbitrary operators $\{A_x\}_{0\le x\le t_0}$ and $B$.
By relying on the similar analyses to Lemma~\ref{lem:quasi-locality_mL}, we can obtain 
\begin{align}
\label{error_between_L_i_L'_i}
\norm{\mathfrak{L}_i- \mathfrak{L}'_i }_{1\to1}  \le \norm{v_{i_0}}   
\Theta(1) e^{-r},
\end{align}
which gives the main inequality~\eqref{lem:Perturbed Liouvillian_locality/main} by using $\norm{v_{i_0}}\le g_0$ from Eq.~\eqref{v_i_0_definition}.
This completes the proof. $\square$

{~}

\hrulefill{\bf [ End of Proof of Lemma~\ref{lem:Perturbed Liouvillian_locality}]}

{~}

\subsection{Convergence to the steady state}  \label{Sec:Convergence to the steady state}

\subsubsection{Liouvillian gap}

As shown in Ref.~\cite{rouze2024efficient}, the Liouvillian has a spectral gap at high temperatures. 
\begin{lemma} [Theorem~1 in Ref.~\cite{rouze2024efficient}]\label{lemma:Liouvillian_gap}
There exists a threshold temperature $\beta_c=\Theta(1)$ such that the CKG Liouvillian is gapped\footnote{In Ref.~\cite{rouze2024efficient}, the explicit condition for $\beta$ is $\beta < \beta_c < 4/(gk)$.}.
In detail, the Liouvillian gap $\Delta$ is larger than or equal to $1/(2\sqrt{2}e^{1/4})$:
\begin{align}
\Delta \ge \frac{1}{2\sqrt{2}e^{1/4}} > \frac{1}{4}.
\end{align}
\end{lemma}

{\bf Remark.} The explicit parameter dependence of the threshold $\beta_c$ is determined by the gap condition in the perturbed frustration-free Hamiltonian~\cite{Michalakis2013}. 
%Also, in Ref.~\cite{rouze2024efficient}, the finite-range interacting systems are considered, but the generalization is straightforward as has been shown in  Appendix~\ref{appendix:Stability of the gap} (see Proosition~\ref{prop:hamiltonian_gap} and Corollary~\ref{corol_long_range_gap}).

\subsubsection{Convergence rate to a perturbed steady state}
We here consider two quantum Gibbs states $\rho_\beta=e^{\beta H}/Z_\beta$ and $\rho'_\beta=e^{\beta (H + v_{i_0})}/Z_\beta'$, where $v_{i_0}$ was defined in Eq.~\eqref{v_i_0_definition} as 
\begin{align}
v_{i_0}= \sum_{|Z|\le k} v_Z , \quad \sum_{|Z|\le k} \norm{v_Z} \le g_0 ,  
\end{align}
We then consider $\chi^2$ divergence, which is defined as 
\begin{align}
\label{chi_2_definieition}
\chi^2(\rho_\beta' ,\rho_\beta) =\tr\brr{ (\rho_\beta' - \rho_\beta) \Gamma_{\rho_\beta}^{-1} (\rho'_\beta - \rho_\beta)}  ,
\end{align}
where $ \Gamma_{\rho_\beta}^{-1} (X):= \rho^{-1/2}_\beta X \rho_\beta^{-1/2}$.
We then prove the following lemma:
\begin{lemma} \label{prop:chi_2_divergenece_bound}
For the $\chi^2$ divergence in Eq.~\eqref{chi_2_definieition}, we obtain the upper bound of   
\begin{align}
\label{main_ineq:prop:chi_2_divergenece_bound}
\chi^2(\rho_\beta' ,\rho_\beta) \le 2+2 e^{\beta g_0 + \beta g_0 /(1-2gk\beta)}   \le 4 e^{3\beta g_0} , 
\end{align}
where we use $\beta \le 1/(4gk)\Leftrightarrow 1-2gk\beta\geq\frac{1}{2}$ in the last inequality.  
\end{lemma}

Using the lemma, we immediately obtain the following corollary, which is derived from~\cite[Corollary~2 in the appendix]{rouze2024efficient} (see also~\cite{10.1063/1.4804995}):
\begin{corol} \label{corol:chi_2_divergenece_bound}
Let $\Delta$ be the spectral gap of the Liouvillian $\mL^{(H')}$.
Then, the convergence of the time-evolved operator $e^{\mL^{(H')} t} \rho_\beta$ to the steady state $\rho'_\beta$ is given by
\begin{align}
\label{corol:main_ineq:prop:chi_2_divergenece_bound}
\norm{ e^{\mL^{(H')} t} \rho_\beta - \rho'_\beta}_1 \le \chi^2(\rho_\beta' ,\rho_\beta)  e^{-t \Delta} \le 4e^{3\beta g_0-t/4}  .
\end{align}
\end{corol}

{\bf Remark.} From the corollary, we can ensure that the local perturbation to the quantum Gibbs state can be recovered by a short-time Liouville dynamics. 
At this stage, we emphasize that the dynamics $e^{\mL^{(H')} t}$ is not proven to be approximated by a local CPTP map around the perturbed site.  
This problem will be treated in Section~\ref{sec:Dynamics by the perturbed Liouvillian}.

\subsubsection{Proof of Lemma~\ref{prop:chi_2_divergenece_bound}}
We start with the inequality of
\begin{align}
\label{prop:chi_2_divergenece_bound_proof0}
\chi^2(\rho_\beta' ,\rho_\beta) 
=\tr\brr{ (\rho_\beta' - \rho_\beta) \Gamma_{\rho_\beta}^{-1} (\rho'_\beta - \rho_\beta)} 
&\le2 \norm{\Gamma_{\rho_\beta}^{-1} (\rho'_\beta - \rho_\beta)} \notag \\
&=2\norm{ \rho^{-1/2}_\beta (\rho'_\beta - \rho_\beta) \rho_\beta^{-1/2}} \notag \\
&\le 2+ 2\norm{ \rho^{-1/2}_\beta \rho'_\beta  \rho_\beta^{-1/2}} . 
\end{align}

We aim to estimate the upper bound of 
\begin{align}
\label{prop:chi_2_divergenece_bound_proof1}
\norm{ \rho^{-1/2}_\beta \rho'_\beta \rho_\beta^{-1/2}} 
= \frac{Z_\beta}{Z'_\beta} \norm{e^{-\beta H/2}e^{\beta (H+v_{i_0})} e^{-\beta H/2}} .
\end{align}
First, using the Golden-Thompson inequality, we obtain 
\begin{align}
\label{prop:chi_2_divergenece_bound_proof2}
Z_\beta = \tr \br{e^{\beta H}} \le  \tr \br{e^{\beta v_{i_0}}e^{\beta (H+v_{i_0}) }}  \le  \tr \br{e^{\beta g_0 }e^{\beta (H+v_{i_0})}} = Z'_\beta e^{\beta g_0} .
\end{align} 
Second, we expand
\begin{align}
e^{\beta (H+v_{i_0})/2} &= \mathcal{T} e^{- \int_0^{\beta/2} e^{x H} v_{i_0} e^{-x H}dx} e^{\beta H/2} ,
\end{align}
which yields
\begin{align}
e^{\beta (H+v_{i_0})/2}e^{-\beta H/2} &= \mathcal{T} e^{- \int_0^{\beta/2} e^{x H} v_{i_0} e^{-x H}dx} .
\end{align}
By applying the above upper bounds to Eq.~\eqref{prop:chi_2_divergenece_bound_proof1}, we derive 
\begin{align}
\label{prop:chi_2_divergenece_bound_proof1.5}
\norm{ \rho^{-1/2}_\beta \rho'_\beta \rho_\beta^{-1/2}}
\le e^{\beta g_0}  \norm{\mathcal{T} e^{- \int_0^{\beta/2} e^{x H} v_{i_0} e^{-x H}dx}}^2 .
\end{align}

We use the above form to obtain the upper bound of 
\begin{align}
\label{prop:chi_2_divergenece_bound_proof3}
\norm{e^{\beta (H+v_{i_0})/2}e^{-\beta H/2} }
& \le e^{\int_0^{\beta/2} \norm{e^{x H} v_{i_0} e^{-x H}}dx} \le e^{(\beta g_0/2) /(1-gk\beta)} ,
\end{align}
where we use $\sum_Z\norm{v_Z}\le g_0$ from Eq.~\eqref{v_i_0_definition} and the following inequality
\begin{align}
\norm{ e^{-x H} v_Z e^{x H}} &\le \sum_{m=0}^\infty \frac{x^m}{m!} \norm{\ad_{H}^m (v_Z)}  \notag \\
&\le \norm{v_Z} \sum_{m=0}^\infty \frac{x^m}{m!} (2gk)^mm!=\frac{1}{1-2gkx}  \norm{v_Z} . 
\end{align} 
Here, the upper bound $\norm{\ad_{H}^m (v_Z)}\le  \norm{v_Z} (2gk)^m m!$ for $|Z|\le k$ is derived in Ref.~\cite[Lemma~3]{KUWAHARA201696}.
By applying the inequality~\eqref{prop:chi_2_divergenece_bound_proof3} to Eq.~\eqref{prop:chi_2_divergenece_bound_proof1.5} and using~\eqref{prop:chi_2_divergenece_bound_proof0}, we prove the desired inequality~\eqref{main_ineq:prop:chi_2_divergenece_bound}.
This completes the proof. $\square$

\section{Why Lindblad dynamics is required?}\label{Sec:Why Lindblad dynamics is required}

In this section, we discuss the possibility of using the purified dynamics to construct the BP channel.  
In conclusion, the purified dynamics can NOT be used for our purpose.
Using the purification of the quantum state, the quantum Gibbs state is given by 
\begin{align}
\label{ket_e_beta_H}
\ket{e^{\beta H}} :=e^{\beta H/2} \ket{\Phi_{\Lambda,\Lambda'}}  ,
\end{align}
where $ \ket{\Phi_{\Lambda,\Lambda'}}$ is the maximally entangled state between the total system $\Lambda$ and the copied total system $\Lambda'$. 
The above state gives 
\begin{align}
\tr_{\Lambda'} \br{\ket{e^{\beta H}} \bra{e^{\beta H}} } = e^{\beta H}.
\end{align}

As a convenient property of the purified state~\eqref{ket_e_beta_H}, it has a quasi-local parent Hamiltonian which has a constant spectral gap above a temperature threshold~\cite[Supplemetnary Theorem~14]{PhysRevLett.116.080503}. 
Then, by using the quasi-adiabatic continuation technique~\cite{PhysRevB.72.045141} with the Lieb--Robinson bound~\cite{PhysRevLett.97.050401}, we can easily derive
\begin{align}
\label{local_perturbation_local_uni}
\ket{e^{\beta (H_0+V_{i_0})}} =  U_{i_0[r]} \ket{e^{\beta H_0}}  + e^{-\Omega(r)},
\end{align}
where we assume the exponentially decaying interaction and $U_{i_0[r]}$ is constructed from the adiabatic continuation operators acting on $\Lambda \cup \Lambda'$. 
At first glance, this allows us to construct the CPTP map $\tau^{(1)}_{B_1B_2}$ in Eq.~\eqref{def:tau_1_B1_B2}.
Indeed, this formalism helps to efficiently prepare the high-temperature quantum Gibbs state on a quantum computer. 

However, it is not helpful for our purpose, i.e., construction of the CPTP map $\tau^{(1)}_{B_1B_2}$.
To see the point, following Eq.~\eqref{local_perturbation_local_uni}, we construct a unitary operator $U_{B_1B_2, B_1'B_2'}$ such that 
\begin{align}
\ket{e^{\beta (H_{AB_1} +H_{B_2C})}}  \approx U_{BB'}   \ket{e^{\beta H}}
\end{align}
with an approximation error of $e^{-\Omega(r)}$. 
By taking the trace of the copy system $\Lambda'$, we have 
\begin{align}
e^{\beta (H_{AB_1} +H_{B_2C})} \approx  \tr_{\Lambda'} \br{U_{BB'}   \ket{e^{\beta H}} \bra{e^{\beta H}}  U_{BB'}^\dagger } .
\end{align}
Then, can we prove the following relation using a CPTP map $\tau_B$ on the subset $B$? 
\begin{align}
\label{tau_rho_beta_AB}
 \tr_{\Lambda'} \br{U_{BB'}   \ket{e^{\beta H}} \bra{e^{\beta H}}  U_{BB'}^\dagger } \overset{?}{=} 
 \tau_B\br{e^{\beta H}}
\end{align}

On this point, we can consider a counterexample.
In general, one can consider the CPTP map $\tau_{L_1 L_2}$ such that 
\begin{align}
\tau_{L_1 L_2} \br{\rho_{L_1L_2}} : = \tr_{L_3} \br{U_{L_2L_3} \rho_{L_1L_2L_3}U_{L_2L_3}^\dagger} ,
\end{align}
where, in Eq.~\eqref{tau_rho_beta_AB}, we let $L_1 \to AC$, $L_2 \to B$ and $L_3\to \Lambda'$. 
Our problem is whether we can reduce the CPTP map $\tau_{L_1 L_2}$ to a local form $\tau_{L_2}$. 
We here consider the three-qubits systems where $\rho_{L_1L_2L_3}$ is given by $2^{-1/2}(\ket{000}+\ket{111})$ and $U_{L_2L_3}$ be the CNOT operation between $L_2$ and $L_3$, which makes $U_{L_2L_3} 2^{-1/2}(\ket{000}+\ket{111})=2^{-1/2}(\ket{00}+\ket{11}) \otimes \ket{0}$.
Hence, we have 
\begin{align}
 \tr_{L_3} \br{U_{L_2L_3} \rho_{L_1L_2L_3}U_{L_2L_3}^\dagger} = \frac{1}{2} (\ket{00}+\ket{11})(\bra{00}+\bra{11}) ,
\end{align}
which is the Bell state.
On the other hand, the state $\rho_{L_1L_2}$ is given by zero entangled state as $(\ket{00}\bra{00} + \ket{11}\bra{11})/2$. 
Therefore, because the local CPTP map $\tau_{L_2}$ cannot create entanglement,  the map $\tau_{L_1 L_2}$ from $(\ket{00}\bra{00} + \ket{11}\bra{11})/2$ to $\frac{1}{2} (\ket{00}+\ket{11})(\bra{00}+\bra{11}) $ cannot be reduced to the local form $\tau_{L_2}$. 

Therefore, for our purpose, it is necessary to work directly with Lindblad dynamics rather than purification-based approaches.

\section{Multi-derivative of the operator logarithm} \label{app:Multi-derivative of the operator logarithm}

The purpose of this appendix is to show the explicit form of the multi-derivative appearing in Eq.~\eqref{parameterized_Hamiltonian_log}, that is, 
\begin{align}
\label{parameterized_Hamiltonian_log_re}
\tilde{H}_{\vec{a}} := \log\brr{ \tr_{L^\co} \br{e^{-\beta H_{\vec{a}}} } } = \sum_{m=0}^\infty \frac{(-\beta)^m}{m!} \frac{d^m}{d\vec{a}^m} \log\brr{ \tr_{L^\co} \br{e^{-\beta H_{\vec{a}}} } } ,
\end{align}
where $L\subset \Lambda$ is arbitrarily chosen.
Note that we adopt the standard quantum Gibbs state $e^{-\beta H}$, rather than $e^{\beta H}$, for consistency with the notation in Ref.~\cite{PhysRevLett.124.220601}.
In the following, we parameterize the Hamiltonian in the form of 
\begin{align}
\label{parameterized_Hamiltonian_re}
H_{\vec{a}}= \sum_{s} a_s h_s ,
\end{align}
where each of $\{h_s\}_s$ denotes an interaction operator involving at most $\orderof{1}$ sites. 

In the case where we take the trace operation $\tr\br{\cdots}$ instead of the partial trace $\tr_{L^\co}\br{\cdots}$, one can efficiently compute the multi-derivative~\cite[Proposition~2 therein]{KUWAHARA2020168278} and compute its upper bound to ensure the convergence of the cluster expansion~\cite{PRXQuantum.4.020340}:
\begin{lemma}[Proposition~2 in Ref.~\cite{KUWAHARA2020168278}] \label{lem:L=emptyset}
Let us assume $L=\emptyset$. 
We here take additional $m-1$ copies of the total Hilbert space $\mathcal{H}$ and distinguish them by $\{\mathcal{H}_j\}_{j=1}^m$. 
Then, we define the extended Hilbert space as $\mathcal{H}_{1:m}$ with
\begin{align}
\mathcal{H}_{1:m}:=\mathcal{H}_{1} \otimes \mathcal{H}_{2} \otimes \cdots \otimes\mathcal{H}_{m}.  \label{copy_hilbert_space_log_0}
\end{align}
For an arbitrary operator $O\in\mathcal{H}$, 
we extend the domain of definition and denote $O_{\mathcal{H}_s}\in \ban( \mathcal{H}_{1:m})$ by the operator which non-trivially acts only on the space $ \mathcal{H}_s$.
Now, for an arbitrary set $w=\{a_{s_1},a_{s_2},\ldots,a_{s_m}\}$, we have 
\begin{align}
 &\Der_{w}\log \left[ \tr (e^{-\beta H_{\vec{a}}}/\mD_{\Lambda}) \right]  \bigl|_{\vec{a}=\vec{0}} =\frac{(-\beta)^m}{\mD_{\Lambda}^{m}}  \mathcal{P}_m \tr_{\Lambda_{1:m}} \left ( h_{s_1}^{(0)} h_{s_2}^{(1)}\cdots h_{s_m}^{(m-1)} \right)  \label{simple_expression_der_w_log_0}
\end{align}
with $\mD_{\Lambda}$ the Hilbert space dimension on the total system $\Lambda$.
Here, $\tr_{\Lambda_{1:m}} $ denotes the trace with respect to the Hilbert space $\mathcal{H} _{1:m}$ and we define
\begin{align}
O^{(0)} :=O_{\mathcal{H}_1} ,\quad O^{(s)} := O_{\mathcal{H}_1} + O_{\mathcal{H}_2} + \cdots +O_{\mathcal{H}_{s}} -  s O_{\mathcal{H}_{s+1}} \label{Def:tilde_O_s_0}
\end{align}
for $s=1,2, \ldots,m$.
Finally, $\mathcal{P}_m$ means the symmetrization operator as
\begin{align}
 \mathcal{P}_m  h_{s_1}^{(0)} h_{s_2}^{(1)}\cdots h_{s_m}^{(m-1)}
 =\frac{1}{m!}\sum_{\sigma} h_{s_{\sigma_1}}^{(0)} h_{s_{\sigma_2}}^{(1)} \cdots h_{s_{\sigma_m}}^{(m-1)} ,  \label{Def:math_P_m_0}
\end{align}
where $\sum_\sigma$ denotes the summation of $m!$ terms which come from all the permutations. 
\end{lemma}

\subsection{Difficulty in the partial trace}

In Ref.~\cite[Supplementary Proposition~3]{PhysRevLett.124.220601}, Lemma~\ref{lem:L=emptyset} is generalized to arbitrary $L\neq \emptyset$. 
As a natural generalization, the following notations are utilized:
\begin{definition}[Extended Hilbert space] \label{partial_trace_space}
We here take additional $m-1$ copies of the partial Hilbert space $\mathcal{H}^{L^\co}$ and distinguish them by $\{\mathcal{H}^{L^\co}_j\}_{j=1}^m$. 
Then, we define the extended Hilbert space as $\mathcal{H}^L \otimes \mathcal{H}^{L^\co} _{1:m}$ with
\begin{align}
\mathcal{H}^{L^\co} _{1:m}:=\mathcal{H}^{L^\co}_{1} \otimes \mathcal{H}^{L^\co}_{2} \otimes \cdots \otimes\mathcal{H}^{L^\co}_{m}.  \label{copy_hilbert_space_log}
\end{align}
For an arbitrary operator $O\in\mathcal{H}$, 
we extend the domain of definition and denote $O_{\tilde{\mathcal{H}}_s}\in \ban( \mathcal{H}^L \otimes \mathcal{H}^{L^\co} _{1:m})$ by the operator which non-trivially acts only on the space $\mathcal{H}^L \otimes \mathcal{H}_s^{L^\co}$.
We also redefine the notations of $\{O^{(s)}\}_s$ as follows:
\begin{align}
O^{(0)} :=O_{\tilde{\mathcal{H}}_1} ,\quad O^{(s)} := O_{\tilde{\mathcal{H}}_1} + O_{\tilde{\mathcal{H}}_2} + \cdots +O_{\tilde{\mathcal{H}}_{s}} -  s O_{\tilde{\mathcal{H}}_{s+1}} \label{Def:tilde_O_s}
\end{align}
for $s=1,2, \ldots,m$.

We denote the Hilbert space dimension on $L^\co$ by $d_{L^\co}$. 
Moreover, $\tr_{L^\co_{1:m}}(\cdots)$ is defined the partial trace with respect to the Hilbert space $\mathcal{H}^{L^\co} _{1:m}$;  
that is, for an arbitrary operator $\Phi$ defined on $\mathcal{H}^L \otimes \mathcal{H}^{L^\co} _{1:m}$, one can ensure 
\begin{align}
\tr_{L^\co_{1:m}}( \Phi ) \in  \ban( \mathcal{H}^L ) .
\end{align}
\end{definition}

Using the above notations, the authors in Ref.~\cite[Supplementary Proposition~3]{PhysRevLett.124.220601} gave the same equation as Eq.~\eqref{simple_expression_der_w_log_0} for $L\neq \emptyset$, which turned out to be not justified in general.  
The authors compared two expansions~\cite[Supplementary Ineqs.~(S.49) and (S.50)]{PhysRevLett.124.220601}.
The first one is about $\log \left[ \tr_{L^\co}  (e^{-\beta H_{\vec{a}}}/\mD_{L^\co}) \right] \Bigl|_{\beta=0}$, which is directly given using the Taylor expansion as follows:
\begin{align}
& \frac{\partial^m}{\partial \beta^m} \log \left[ \tr_{L^\co}  (e^{-\beta H_{\vec{a}}}/\mD_{L^\co}) \right] \Bigl|_{\beta=0}  \notag \\
 =& \sum_{q=1}^m\frac{(-1)^{q-1}}{q}\sum_{\substack{m_1+m_2+\cdots+ m_q=m\\ m_1\ge1,m_2\ge1,\ldots,m_q\ge1} } \frac{m! (-1)^{m}}{m_1!m_2!\cdots m_q!} 
\frac{\mathcal{P}_q \tr_{L^\co}(H_{\vec{a}}^{m_1}) \tr_{L^\co}(H_{\vec{a}}^{m_2})   \cdots \tr_{L^\co}(H_{\vec{a}}^{m_q})}{q!\mD_{L^\co}^q}  ,\label{beta_m_term_complex_re}
\end{align}
where $\mathcal{P}_q$ is the symmetrization operator with respect to $\{m_1,m_2,\ldots,m_q\}$.
The second expansion is about $\tr_{L_{1:m}^\co}\left( H_{\vec{a}}^{(0)}H_{\vec{a}}^{(1)}\cdots H_{\vec{a}}^{(m-1)}  \right)$, which was supposed to decomposed in the form of
\begin{align}
&\frac{(-1)^{m}}{\mD_{L^\co}^{m}}  \tr_{L_{1:m}^\co}\left( H_{\vec{a}}^{(0)}H_{\vec{a}}^{(1)}\cdots H_{\vec{a}}^{(m-1)}  \right)  \notag \\
=&\sum_{q=1}^m \sum_{\substack{m_1+m_2+\cdots+ m_q=m\\ m_1\ge1,m_2\ge1,\ldots,m_q\ge1} } C^{(q)}_{m_1,m_2,\ldots,m_q} \mathcal{P}_q \tr_{L^\co}(H_{\vec{a}}^{m_1}) \tr_{L^\co}(H_{\vec{a}}^{m_2})   \cdots \tr_{L^\co}(H_{\vec{a}}^{m_q}) ,
\label{derivative_log_beta_general_proof_re}
\end{align}
where $C^{(q)}_{m_1,m_2,\ldots,m_q}$ is an appropriate coefficient calculated from the definition~\eqref{Def:tilde_O_s}.

The problem is that Eq.~\eqref{derivative_log_beta_general_proof_re} is NOT correct because the operators $H_{\vec{a},\tilde{\mathcal{H}}_s}$ and $H_{\vec{a},\tilde{\mathcal{H}}_{s'}}$ do not commute with each other unless $L=\emptyset$. 
For example, we can obtain for $m=3$
\begin{align}
\label{tilde_H_0_to_3}
\tr_{L_{1:3}^\co}\left( H_{\vec{a}}^{(0)}H_{\vec{a}}^{(1)}H_{\vec{a}}^{(2)}  \right) 
&= \tr_{L_{1:3}^\co}\left( H_{\vec{a},\tilde{\mathcal{H}}_1}^3  \right) 
-\tr_{L_{1:3}^\co}\left( H_{\vec{a},\tilde{\mathcal{H}}_1}H_{\vec{a},\tilde{\mathcal{H}}_2}H_{\vec{a},\tilde{\mathcal{H}}_1}  \right) 
+\tr_{L_{1:3}^\co}\left( H_{\vec{a},\tilde{\mathcal{H}}_1}^2H_{\vec{a},\tilde{\mathcal{H}}_2}  \right)  \notag \\
&
-\tr_{L_{1:3}^\co}\left( H_{\vec{a},\tilde{\mathcal{H}}_1}H_{\vec{a},\tilde{\mathcal{H}}_2}^2  \right) 
-2\tr_{L_{1:3}^\co}\left( H_{\vec{a},\tilde{\mathcal{H}}_1}^2H_{\vec{a},\tilde{\mathcal{H}}_3} \right) 
+2\tr_{L_{1:3}^\co}\left( H_{\vec{a},\tilde{\mathcal{H}}_1} H_{\vec{a},\tilde{\mathcal{H}}_2} H_{\vec{a},\tilde{\mathcal{H}}_3}  \right)  \notag \\
&=
\mD_{L^\co}^2\tr_{L^\co}(H_{\vec{a}}^3) 
-\tr_{L_{1:3}^\co}\left( H_{\vec{a},\tilde{\mathcal{H}}_1}H_{\vec{a},\tilde{\mathcal{H}}_2}H_{\vec{a},\tilde{\mathcal{H}}_1}  \right) 
+\mD_{L^\co}\tr_{L^\co}(H_{\vec{a}}^2) \tr_{L^\co}(H_{\vec{a}}) \notag \\
&-\mD_{L^\co}\tr_{L^\co}(H_{\vec{a}}) \tr_{L^\co}(H_{\vec{a}}^2) 
-2\mD_{L^\co}\tr_{L^\co}(H_{\vec{a}}^2) \tr_{L^\co}(H_{\vec{a}}) 
+2\tr_{L^\co}(H_{\vec{a}}) \tr_{L^\co}(H_{\vec{a}}) \tr_{L^\co}(H_{\vec{a}})  ,
\end{align}
To reduce the above equation to the form of Eq.~\eqref{derivative_log_beta_general_proof_re}, we need the following conditions, which cannot be satisfied in general;
\begin{align}
\label{cond_1_revise}
\tr_{L_{1:3}^\co}\left( H_{\vec{a},\tilde{\mathcal{H}}_1}H_{\vec{a},\tilde{\mathcal{H}}_2}H_{\vec{a},\tilde{\mathcal{H}}_1}  \right)  
\overbrace{=}^{\textrm{not satisfied!}}\tr_{L_{1:3}^\co}\left( H_{\vec{a},\tilde{\mathcal{H}}_1}^2 H_{\vec{a},\tilde{\mathcal{H}}_2}\right)  
= \mD_{L^\co}\tr_{L^\co}(H_{\vec{a}}^2) \tr_{L^\co}(H_{\vec{a}}) 
\end{align}
and 
\begin{align}
\label{cond_2_revise}
\tr_{L^\co}(H_{\vec{a}}^2) \tr_{L^\co}(H_{\vec{a}})  \overbrace{=}^{\textrm{not satisfied!}} \tr_{L^\co}(H_{\vec{a}}) \tr_{L^\co}(H_{\vec{a}}^2) .
\end{align}

\subsection{Ordering operator and symmetrizing operator}

To resolve the error, we have to modify Lemma~\ref{lem:L=emptyset} so that we can utilize the conditions~\eqref{cond_1_revise} and \eqref{cond_2_revise}. 
We here define two super-operators $\mathcal{W}_O$ and $\mathcal{W}_S$.

First, the super-operator $\mathcal{W}_O$ puts the operators in the same Hilbert space together.
For arbitrary operators $\{O_{1,\tilde{\mathcal{H}}_{i_1}},O_{2,\tilde{\mathcal{H}}_{i_2}},\ldots,O_{m,\tilde{\mathcal{H}}_{i_m}}\}$ with $i_1,i_2,\ldots,i_m \in [1,q]$, the super-operator $\mathcal{W}_O$ acts as
\begin{align}
\label{cond_1_super}
&\mathcal{W}_O O_{1,\tilde{\mathcal{H}}_{i_1}} O_{2,\tilde{\mathcal{H}}_{i_2}} \cdots O_{m,\tilde{\mathcal{H}}_{i_m}}
=O_{\tilde{\mathcal{H}}_1}^{(1)} \cdots O_{\tilde{\mathcal{H}}_q}^{(q)} \notag \\
&O_{\tilde{\mathcal{H}}_s}^{(s)} = \overline{O_{i_1, \tilde{\mathcal{H}}_{s}}  O_{i_2, \tilde{\mathcal{H}}_{s}}\cdots  O_{i_k, \tilde{\mathcal{H}}_{s}}}  \for s=1,2,\ldots,q ,
\end{align}
where $\overline{O_1O_2\cdots O_m}$ means the symmetrization of the operators, e.g., $\overline{O_1O_2}=(O_1O_2+O_2O_1)/2!$, 
$\overline{O_1O_2O_3}=(O_3O_1O_2+O_1O_3O_2+O_1O_2O_3+O_3O_2O_1+O_2O_3O_1+O_2O_1O_3)/3!$, and so on.
Note that we have 
\begin{align}
\label{overline_notation}
\overline{\overline{O_1O_2\cdots O_m}  \ \overline{O_{m+1}O_{m+2}\cdots O_n}}=\overline{O_1O_2\cdots O_n}. 
\end{align}
By applying $\mathcal{W}_O$ to~\eqref{cond_1_revise}, we have 
\begin{align}
\tr_{L_{1:3}^\co}\left( \mathcal{W}_O H_{\vec{a},\tilde{\mathcal{H}}_1}H_{\vec{a},\tilde{\mathcal{H}}_2}H_{\vec{a},\tilde{\mathcal{H}}_1}  \right)  
= \tr_{L_{1:3}^\co}\left( \overline{H_{\vec{a},\tilde{\mathcal{H}}_1}^2} \overline{ H_{\vec{a},\tilde{\mathcal{H}}_2}}\right) 
= \tr_{L_{1:3}^\co}\left( H_{\vec{a},\tilde{\mathcal{H}}_1}^2 H_{\vec{a},\tilde{\mathcal{H}}_2}\right)   , \label{cond_rec_1}
\end{align}
which resolves the first problem~\eqref{cond_1_revise}. 

Second, we define $\mathcal{W}_S$ as a superoperator that takes the average for all the patterns of the swapping of the Hilbert spaces $\{\tilde{\mathcal{H}}_s\}$:
\begin{align}
\label{cond_2_super}
\mathcal{W}_SO_{1,\tilde{\mathcal{H}}_1} \cdots O_{q,\tilde{\mathcal{H}}_q}
=\frac{1}{q!} \sum_{\sigma} O_{\sigma(1),\tilde{\mathcal{H}}_{\sigma(1)}} \cdots O_{\sigma(q),\tilde{\mathcal{H}}_{\sigma(q)}} ,
\end{align}
where the summation takes all the permutations $\sigma$ for $\{1,2,\ldots,q\}$.
By applying $\mathcal{W}_S$ to Eq.~\eqref{cond_rec_1}, we have 
\begin{align}
 \tr_{L_{1:3}^\co}\left(\mathcal{W}_S H_{\vec{a},\tilde{\mathcal{H}}_1}^2 H_{\vec{a},\tilde{\mathcal{H}}_2}\right)
&=\frac{1}{2}  \tr_{L_{1:3}^\co}\left(H_{\vec{a},\tilde{\mathcal{H}}_1}^2 H_{\vec{a},\tilde{\mathcal{H}}_2}\right) + \frac{1}{2} \tr_{L_{1:3}^\co}\left(H_{\vec{a},\tilde{\mathcal{H}}_2}H_{\vec{a},\tilde{\mathcal{H}}_1}^2\right)   \notag \\
&= \frac{1}{2} \tr_{L^\co}(H_{\vec{a}}^2) \tr_{L^\co}(H_{\vec{a}}) +\frac{1}{2}  \tr_{L^\co}(H_{\vec{a}}) \tr_{L^\co}(H_{\vec{a}}^2) , 
\end{align}
which resolves the second problem in \eqref{cond_2_revise}.
We here note that these super-operators satisfy the linearity condition, i.e., 
\begin{align}
\label{Linearlity_W_S_WO}
\mathcal{W}_S\mathcal{W}_O (A+B) = \mathcal{W}_S\mathcal{W}_O A+\mathcal{W}_S\mathcal{W}_O B \quad {\rm and} \quad 
 \mathcal{W}_S\mathcal{W}_O (a A) =  a \mathcal{W}_S\mathcal{W}_O (A)  \quad (a\in \mathbb{C}) 
\end{align}
for arbitrary operators $A$ and $B$ in the form of $O_{1,\mathcal{H}_{i_1}} O_{2,\mathcal{H}_{i_2}} \cdots O_{m,\mathcal{H}_{i_m}}$.   

Therefore, by combining $\mathcal{W}_O$ and $\mathcal{W}_S$, we reduce Eq.~\eqref{tilde_H_0_to_3} to
\begin{align}
&\tr_{L_{1:3}^\co}\left(\mathcal{W}_S\mathcal{W}_O H_{\vec{a}}^{(0)}H_{\vec{a}}^{(1)}H_{\vec{a}}^{(2)}  \right)  \notag \\
&= \tr_{L_{1:3}^\co}\left(\mathcal{W}_S H_{\vec{a},\tilde{\mathcal{H}}_1}^3  \right) 
-\tr_{L_{1:3}^\co}\left( \mathcal{W}_SH_{\vec{a},\tilde{\mathcal{H}}_1}H_{\vec{a},\tilde{\mathcal{H}}_2}^2  \right) 
-2\tr_{L_{1:3}^\co}\left(\mathcal{W}_S H_{\vec{a},\tilde{\mathcal{H}}_1}^2H_{\vec{a},\tilde{\mathcal{H}}_3} \right) 
+2\tr_{L_{1:3}^\co}\left(\mathcal{W}_S H_{\vec{a},\tilde{\mathcal{H}}_1} H_{\vec{a},\tilde{\mathcal{H}}_2} H_{\vec{a},\tilde{\mathcal{H}}_3}  \right)  \notag \\
&=
\mD_{L^\co}^2\tr_{L^\co}(H_{\vec{a}}^3) 
- \frac{3}{2}\mD_{L^\co} \brr{ \tr_{L^\co}(H_{\vec{a}}^2) \tr_{L^\co}(H_{\vec{a}}) +\tr_{L^\co}(H_{\vec{a}}) \tr_{L^\co}(H_{\vec{a}}^2)} 
+2\tr_{L^\co}(H_{\vec{a}}) \tr_{L^\co}(H_{\vec{a}}) \tr_{L^\co}(H_{\vec{a}})  ,
\end{align}
which is equal to the terms in Eq.~\eqref{beta_m_term_complex_re} with $m=3$.
In this way, by inserting $\mathcal{W}_S\mathcal{W}_O $ to $\tr_{L_{1:m}^\co}\left( H_{\vec{a}}^{(0)}H_{\vec{a}}^{(1)}\cdots H_{\vec{a}}^{(m-1)}  \right)  $, we reduce Eq.~\eqref{derivative_log_beta_general_proof_re} to 
\begin{align}
&\frac{(-1)^{m}}{\mD_{L^\co}^{m}}  \tr_{L_{1:m}^\co}\left(\mathcal{W}_S\mathcal{W}_O H_{\vec{a}}^{(0)}H_{\vec{a}}^{(1)}\cdots H_{\vec{a}}^{(m-1)}  \right)  \notag \\
=&\sum_{q=1}^m \sum_{\substack{m_1+m_2+\cdots+ m_q=m\\ m_1\ge1,m_2\ge1,\ldots,m_q\ge1} } C^{(q)}_{m_1,m_2,\ldots,m_q} \mathcal{P}_q \tr_{L^\co}(H_{\vec{a}}^{m_1}) \tr_{L^\co}(H_{\vec{a}}^{m_2})   \cdots \tr_{L^\co}(H_{\vec{a}}^{m_q}) .
\label{derivative_log_beta_general_proof_re_modified}
\end{align}
By proving equivalence between the modified expansion~\eqref{derivative_log_beta_general_proof_re_modified} with Eq.~\eqref{beta_m_term_complex_re}, we prove the correct expression of the multi-derivative.
This equivalence can be proven in the same way as in Ref.~\cite{PhysRevLett.124.220601}, which utilized the equivalence in the case of $L=\emptyset$.  
We then prove the following lemma\footnote{In Section~\ref{appA}, we show an explicit formula for calculating $C^{(q)}_{m_1,m_2,\ldots,m_q}$ from Eq.~\eqref{derivative_log_beta_general_proof_re_modified} and numerically demonstrate that it indeed gives the same expression as in Eq.~\eqref{beta_m_term_complex_re}.}:
\begin{lemma} [Multi-derivative of the generalized cluster expansion]
Let us adopt the notations in Def.~\ref{partial_trace_space}. 
Then, using the super-operators $\mathcal{W}_S$ and $\mathcal{W}_O$ in Eqs.~\eqref{cond_1_super} and \eqref{cond_2_super}, respectively, we obtain 
\begin{align}
 &\Der_{w}\log \tilde{\rho}^{L}_{\vec{a}} \bigl|_{\vec{a}=\vec{0}} =\frac{(-\beta)^m}{\mD_{L^\co}^{m}} \mathcal{P}_m   \tr_{L^\co_{1:m}} \left (\mathcal{W}_S\mathcal{W}_Oh_{s_1}^{(0)} h_{s_2}^{(1)}\cdots h_{s_m}^{(m-1)} \right),  \label{simple_expression_der_w_log}
\end{align}
where $\mathcal{P}_m$ was defined as the symmetrization operator in Eq.~\eqref{Def:math_P_m_0}. 
\end{lemma} 

Unfortunately, the norm of the new expression~\eqref{simple_expression_der_w_log} cannot be upper-bounded in a simple way as in the case of $L\neq \emptyset$.  
The most straightforward estimation yields an upper bound of $\orderof{m!\beta^m}$ and breaks the convergence of the cluster expansion. 
If the Hamiltonian is commuting, a similar analysis to the case of $L=\emptyset$ is employed, and the convergence issue can be resolved~\cite{bluhm2024}. 
For general non-commuting Hamiltonians, we conjecture from the argument in Section~\ref{Seec:Divergence problem} that qualitative improvement is not impossible in principle.

\subsection{Calculation of coefficient $C_{m_1,...,m_q}^{(q)}$} \label{appA}
Here we show an explicit calculation of coefficient $C_{m_1,...,m_q}^{(q)}$ in Eq.~\eqref{derivative_log_beta_general_proof_re}. 
We begin with a calculation
\begin{align}
\label{eq.A1}
{\tilde H}_{\vec{a}}^{(0)}{\tilde H}_{\vec{a}}^{(1)}\ldots{\tilde H}_{\vec{a}}^{(m-1)}&=\left(H_{\vec{a},{\tilde {\mathcal H}}_1}\right)\left(H_{\vec{a},{\tilde {\mathcal H}}_1}-H_{\vec{a},{\tilde {\mathcal H}}_2}\right)\ldots\left(\sum_{i=1}^{m-1}H_{\vec{a},{\tilde {\mathcal H}}_i}-(m-1)H_{\vec{a},{\tilde {\mathcal H}}_m}\right)\notag \\
&=\sum_{\iota\in\Delta_m}c^{(m)}_\iota H_{\vec{a},{\tilde {\mathcal H}}_{i_0}}\cdots H_{\vec{a},{\tilde {\mathcal H}}_{i_{m-1}}},
\end{align}
where  $\iota=(i_0,i_1,...,i_{m-1})$ and $\Delta_m:=\{1\}\times\{1,2\}\times\{1,2,3\}\times\cdots\times\{1,2,...,m\}$. The coefficient $c^{(m)}_\iota$ is given by
\begin{equation}
c^{(m)}_\iota=\prod_{k=1}^{m-1}\left(1-(k+1)\delta_{i_k, k+1} \right).
\end{equation}
For instance, $c^{(3)}_{(1,2,3)}=(-1)\cdot(-2)=2$.

Let $N(\iota,k)$ be the number of $k\in[m]$ appearing in the sequence $\iota=(i_0,i_1,...,i_{m-1})$. By applying the ordering operator ${\mathcal W}_O$ to Eq.~\eqref{eq.A1}, we obtain
\begin{align}
{\mathcal W}_O{\tilde H}_{\vec{a}}^{(0)}{\tilde H}_{\vec{a}}^{(1)}\ldots{\tilde H}_{\vec{a}}^{(m-1)}&=\sum_{\iota\in\Delta_m}c^{(m)}_\iota {\mathcal W}_OH_{\vec{a},{\tilde {\mathcal H}}_{i_0}}...H_{\vec{a},{\tilde {\mathcal H}}_{i_{m-1}}}\notag \\
&=\sum_{\substack{{\bar m}_1+{\bar m}_2+\cdots+{\bar m}_m=m\\1\leq{\bar m}_1\leq m\\0\leq{\bar m}_2\leq m-1\\\vdots\\0\leq{\bar m}_m\leq1}}\left(\sum_{\substack{\iota\in\Delta_m\\ N(\iota,k)={\bar m}_k,\forall k}}c^{(m)}_\iota\right) H^{{\bar m}_1}_{\vec{a},{\tilde {\mathcal H}}_1}\cdots H^{{\bar m}_m}_{\vec{a},{\tilde {\mathcal H}}_m}.\label{eq:KKcoef1}
\end{align}

To further proceed, define $\{{\bar m}_i\}^\times$ as the sequence of all the nonzero elements of $\{{\bar m}_i\}=\{{\bar m}_1,...,{\bar m}_m\}$\footnote{For example, $\{1,0,2,3\}^\times$, $\{1,2,0,3\}^\times$ and $\{1,2,3,0\}^\times$ gives the same sequence $\{1,2,3\}$.}. 
Then, we denote the term $H^{{\bar m}_1}_{\vec{a},{\tilde {\mathcal H}}_1}\cdots H^{{\bar m}_m}_{\vec{a},{\tilde {\mathcal H}}_m}$ by 
\begin{align}
\label{tr_W_O_0}
H^{{\bar m}_1}_{\vec{a},{\tilde {\mathcal H}}_1}H^{{\bar m}_2}_{\vec{a},{\tilde {\mathcal H}}_2}\cdots H^{{\bar m}_m}_{\vec{a},{\tilde {\mathcal H}}_m}
=  H^{m_1}_{\vec{a},{\tilde {\mathcal H}}_{i_1}} H^{m_2}_{\vec{a},{\tilde {\mathcal H}}_{i_2}}\cdots H^{m_q}_{\vec{a},{\tilde {\mathcal H}}_{i_q}} ,
\end{align}
where $\{\bar{m}_{i_1},\bar{m}_{i_2},\ldots, \bar{m}_{i_q}\}= \{m_1,m_2,\ldots,m_q\}$ and $\bar{m}_s = 0$ for $s\notin\{i_1,i_2,\ldots,i_q\}$.
At this stage, the sum over $m_i$ cannot be simply taken due to the restriction following ${\bar m}_i$, i.e., $1\leq{\bar m}_1\leq m$, $0\leq{\bar m}_2\leq m-1$, ..., $0\leq{\bar m}_m\leq1$. 
By further applying ${\mathcal W}_S$ to Eq.~\eqref{tr_W_O_0} with the partial trace over the copies of $L^c$, we obtain 
\begin{align}
\label{tr_W_O_0_2}
\frac{(-1)^m}{d_{L^c}^m}{\rm tr}_{L_{1:m}^c}\br{{\mathcal W}_S H^{{\bar m}_1}_{\vec{a},{\tilde {\mathcal H}}_1}H^{{\bar m}_2}_{\vec{a},{\tilde {\mathcal H}}_2}\cdots H^{{\bar m}_m}_{\vec{a},{\tilde {\mathcal H}}_m}}
=\mathcal{P}_q  \frac{{\rm tr}_{L^\co} \br{H^{m_1}_{\vec{a}}}{\rm tr}_{L^\co} \br{H^{m_2}_{\vec{a}}}  \cdots {\rm tr}_{L^\co} \br{H^{m_q}_{\vec{a}}}}{q! d_{L^c}^q}  .
\end{align}
The sum over $m_i$ is now no longer restricted except $\sum_{i=1}^q m_i=m$ owing to the symmetrization.
Here, from Eq.~\eqref{eq:KKcoef1}, summing up the coefficients of $\{{\bar m}_i\}=\{{\bar m}_1,...,{\bar m}_m\}$ with $\{{\bar m}_i\}^\times=\{m_1,m_2,\ldots,m_q\}$ gives 
\begin{equation}
\label{tr_W_O_0_coeff}
{\tilde C}^{(q)}_{m_1,m_2,...,m_q}=\left(\sum_{\{\bar m_i\}^\times=\{m_i\}}\sum_{\substack{\iota\in\Delta_m\\ N(\iota,k)={\bar m}_k,\forall k}}c^{(m)}_\iota\right).
\end{equation}

By combining Eqs.~\eqref{eq:KKcoef1}, \eqref{tr_W_O_0_2} and \eqref{tr_W_O_0_coeff}, we obtain 
\begin{align}
&\frac{(-1)^m}{d_{L^c}^m}{\rm tr}_{L_{1:m}^c}\left({\mathcal W}_S{\mathcal W}_O{\tilde H}_{\vec{a}}^{(0)}...{\tilde H}_{\vec{a}}^{(m-1)}\right) \notag \\
&=(-1)^m\sum_{q=1}^m\sum_{m_1+\cdots+m_q=m}\brr{\frac{{\tilde C}^{(q)}_{m_1,m_2,...,m_q}}{q!} {\mathcal P}_q \frac{{\rm tr}_{L^\co} \br{H^{m_1}_{\vec{a}}}{\rm tr}_{L^\co} \br{H^{m_2}_{\vec{a}}}  \cdots {\rm tr}_{L^\co} \br{H^{m_q}_{\vec{a}}}}{d_{L^c}^q} } \notag \\
&=:\sum_{q=1}^m\sum_{m_1+\cdots+m_q=m} \tilde{\tilde{C}}^{(q)}_{m_1,m_2,...,m_q} {\mathcal P}_q{\rm tr}_{L^\co} \br{H^{m_1}_{\vec{a}}}{\rm tr}_{L^\co} \br{H^{m_2}_{\vec{a}}}  \cdots {\rm tr}_{L^\co} \br{H^{m_q}_{\vec{a}}}
\end{align}
with 
\begin{equation}
\tilde{\tilde{C}}^{(q)}_{m_1,m_2,...,m_q}=\frac{(-1)^m}{q!d_{L^c}^q} \sum_{\{\bar m_i\}^\times=\{m_i\}}\sum_{\substack{\iota\in\Delta_m\\ N(\iota,k)={\bar m}_k,\forall k}}\prod_{k=1}^{m-1}\left(1-(k+1)\delta_{i_k,k+1} \right),
\end{equation}
Finally, ${\mathcal P}_q{\rm tr}_{L^\co} \br{H^{m_1}_{\vec{a}}}{\rm tr}_{L^\co} \br{H^{m_2}_{\vec{a}}}  \cdots {\rm tr}_{L^\co} \br{H^{m_q}_{\vec{a}}}$ is invariant under the permutation of $\{m_i\}$, and hence we can replace the final form of the coefficient ${\mathcal C}^{(q)}_{m_1,m_2...,m_q}$ in Eq.~\eqref{derivative_log_beta_general_proof_re} with the symmetric coefficient
\begin{align}
\label{final1}
C^{(q)}_{m_1,m_2...,m_q}
&=\frac{1}{\mathcal{N}_{\sigma(\{m_i\})}}\sum_{\substack{\sigma(\{m_i\})\\ \sigma\in S_q}}\tilde{\tilde{C}}^{(q)}_{m_1,m_2,...,m_q} \notag \\
&=\frac{(-1)^m}{q!d_{L^c}^q}\frac{1}{\mathcal{N}_{\sigma(\{m_i\})}}\sum_{\substack{\sigma(\{m_i\})\\\sigma\in S_q}}\sum_{\{\bar m_i\}^\times=\{m_i\}}\sum_{\substack{\iota\in\Delta_m\\ N(\iota,k)={\bar m}_k,\forall k}}\prod_{k=1}^{m-1}\left(1-(k+1)\delta_{i_k,k+1} \right),
\end{align}
where $\sum_{\substack{\sigma(\{m_i\})\\\sigma\in S_q}}$ takes the summations for all the permutations of $\{m_i\}_{i=1}^q$ and 
$\displaystyle \mathcal{N}_{\sigma(\{m_i\})}:= \sum_{\substack{\sigma(\{m_i\}), \sigma\in S_q}}1$.  
Moreover, the argument in the main text of the paper shows that this coefficient matches the one in Eq.~\eqref{beta_m_term_complex_re}, thus it also holds that
\begin{equation}
\label{final2}
{\mathcal C}^{(q)}_{m_1,m_2...,m_q}=\frac{(-1)^{m+q-1}}{q\cdot q! d_{L^c}^q} \frac{m!}{m_1!m_2!...m_q!}.
\end{equation}

\subsubsection{List of ${\mathcal C}^{(q)}_{m_1,m_2...,m_q}$}
Here, we show some explicit values of ${\mathcal C}^{(q)}_{m_1,m_2...,m_q}$ for $m=3,4,5$. 
For simplicity, we multiply $d_{L^c}^q$ in the list.
Note that the coefficient ${\mathcal C}^{(q)}_{m_1,m_2...,m_q}$ is invariant under the permutation of $\{m_1,m_2...,m_q\}$.
We can quickly check that the two expressions~\eqref{final1} and \eqref{final2} give the same values. 

{~}\\

$m=3$\\
\begin{tabular}{lc} \hline
  $\{m_i\}$ & ${\mathcal C}^{(q)}_{m_1,m_2...,m_q}$ \\ \hline
   \{3\} & $-1$ \\
  \{1,2\}, \{2,1\} & $3/4$ \\
  \{1,1,1\} & $-1/3$ \\
\hline
\end{tabular}

{~}\\

$m=4$\\
\begin{tabular}{lc} \hline
  $\{m_i\}$ & ${\mathcal C}^{(q)}_{m_1,m_2...,m_q}$ \\ \hline
   \{4\} & $1$ \\
  \{1,3\}, \{3,1\} & $-1$ \\
  \{2,2\} & $-3/2$ \\
  \{1,1,2\},\{1,2,1\}, \{2,1,1\} & $2/3$ \\
  \{1,1,1,1\} & $-1/4$ \\
\hline
\end{tabular}

{~}\\

$m=5$\\
\begin{tabular}{lc} \hline
  $\{m_i\}$ & ${\mathcal C}^{(q)}_{m_1,m_2...,m_q}$ \\ \hline
   \{5\} & $-1$ \\
  \{1,4\},\{4,1\} & $\frac{5}{4}$ \\
  \{2,3\},\{3,2\} & $5/2$ \\
  \{1,1,3\},\{1,3,1\}, \{3,1,1\}  & $-\frac{10}{9}$ \\
  \{1,2,2\},\{2,1,2\}, \{2,2,1\} & $-5/3$ \\
  \{1,1,1,2\}, \{1,1,2,1\}, \{1,2,1,1\},\{2,1,1,1\} & $5/8$ \\
  \{1,1,1,1,1\} & $-1/5$ \\
\hline
\end{tabular}

{~}\\

\def\bibsection{\section*{References}}

\bibliography{Quantum_Markov_high.bib}

\end{document}